\documentclass[a4paper,12pt]{article}
\usepackage{a4wide}

\usepackage{epsfig}
\usepackage{cite}
\usepackage{amsmath}
\usepackage{amssymb}
\usepackage{feynmf}
\usepackage{graphicx}
\usepackage{rotate}
\usepackage{latexsym}
\usepackage{array}
\usepackage{verbatim}
\usepackage{rotating}
\usepackage{color}

\usepackage[utf8]{inputenc}

\def\lsim{\mathrel{\raise.3ex\hbox{$<$\kern-.75em\lower1ex\hbox{$\sim$}}}}
\def\gsim{\mathrel{\raise.3ex\hbox{$>$\kern-.75em\lower1ex\hbox{$\sim$}}}}

\newcommand{\Eq}[1]{Eq.~(\ref{#1})}

\newcommand{\be}{\begin{equation}}
\newcommand{\ee}{\end{equation}}

\newlength{\absize}
\def\lsim{\mathrel{\rlap{\raise 2.5pt \hbox{$<$}}\lower 2.5pt
\hbox{$\sim$}}}

\renewcommand{\Re}{\mbox{Re\thinspace}}
\renewcommand{\Im}{\mbox{Im\thinspace}}

\newcommand{\gev}{\text{ GeV}}
\newcommand*{\Ave}[1]{\mathinner{\left\langle{#1}\right\rangle}}

\allowdisplaybreaks

\begin{document}

\thispagestyle{empty}
\renewcommand{\thefootnote}{\fnsymbol{footnote}}
\pagestyle{plain}
\setlength{\baselineskip}{4ex}\par
\setcounter{footnote}{0}
\renewcommand{\thefootnote}{\alph{footnote}}
\newcommand{\preprint}[1]{%
\begin{flushright}
\setlength{\baselineskip}{3ex} #1
\end{flushright}}
\renewcommand{\title}[1]{%
\begin{center}
\LARGE #1
\end{center}\par}

\renewcommand{\thanks}[1]{\footnote{#1}}
\renewcommand{\abstract}[1]{%
\vspace{2ex}
\normalsize
\begin{center}
\centerline{\bf Abstract}\par
\vspace{2ex}
\parbox{\absize}{#1\setlength{\baselineskip}{2.5ex}\par}
\end{center}}


\title{Prospects for charged Higgs searches at the LHC} 

\vspace{0.5cm}
{\sc
A.G.~Akeroyd$^{1}$,
M.~Aoki$^{2}$,
A.~Arhrib$^{3,4}$,
L.~Basso$^{5}$, 
I.F.~Ginzburg$^{6}$, 
R.~Guedes$^{7}$,
J.~Hernandez-Sanchez$^{8}$,
K.~Huitu$^{9}$, 
T.~Hurth$^{10}$,
M.~Kadastik$^{11}$, 
S.~Kanemura$^{12}$,
K.~Kannike$^{11}$, 
W.~Khater$^{13}$,
M.~Krawczyk$^{14,}$\footnote{Corresponding authors: Maria.Krawczyk@fuw.edu.pl, Per.Osland@uib.no},
F.~Mahmoudi$^{15,16}$, 
S.~Moretti$^{1}$,
S.~Najjari$^{14}$, 
P.~Osland$^{17,a}$,
G.M.~Pruna$^{18}$, 
M.~Purmohammadi$^{17}$, 
A.~Racioppi$^{11}$,
M.~Raidal$^{11}$,
R.~Santos$^{19,20}$,
P.~Sharma$^{21}$,
D.~Soko\l owska$^{14}$, 
O.~St{\aa}l$^{22}$, 
K.~Yagyu$^{1}$,
E.~Yildirim$^{1}$
}
\vspace*{0.5cm}

\begin{center}
\noindent
{\sl \small
$^{1}$School of Physics and Astronomy, University of Southampton,
Highfield, Southampton SO17 1BJ, United Kingdom, \\
$^{2}$Institute for Theoretical Physics, Kanazawa University, Kanazawa 920-1192, Japan, \\
$^{3}$D\'epartement de Math\'ematique, Facult\'e des Sciences et
Techniques, Universit\'e Abdelmalek Essa\^adi, B.~416, Tangier,
Morocco, \\
$^{4}$LPHEA, Facult\'e des Sciences-Semlalia, B.P.~2390 Marrakesh,
Morocco, \\
$^{5}$CPPM, Aix-Marseille Universit\'e, CNRS-IN2P3, UMR 7346, 163
avenue de Luminy, 13288 Marseille Cedex 9, France, \\
$^{6}$Sobolev Inst. of Mathematics SB RAS and Novosibirsk University,
630090 Novosibirsk, Russia, \\
$^{7}$IHC, Instituto de Hist\'oria Contemporanea, FCSH - New
University of Lisbon, Portugal, \\
$^{8}$Facultad de Ciencias de la Electr\'onica, 
Benem\'erita Universidad Aut\'onoma de Puebla, \\ 
Apdo. Postal 542, C.P. 72570 Puebla, Puebla, M\'exico \\
and Dual C-P Institute of High Energy Physics, M\'exico, \\
$^{9}$Department of Physics, and Helsinki Institute of Physics,
P.O.Box 64 (Gustaf H\"allstr\"omin katu 2), FIN-00014 University of
Helsinki, Finland, \\
$^{10}$PRISMA Cluster of Excellence and Institute for Physics (THEP),
Johannes Gutenberg University, D-55099 Mainz, Germany, \\
$^{11}$National Institute of Chemical Physics and Biophysics, R\"avala
10, 10143 Tallinn, Estonia, \\
$^{12}$Department of Physics, University of Toyama, 3190 Gofuku,
Toyama 930-8555, Japan, \\
$^{13}$Department of Physics, Birzeit University, Palestine, \\
$^{14}$Faculty of Physics, University of Warsaw, 
   Pasteura 5, 02-093 Warsaw, Poland, \\
$^{15}$Univ Lyon, Univ Lyon 1, ENS de Lyon, CNRS, Centre de Recherche 
Astrophysique de Lyon UMR5574, F-69230 Saint-Genis-Laval, France, \\
$^{16}$Theoretical Physics Department, CERN, CH-1211 Geneva 23, Switzerland, \\
$^{17}$Department of Physics and Technology, University of Bergen, Postboks
7803, N-5020  Bergen, Norway, \\
$^{18}$Paul Scherrer Institute, CH-5232 Villigen PSI, Switzerland, \\
$^{19}$Centro de F\'isica Te\'{o}rica e Computacional, Faculdade de Ci\^encias,
Universidade de Lisboa, Campo Grande, Edif\'icio C8 1749-016 Lisboa, Portugal, \\
$^{20}$Instituto Superior de Engenharia de Lisboa - ISEL, 1959-007
Lisboa, Portugal, \\
$^{21}$Center of Excellence in Particle Physics (CoEPP),
The University of Adelaide, South Australia, \\
$^{22}$The Oskar Klein Centre, Department of Physics, Stockholm
University, SE-106 91 Stockholm, Sweden
}
\end{center}
\vspace{5mm}

\abstract{The goal of this report is to summarize the current situation and discuss possible search strategies for charged scalars, in non-supersymmetric extensions of the Standard Model at the LHC.
Such scalars appear in Multi-Higgs-Doublet models (MHDM), in particular in the popular Two-Higgs-Doublet model (2HDM), allowing for charged and additional neutral Higgs bosons. These models have the attractive property that electroweak precision observables are automatically in agreement with the Standard Model at the tree level.
For the most popular version of this framework, Model~II, a discovery of a charged Higgs boson remains challenging, since the parameter space is becoming very constrained, and the QCD background is very high.
We also briefly comment on models with dark matter which constrain the corresponding charged scalars that occur in these models.
The stakes of a possible discovery of an extended scalar sector are very high, and these searches should be pursued in all conceivable channels, at the LHC and at future colliders.
}

\setcounter{footnote}{0}
\renewcommand{\thefootnote}{\arabic{footnote}}

\vspace{10mm}
\section{Introduction} \label{sect:introduction}

In the summer of 2012 an SM-like Higgs particle ($h$) was found at the LHC \cite{Aad:2012tfa,Chatrchyan:2012ufa}. As of today its properties agree with the SM predictions at the 20\% level \cite{Khachatryan:2014jba,Aad:2015ona}. Its mass  derived from the $\gamma \gamma$ and $ZZ$ channels is $125.09 \pm 0.24~\text{GeV}$ \cite{Aad:2015zhl}.  However, the SM-like limit exists in various models with extra neutral Higgs scalars.  
A charged Higgs boson ($H^+$) would be the most striking signal of an extended Higgs sector, for example with
more than one Higgs doublet. Such a discovery at the LHC is a distinct
possibility, with or without supersymmetry. However, a charged Higgs particle
might be rather hard to find, even if it is abundantly produced.

We here survey existing results on charged scalar phenomenology, and discuss possible strategies for further searches at the LHC.
Such scalars appear in Multi-Higgs-Doublet models (MHDM), in particular in the popular Two-Higgs-Doublet model (2HDM) \cite{Gunion:1989we,Branco:2011iw}, allowing for charged and more neutral Higgs bosons. We focus on these models, since they have the attractive property that electroweak precision observables are automatically in agreement with the Standard Model at the tree level, in particular, $\rho=1$ \cite{Ross:1975fq,Veltman:1976rt,Veltman:1977kh}. 

The production rate and the decay pattern would depend on details of the
theoretical model \cite{Gunion:1989we}, especially the Yukawa interaction.
It is useful to distinguish two cases, depending on whether the mass of the charged scalar ($M_{H^\pm}$) is below or above the top mass.
Since an extended Higgs sector naturally leads to Flavor-Changing
Neutral Currents (FCNC), these would have to be suppressed \cite{Glashow:1976nt,Paige:1977nz}. This is
normally achieved by imposing discrete symmetries in modeling the
Yukawa interactions. For example, in the 2HDM with Model~II Yukawa interactions a
$Z_2$ symmetry under the transformation $\Phi_1 \to \Phi_1$, $\Phi_2
\to -\Phi_2$ is assumed.  
In this case, the $B \to X_s \gamma$ data constrain the mass of $H^+$ to be above approximately 480~GeV \cite{Misiak:2015xwa}. A recent study concludes that this limit is even higher, in the range 570--800~GeV \cite{Misiak:2017bgg}. Our results can easily be re-interpreted for this new limit.
Alternatively, if all fermion masses are generated
by only one doublet ($\Phi_2$, Model I) there is no enhancement in
the Yukawa coupling of $H^+$ with down-type quarks and 
the allowed mass range is less constrained. The same is true for
the Model X (also called Model IV or lepton-specific 2HDM)
\cite{Akeroyd:1994ga,Logan:2009uf}, where the second doublet is responsible for the
mass of all quarks, while the first doublet deals with leptons.
Charged Higgs mass below ${\cal O}(M_Z)$ has been excluded at LEP \cite{Abbiendi:2013hk}. Low and high values of $\tan\beta$ are excluded by various theoretical and experimental model-dependent constraints.

An extension of the scalar sector also offers an opportunity to introduce additional CP violation \cite{Lee:1973iz}, which may facilitate baryogenesis \cite{Riotto:1999yt}.

Charged scalars may also appear in models explaining dark matter (DM). These are charged scalars not involved in the spontaneous symmetry breaking, and we will denote them as $S^+$. Such charged particles will typically be members of an ``inert'' or ``dark'' sector, the lightest neutral member of which is the DM particle ($S$).
In these scenarios a $Z_2$ symmetry
will make the scalar DM stable and forbid any charged-scalar Yukawa coupling. 
Consequently, the phenomenology of the  $S^+$, the charged component of a $Z_2$-odd doublet, is rather different from the one in usual 2HDM models.
In particular, $S^+$ may become long-lived and induce observable displaced vertices in its leptonic decays.  
This is a background-free experimental signature and would allow one to discover the $S^+$ at  the LHC.

The SM-like scenario (also referred to as the ``alignment limit'') observed at the LHC corresponds to the case when the relative couplings of the 125 GeV Higgs particle  to the electroweak gauge bosons $W/Z$  with respect to the ones in the SM are close to unity.  We will assume that this applies to the lightest neutral, mainly CP-even Higgs particle, denoted $h$. Still there are two distinct options  possible---with and without decoupling of other scalars in the model. In the case of decoupling, very high masses of other Higgs particles (both neutral and charged) arise from the soft  $Z_2$ breaking term in the potential without any conflict with unitarity.  
 
The focus of this paper will be the $Z_2$-softly-broken 2HDM, but we will also briefly
discuss models with more doublets.
In such models, one pair of charged Higgs-like scalars $(H^+H^-)$
would occur for each additional doublet. We also briefly describe scalar dark matter models.

This work arose as a continuation of activities around the workshops ``Prospects for Charged Higgs Discovery at Colliders'', taking place every two years in Uppsala. The paper is organized as follows.
In sections~\ref{sect:notation}--\ref{sect:Yukawa} we review the basic theoretical framework. Then, in section~\ref{sect:decays} we review charged Higgs decays, and in section~\ref{sect:production} we review charged-Higgs production at the LHC. Section~\ref{sect:ex-constraints} is devoted to an overview of different experimental constraints.
Proposed search channels for the 2HDM are presented in section~\ref{sect-benchmarks}, whereas in sections~\ref{sect:other-models} and \ref{sect:inert-models} we discuss models with several doublets, and models with dark matter, respectively. 
Section~\ref{sect:summary} contains a brief summary.
Technical details are collected in appendices.

\section{Potential and states} 
\label{sect:notation}
\setcounter{equation}{0}
The general 2HDM potential 
allows for various vacua, including CP violating, charge breaking and inert ones, leading to distinct phenomenologies. Here we consider the case when both doublets have non-zero vacuum expectation values.  CP violation, explicit or spontaneous, is possible in this case. 
\subsection{The potential}

We limit ourselves to studying the softly $Z_2$-violating 2HDM potential, which  reads
\begin{align} \label{Eq:fullpot}
V(\Phi_1,\Phi_2)
&= -\frac12\left\{m_{11}^2\Phi_1^\dagger\Phi_1
+ m_{22}^2\Phi_2^\dagger\Phi_2 + \left[m_{12}^2 \Phi_1^\dagger \Phi_2
+ \text{h.c.}\right]\right\} \nonumber \\
& + \frac{\lambda_1}{2}(\Phi_1^\dagger\Phi_1)^2
+ \frac{\lambda_2}{2}(\Phi_2^\dagger\Phi_2)^2
+ \lambda_3(\Phi_1^\dagger\Phi_1)(\Phi_2^\dagger\Phi_2)
+ \lambda_4(\Phi_1^\dagger\Phi_2)(\Phi_2^\dagger\Phi_1)
\nonumber \\
& + \frac12\left[\lambda_5(\Phi_1^\dagger\Phi_2)^2 +\text{h.c.}\right].
\end{align}
Apart from the term $m_{12}^2$, this potential exhibits a $Z_2$ symmetry,
\begin{equation} \label{Eq:Z2-symmetries}
(\Phi_1,\Phi_2) \leftrightarrow (\Phi_1,-\Phi_2) \quad\text{or} \quad
(\Phi_1,\Phi_2) \leftrightarrow (-\Phi_1,\Phi_2).
\end{equation}
The most general potential contains in addition two more quartic terms, with coefficients $\lambda_6$ and $\lambda_7$,  and violates $Z_2$ symmetry in a hard way \cite{Gunion:1989we}.
The
parameters $\lambda_1$--$\lambda_4$, $m^2_{11}$ and $m^2_{22}$ are real.  There
are various bases in which this potential can be written, often they are
defined by fixing properties of the vacuum state.  The potential (\ref{Eq:fullpot}) can lead to
CP violation, provided $m_{12}^2\neq 0$.
\subsection{Mass eigenstates}
\label{subsect:masses}
We use the following decomposition of the doublets (see Appendix~A):
\begin{equation}
\Phi_1=
\begin{pmatrix}
\varphi_1^+\\ (v_1+\eta_1+i\chi_1)/\sqrt{2}
\end{pmatrix},
\quad
\Phi_2=
\begin{pmatrix}
\varphi_2^+\\ (v_2+\eta_2+i\chi_2)/\sqrt{2}
\end{pmatrix},
\label{Obasis}
\end{equation}
which corresponds to a basis where both have a non-zero, real and
positive, vacuum expectation value (vev).
Here $v_1=\cos\beta\, v$,  $v_2=\sin\beta\, v$,  $v=2\,m_W/g$, with $\tan\beta=v_2/v_1$.

We adopt the mixing matrix $R$, between the scalar fields $\eta_1, \eta_2,
\eta_3 $ and mass eigenstates $H_1,H_2,H_3$ (for the CP conserving case
CP-even $h$, $H$ and CP-odd $A$, respectively) defined by
\begin{equation} \label{Eq:R-def}
\begin{pmatrix}
H_1 \\ H_2 \\ H_3
\end{pmatrix}
=R
\begin{pmatrix}
\eta_1 \\ \eta_2 \\ \eta_3
\end{pmatrix},
\end{equation}
satisfying
\begin{equation}
\label{Eq:cal-M}
R{\cal M}^2R^{\rm T}={\cal M}^2_{\rm diag}={\rm diag}(M_1^2,M_2^2,M_3^2),
\quad M_1\leq M_2\leq M_3.
\end{equation}

The rotation matrix $R$ is parametrized in terms of three rotation angles $\alpha_i$ as \cite{Accomando:2006ga}
\begin{equation}     \label{Eq:R-angles}
R
=\begin{pmatrix}
c_1\,c_2 & s_1\,c_2 & s_2 \\
- (c_1\,s_2\,s_3 + s_1\,c_3)
& c_1\,c_3 - s_1\,s_2\,s_3 & c_2\,s_3 \\
- c_1\,s_2\,c_3 + s_1\,s_3
& - (c_1\,s_3 + s_1\,s_2\,c_3) & c_2\,c_3
\end{pmatrix}
\end{equation}
with $c_i=\cos\alpha_i$, $s_i=\sin\alpha_i$, and $\alpha_{1,2,3}\in(-\pi/2,\pi/2]$.
In Eq.~(\ref{Eq:R-def}), $\eta_3 \equiv -\sin\beta\chi_1+\cos\beta\chi_2$
is the combination of $\chi_i$'s which is
orthogonal to the neutral Nambu--Goldstone boson.
In terms of these angles, the limits of CP conservation correspond to \cite{ElKaffas:2007rq}
\begin{align}
H_1\text{ odd } (H_1\equiv A):&\quad \alpha_2=\pm\pi/2, \nonumber \\
H_2\text{ odd } (H_2\equiv A):&\quad \alpha_2=0,\alpha_3=\pm\pi/2, \nonumber \\
H_3\text{ odd } (H_3\equiv A):&\quad \alpha_2=0,\alpha_3=0.
\end{align}

The charged Higgs bosons are the combination orthogonal to the
charged Nambu--Goldstone bosons: $H^\pm = -\sin \beta \varphi_1^\pm +\cos
\beta \varphi_2^\pm$, and their mass is given by
\begin{equation}
M_{H^\pm}^2=\mu^2-\frac{v^2}{2}(\lambda_4+\Re\lambda_5),
\end{equation}
where we define a mass parameter $\mu$ by
\begin{equation}
\mu^2\equiv (v^2/2 v_1 v_2) \Re m_{12}^2.
\end{equation}
Note also
the following relation arising from the extremum condition:
\begin{equation}
\Im m_{12}^2=\Im\lambda_5v_1v_2.
\label{Eq:m12}
\end{equation}

\subsection{Gauge couplings}
With all momenta incoming, we have the $H^\mp W^\pm H_j$ gauge couplings \cite{ElKaffas:2006nt}:
\begin{equation} \label{Eq:gauge-HH}
H^\mp W^\pm H_j: \qquad
\frac{g}{2}
[\pm i(\sin\beta R_{j1}-\cos\beta R_{j2})+ R_{j3}]
(p_\mu^j-p_\mu^\mp).
\end{equation}
Specifically, for coupling to the lightest neutral Higgs boson, the $R$-matrix (\ref{Eq:R-angles}) gives:
\begin{equation} \label{Eq:H_chWH_1}
H^\mp W^\pm H_1: \qquad
\frac{g}{2}
[\pm i\cos\alpha_2\sin(\beta-\alpha_1)+ \sin\alpha_2]
(p_\mu-p_\mu^\mp).
\end{equation}

The familiar CP-conserving limit is obtained by evaluating $R$
for $\alpha_2=0$, $\alpha_3=0$, $\alpha_1=\alpha+\pi/2$, with the mapping
$H_1\to h$, $H_2\to -H$ and $H_3\to A$.
In that limit, we recover the results of \cite{Gunion:1989we}:
\begin{alignat}{2}
&H^\mp W^\pm h: &\quad
&\frac{\mp ig}{2}
\cos(\beta -\alpha)
(p_\mu-p_\mu^\mp), \nonumber \\
&H^\mp W^\pm H: &\quad
&\frac{\pm ig}{2}
\sin(\beta -\alpha)
(p_\mu-p_\mu^\mp),  \nonumber \\
&H^\mp W^\pm A: &\quad
&\frac{g}{2}
(p_\mu-p_\mu^\mp).
\label{Eq:CPC-gauge-couplings}
\end{alignat}

The strict SM-like limit corresponds to $\sin(\beta-\alpha)=1$,
however the experimental data from the LHC
\cite{Khachatryan:2014jba,Aad:2015ona} allow for a departure from this
limit\footnote{Note that in the 2HDM, this factor cannot exceed 1.}
down to approximately 0.7, which we  are going to allow in our study.

In the following analysis, the gauge couplings to neutral Higgs bosons are also involved. They differ from the SM coupling by the factor ($V=W^\pm,Z$):
\begin{equation} \label{Eq:gauge-VVH}
VVH_j: \quad \cos\beta R_{j1}+\sin\beta R_{j2}.
\end{equation}
In particular, for $H_1$, this factor becomes $\cos(\beta-\alpha_1)\cos\alpha_2$.
In the CP-conserving case, we have
\begin{alignat}{2}
&VVh: &\quad &\sin(\beta-\alpha), \nonumber \\
&VVH: &\quad &\cos(\beta-\alpha), \nonumber \\
&VVA: &\quad &0.
\end{alignat}
Note that the couplings (\ref{Eq:gauge-HH}) and (\ref{Eq:gauge-VVH}) are given by unitary matrices, and hence satisfy sum rules. Furthermore, for any $j$, the relative couplings of (\ref{Eq:gauge-HH}) (the expression in the square brackets) and (\ref{Eq:gauge-VVH}) satisfy the following relation \cite{Ginzburg:2014pra}:
\begin{equation}
|(\ref{Eq:gauge-HH})|^2 + [(\ref{Eq:gauge-VVH})]^2=1.
\end{equation}
These relations are valid for both the CP-conserving and the CP-violating cases.
\section{Theoretical constraints} 
\label{sect:th-constraints}
\setcounter{equation}{0}
The 2HDM is subject to various theoretical constraints. First, it has to have a stable vacuum\footnote{Here we perform an analysis at the tree level, for more advanced studies, see \cite{Nie:1998yn,Ferreira:2004yd,Goudelis:2013uca,Swiezewska:2015paa,Khan:2015ipa}.}, what leads to so-called positivity constraints for the potential \cite{Deshpande:1977rw,Nie:1998yn,Kanemura:1999xf}, $V(\Phi_1,\Phi_2)>0$ as $|\Phi_1|, |\Phi_2| \to\infty$.
Second, we should be sure to deal with a particular vacuum (a global minimum) as in some cases various minima can coexist \cite{Barroso:2013awa,Ginzburg:2010wa,Swiezewska:2012ej}. 

Other types of constraints arise from requiring perturbativity of the calculations, tree-level unitarity 
\cite{Kanemura:1993hm,Akeroyd:2000wc,Arhrib:2000is,Ginzburg:2003fe,Ginzburg:2005dt}
and perturbativity of the Yukawa couplings. 
In general, imposing tree-level unitarity has a significant
effect at high values of $\tan\beta$ and $M_{H^\pm}$, by excluding such values. 
These constraints limit the absolute values of the $\lambda$ parameters as
well as $\tan \beta$, the latter both at very low and very high values. This limit is particularly strong for a $Z_2$ symmetric model \cite{WahabElKaffas:2007xd,Gorczyca:2011he,Swiezewska:2012ej}.
The dominant one-loop corrections to the perturbative unitarity constraints for the model with softly-broken $Z_2$ symmetry are also available \cite{Grinstein:2015rtl}.

The electroweak precision data, parametrized in terms of $S,T$ and $U$ \cite{Kennedy:1988sn,Peskin:1990zt,Altarelli:1990zd,Peskin:1991sw,Altarelli:1991fk,Grimus:2007if,Grimus:2008nb}, also provide important constraints on these models.

\section{{Yukawa Interaction}} 
\label{sect:Yukawa}
\setcounter{equation}{0}

There are various models of Yukawa interactions, all of them, except Model~III, lead to
suppression of FCNCs at the tree level, assuming some vanishing Yukawa matrices.
The most popular is Model~II,
in which up-type quarks couple to one (our choice: $\Phi_2$) while
down-type quarks and charged leptons couple to the other scalar
doublet ($\Phi_1$). 
They are presented schematically in Table~\ref{tab:Z2}.
For a self-contained description of the 2HDM
Yukawa sector, see Appendix~B.\footnote{The absence of tree-level
FCNC interactions can also be obtained by imposing
flavor space alignment of the Yukawa couplings of the
two scalar doublets \cite{Jung:2010ik}.}

\begin{table}
\centering
\begin{tabular}{cccc}
\hline
Model & $d$ & $u$ & $\ell$ \\
\hline
I & $\Phi_2$ & $\Phi_2$ & $\Phi_2$ \\[0.1cm]
II & $\Phi_1$ & $\Phi_2$ & $\Phi_1$ \\[0.1cm]
III & $\Phi_1\&\Phi_2$ & $\Phi_1\&\Phi_2$ & $\Phi_1\&\Phi_2$ \\[0.1cm]
X & $\Phi_2$ & $\Phi_2$ & $\Phi_1$ \\[0.1cm]
Y & $\Phi_1$ & $\Phi_2$ & $\Phi_2$ \\[0.1cm]
\hline
\end{tabular}
\caption{The most popular models of the Yukawa interactions in the 2HDM 
  (also referred to as ``Types''). The symbols $u$, $d$, $\ell$ refer to up-
  and down-type quarks, and charged leptons of any generation.
  Here, $\Phi_1$ and $\Phi_2$ refer to the Higgs doublet coupled to the particular fermion.
  Also other conventions are being used in the literature, see Appendix~B.} 
\label{tab:Z2}
\end{table}

For Model~II, and the third generation, the neutral-sector Yukawa couplings are:
\begin{alignat}{2}  \label{Eq:H_j_Yuk}
&H_j  b\bar b: &\qquad
&\frac{-ig\,m_b}{2\,m_W}\frac{1}{\cos\beta}\, [R_{j1}-i\gamma_5\sin\beta R_{j3}], 
\nonumber \\
&H_j  t\bar t: &\qquad
&\frac{-ig\,m_t}{2\,m_W}\frac{1}{\sin\beta}\, [R_{j2}-i\gamma_5\cos\beta R_{j3}].
\end{alignat}

Explicitly, for the charged Higgs bosons in Model~II, we have for the
coupling to the third generation of quarks \cite{Gunion:1989we}
\begin{alignat}{2}  \label{Eq:Yukawa-charged-II}
&H^+ b \bar t: &\qquad
&\frac{ig}{2\sqrt2 \,m_W}\,V_{tb}
[m_b(1+\gamma_5)\tan\beta+m_t(1-\gamma_5)\cot\beta], \nonumber \\
&H^-  t\bar b: &\qquad
&\frac{ig}{2\sqrt2 \,m_W}\,V_{tb}^*
[m_b(1-\gamma_5)\tan\beta+m_t(1+\gamma_5)\cot\beta],
\end{alignat}
where $V_{tb}$ is the appropriate element of the CKM matrix.
For other Yukawa models the factors $\tan\beta$ and $\cos\beta$ will be substituted according to Table~\ref{tab:couplings} in Appendix~B.

As mentioned above, the range in $\alpha$ (or $\alpha_1$) is $\pi$, which can be taken as $[-\pi,0]$, $[-\pi/2,\pi/2]$ or $[0,\pi]$. This is different from the MSSM, where only a range of $\pi/2$ is required \cite{Gunion:1986nh}, $-\pi/2\leq\alpha\leq0$. The spontaneous breaking of the symmetry and the convention of having a positive value for $v$ means that the sign (phase) of the field is relevant. This doubling of the range in the 2HDM as compared with the MSSM is the origin of ``wrong-sign'' Yukawa couplings.

\section{Charged Higgs boson decays} 
\label{sect:decays}
\setcounter{equation}{0}

This section presents an overview of the different $H^+$ decay modes, illustrated with branching ratio plots for parameter sets that are chosen to exhibit the most interesting features. Branching ratios required for modes considered in sections~\ref{sect-benchmarks}--\ref{sect:inert-models} are calculated independently.

As discussed in \cite{Gunion:1989we,Moretti:1994ds,Djouadi:1995gv,Djouadi:1997yw,Kanemura:2009mk,Eriksson:2009ws}, a charged Higgs boson can decay to a fermion-antifermion pair,
\begin{subequations}
 \label{Eq:fermion-decay-channels}
\begin{align}
H^+&\to c\bar s, \\
H^+&\to c\bar b, \label{Eq:to-cb}\\
H^+&\to \tau^+\nu_\tau, \\
H^+&\to t\bar b \label{Eq:t-bbar},
\end{align}
\end{subequations}
(note that (\ref{Eq:to-cb}) refers to a mixed-generation final state), to gauge bosons,
\begin{subequations}
 \label{Eq:gauge-decay-channels}
\begin{align}
H^+&\to W^+\gamma,\\
H^+&\to W^+Z,
\end{align}
\end{subequations}
or to a neutral Higgs boson and a gauge boson:
\begin{equation}
H^+\to H_j W^+, \label{Eq:WHj-light}
\end{equation}
and their charge conjugates.

Below, we consider branching ratios mainly for the CP-conserving case.
For the lightest neutral scalar we take the mass $M_h=125~\text{GeV}$.
Neither experimental nor theoretical constraints are here imposed. (They have significant impacts, as will be discussed in subsequent sections.) 
For the calculation of branching ratios, we use the software 
{\tt 2HDMC} \cite{Eriksson:2009ws} and {\tt HDECAY} \cite{Djouadi:1997yw,Harlander:2013qxa}. 
As discussed in \cite{Harlander:2013qxa},
branching ratios are calculated at leading order in the 2HDM parameters, but include QCD corrections according to
\cite{Mendez:1990jr,Li:1990ag,Djouadi:1994gf}, and
three-body modes via off-shell extensions of $H^+\to
t\bar b$, $H^+\to hW^+$, $H^+\to  HW^+$ and $H^+\to  AW^+$.
The treatment of three-body decays is according to Ref.~\cite{Djouadi:1995gv}.

For light charged Higgs bosons, $M_{H^\pm}<m_t$, Model~II is excluded
by the $B \to X_s \gamma$ constraint discussed in section~\ref{sect:ex-constraints}.
For Model~I (which in this region is {\it not} excluded by $B \to X_s \gamma$), the open channels have fermionic couplings proportional to $\cot\beta$.
The gauge couplings (involving decays to a $W^+$ and a neutral Higgs) are proportional to $\sin(\beta-\alpha)$ or $\cos(\beta-\alpha)$, whereas the corresponding Yukawa couplings depend on the masses involved, together with $\tan\beta$. 

The CP-violating case for the special channel $H^+\to H_1 W^+$ is presented in section \ref{subsec:br-toWH1}.

\begin{figure}[htb]
\refstepcounter{figure}
\label{Fig:cpc-br-ratios-low}
\addtocounter{figure}{-1}
\begin{center}
\includegraphics[scale=0.70]{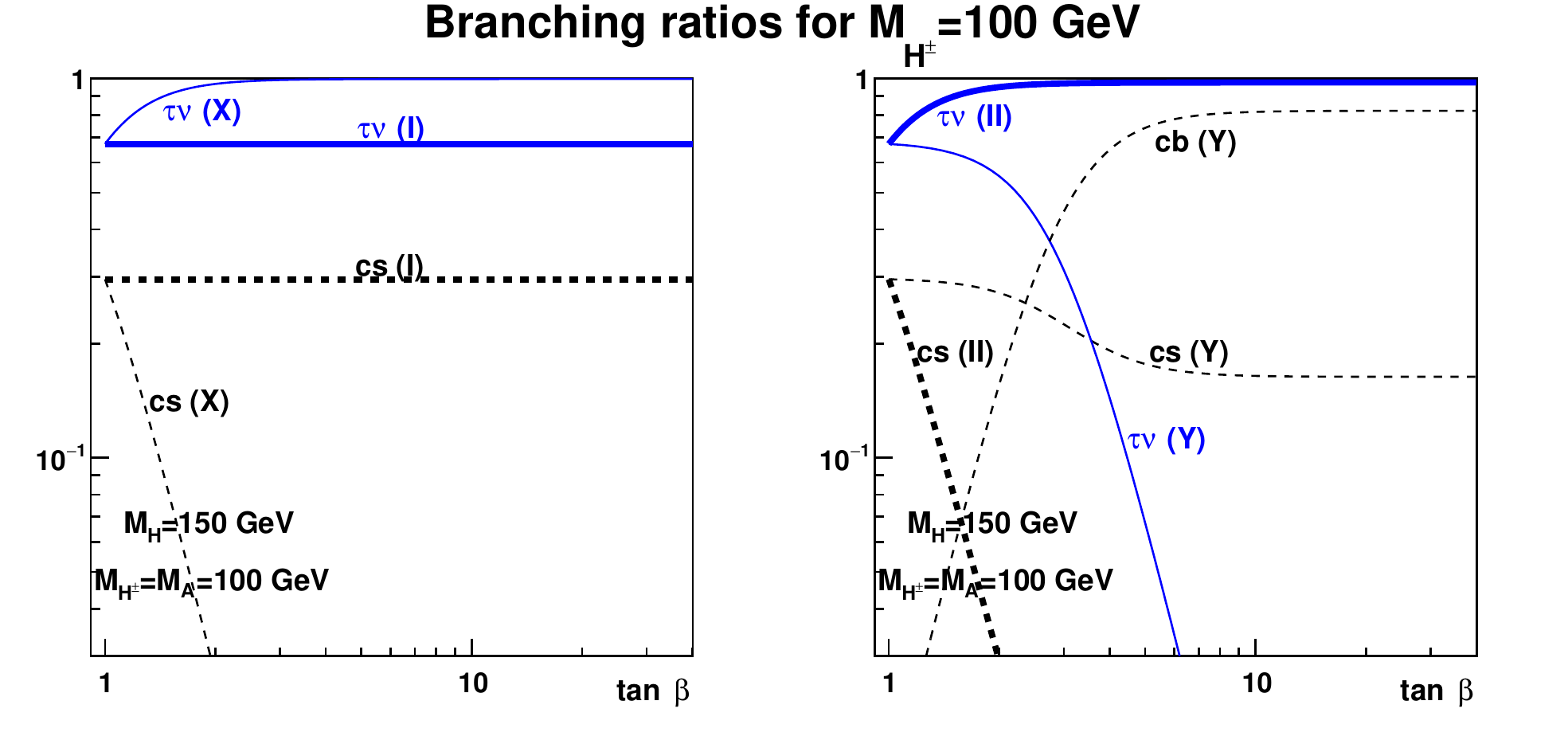}
\end{center}
\vspace*{-4mm}
\caption{Light charged-Higgs branching ratios vs $\tan\beta$. Left: Models~I and X, right: Models~II and Y.
The panel on the right is only for illustration, such a light $H^+$ is
excluded for the models II and Y.}
\end{figure}

\subsection{Branching ratios vs $\tan\beta$} 
Below, we consider branching ratios, assuming for simplicity $M_{H^\pm}=M_A$, in the low and high mass regions.

\subsubsection{Light $H^+$ ($M_{H^\pm}<m_t$)} 

For a light charged Higgs boson, such as might be produced in top decay, the $tb$ and $Wh$ channels would be closed, and the $\tau\nu$ and $cs$ channels would dominate. The relevant Yukawa couplings are given by $\tan\beta$ and the fermion masses involved. 
With scalar masses  taken as follows: 
\begin{equation}
M_{H^\pm}=M_A=100~\text{GeV}, \qquad
M_H=150~\text{GeV},
\end{equation}
we show in Fig.~\ref{Fig:cpc-br-ratios-low} branching ratios for the different Yukawa models. 

Since the $\tau\nu$ and $cs$ couplings for Model~I are the {\it same}, the branching ratios are independent of $\tan\beta$, as seen in the left panel. For Models~X and II the couplings to $c$ and $\tau$ have different dependences on $\tan\beta$, and consequently the branching ratios will depend on $\tan\beta$.  In the case of Model~Y, the $cs$ channel is for $\tan\beta>\sqrt{m_c/m_s}$ controlled by the term $m_s\tan\beta$, which dominates over the $\tau\nu$ channel at high $\tan\beta$.

\subsubsection{Heavy $H^+$ ($M_{H^\pm}>m_t$)}

\begin{figure}[htb]
\refstepcounter{figure}
\label{Fig:cpc-br-ratios-120}
\addtocounter{figure}{-1}
\begin{center}
\vspace*{-4mm}
\includegraphics[scale=0.60]{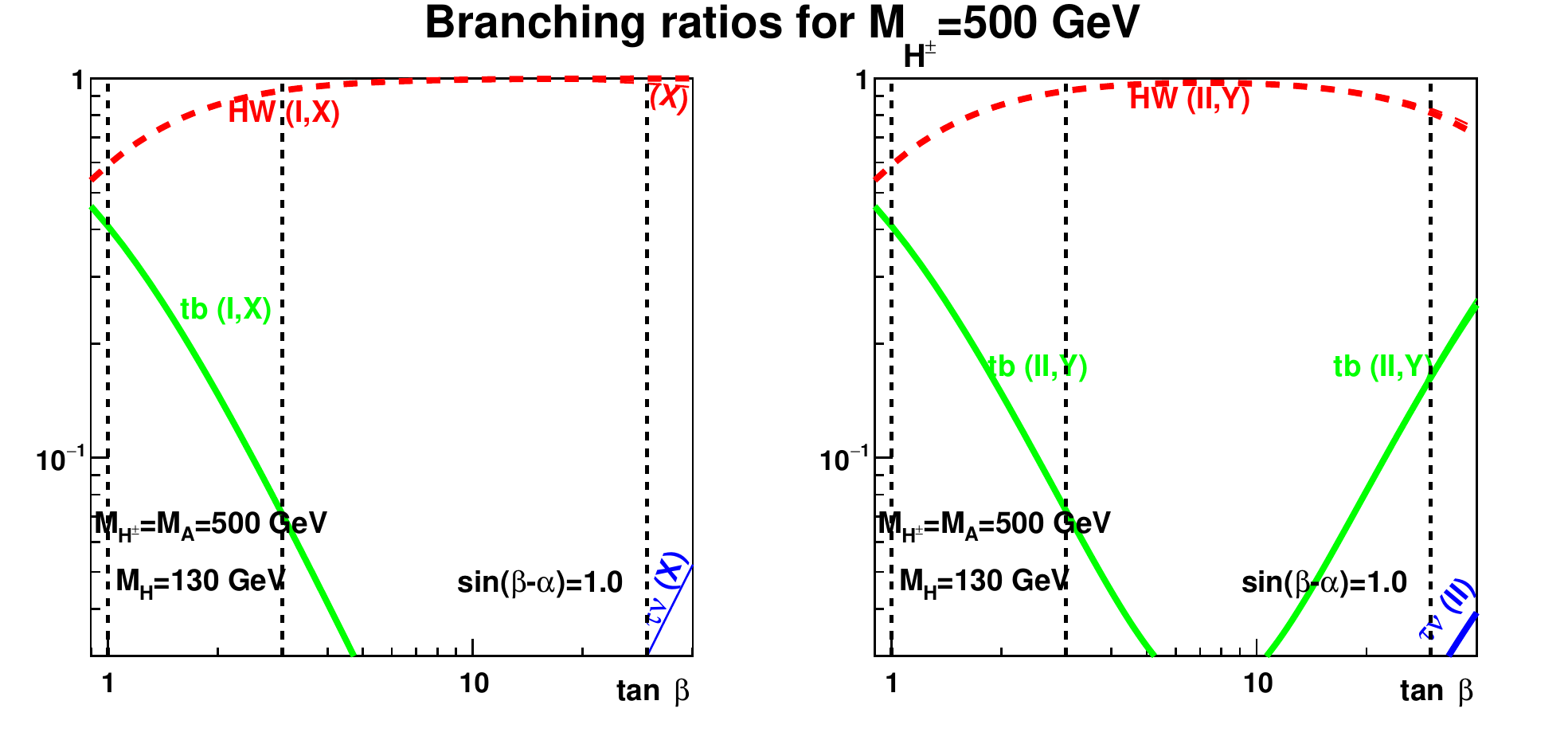} 
\includegraphics[scale=0.60]{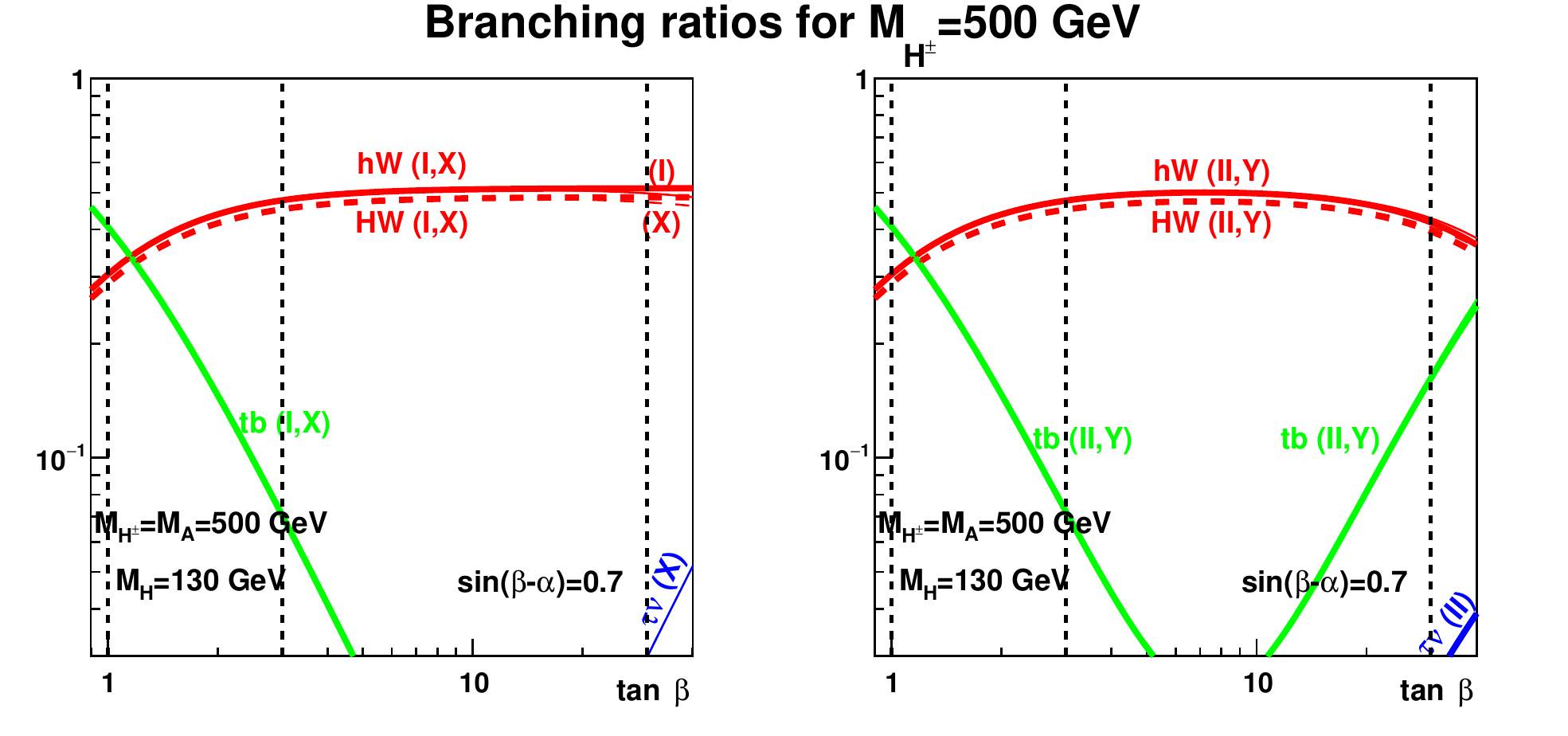}
\end{center}
\vspace*{-4mm}
\caption{Heavy charged-Higgs branching ratios vs $\tan\beta$ for two light neutral Higgs bosons $h$ and $H$. Left: Models~I and X, right: Models~II and Y. Upper two panels: $\sin(\beta-\alpha)=1$, lower panels: $\sin(\beta-\alpha)=0.7$.
The dashed vertical lines are for comparison with Figs.~\ref{Fig:cpc-br-ratios-vs-mass-1}--\ref{Fig:cpc-br-ratios-vs-mass-3-30}.}
\end{figure}

Below, we consider separately the two cases where one more neutral scalar is light, besides $h$, this being either $H$ or $A$.
For a case where both the channels $hW$ and $HW$ are open, whereas $AW$ is not, exemplified by the masses
\begin{equation}
M_{H^\pm}=M_A=500~\text{GeV}, \qquad
M_H=130~\text{GeV},
\end{equation}
we show in Fig.~\ref{Fig:cpc-br-ratios-120} branching ratios for the different Yukawa models. 
Two values of $\sin(\beta-\alpha)$ are considered, 1 and 0.7. 
For comparison with section~\ref{subsec:br-vs.mass}, we have drawn dashed lines at $\tan\beta=1$, 3 and 30.

For Model~I (left part of Fig.~\ref{Fig:cpc-br-ratios-120}), the
dominant decay rates are to the heaviest fermion-antifermion pair and
to $W$ together with $h$ or $H$ (for the considered parameters, both
$h$ and $H$ are kinematically available). Model~X differs in having an
enhanced coupling to tau leptons at high $\tan\beta$, see
Table~\ref{tab:couplings} in Appendix~B. If the decay to $Wh$ is kinematically not accessible, the $\tau\nu$ mode may be accessible at high $\tan\beta$.

For Model~II (right part of Fig.~\ref{Fig:cpc-br-ratios-120}), the dominant decay rates are to the heaviest fermion-antifermion pair at low and high values of $\tan\beta$, with $hW$ or $HW$ dominating at medium $\tan\beta$ (if kinematically available). At high $\tan\beta$ it is the down-type quark that has the dominant  coupling. Hence, modulo phase space effects, the $\tau\nu$ rate is only suppressed by the mass ratio $(m_\tau/m_b)^2$. Model~Y differs from Model~II in {\it not} having enhanced coupling to the tau at high values of $\tan\beta$.

Whereas the couplings and hence the decay rates to $hW$ and $HW$, for fixed values of $\sin(\beta-\alpha)$, are independent of $\tan\beta$, the branching ratios are not. They will depend on the strengths of the competing $tb$ Yukawa couplings.
The strength of the $hW$ channel increases with $\cos^2(\beta-\alpha)$, and is therefore absent in the upper panels where $\sin(\beta-\alpha)=1$.

It should also be noted that if the $Wh$ channel is not kinematically
available, the $tb$ channel would dominate for all values of $\tan\beta$.
The $\tau\nu$ channel, which may offer less background for experimental searches, is only relevant at higher $\tan\beta$, and then only in Models~II and X.

When $A$ is light, such that the channels $H^+\to W^+A$ and $H^+\to W^+h$ are both open, whereas $H^+\to W^+H$ is not, the situation is similar to the previous case, with the $HW$ mode replaced by the $AW$ mode.
The choice $\sin(\beta-\alpha)=1$ turns off the $H^+\to W^+h$ mode (see Eq.~(\ref{Eq:CPC-gauge-couplings})), and there is a competition among the $WA$ and the $tb$ modes, except for the region of high $\tan\beta$, where also the $\tau\nu$ mode can be relevant.

\begin{figure}[htb]
\refstepcounter{figure}
\label{Fig:cpc-br-ratios-vs-mass-1}
\addtocounter{figure}{-1}
\begin{center}
\vspace*{-3mm}
\includegraphics[scale=0.30]{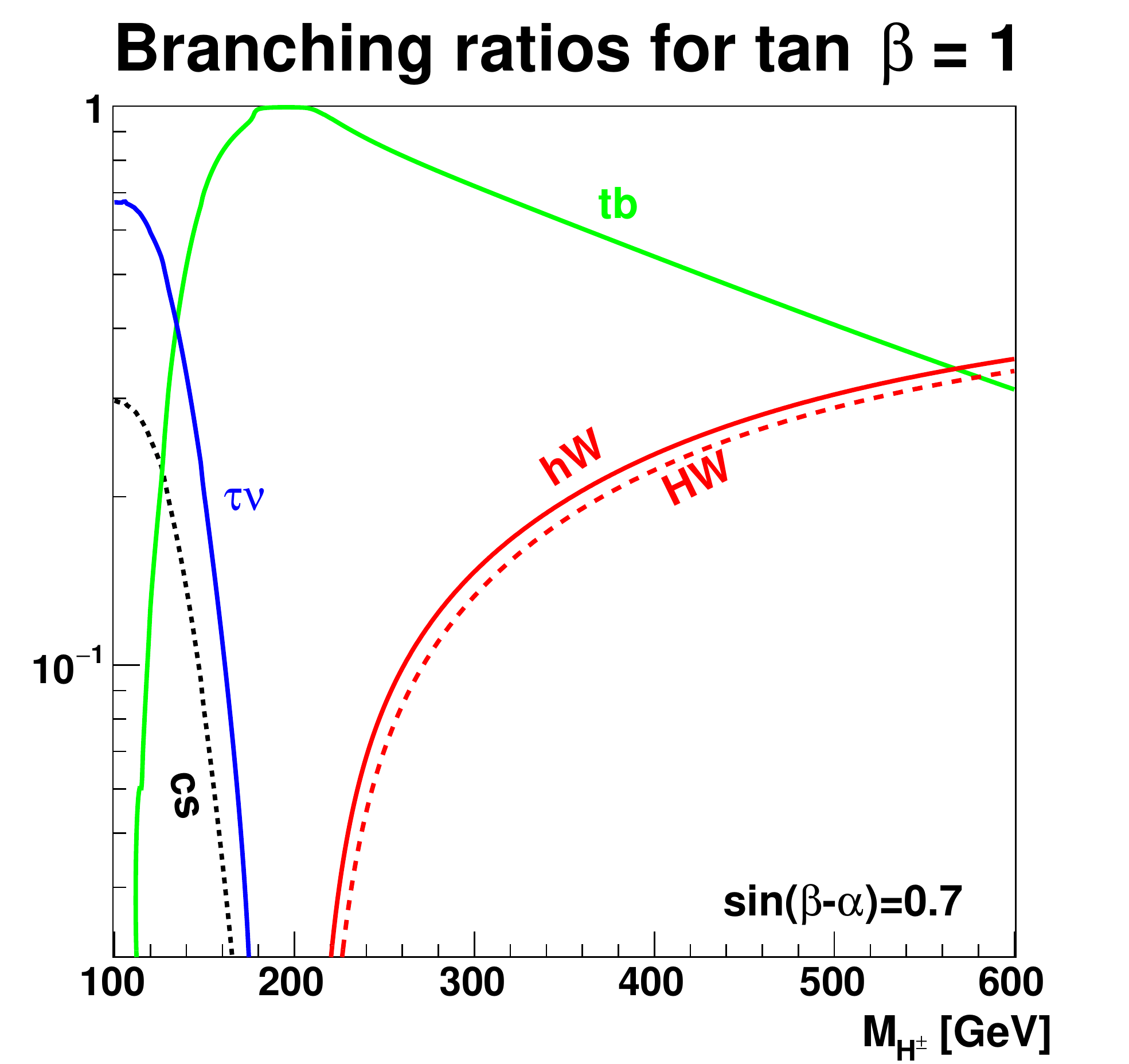}
\end{center}
\vspace*{-6mm}
\caption{Charged-Higgs branching ratios vs $M_{H^\pm}$, for $\tan\beta=1$ and $\sin(\beta-\alpha)=0.7$. Here, two light neutral Higgs bosons $h$ and $H$ (125~GeV and 130~GeV) are considered.}
\end{figure}

\subsection{Branching ratios vs $M_{H^\pm}$} 
\label{subsec:br-vs.mass}

In Figs.~\ref{Fig:cpc-br-ratios-vs-mass-1}--\ref{Fig:cpc-br-ratios-vs-mass-3-30} we show how the branching ratios change with the charged Higgs mass. Here, we have taken $\tan\beta=1$ (Fig.~\ref{Fig:cpc-br-ratios-vs-mass-1}), 3 and 30 (Fig.~\ref{Fig:cpc-br-ratios-vs-mass-3-30}), together with the neutral-sector masses
\begin{equation}
(M_H,M_A)=(130~\text{GeV},M_{H^\pm}),
\end{equation}
(note that here we take $M_{H^\pm}=M_A$)
and consider the two values $\sin(\beta-\alpha)=1$ and 0.7, corresponding to different strengths of the gauge couplings (\ref{Eq:CPC-gauge-couplings}).

\begin{figure}[htb]
\refstepcounter{figure}
\label{Fig:cpc-br-ratios-vs-mass-3-30}
\addtocounter{figure}{-1}
\begin{center}
\vspace*{-2mm}
\includegraphics[scale=0.60]{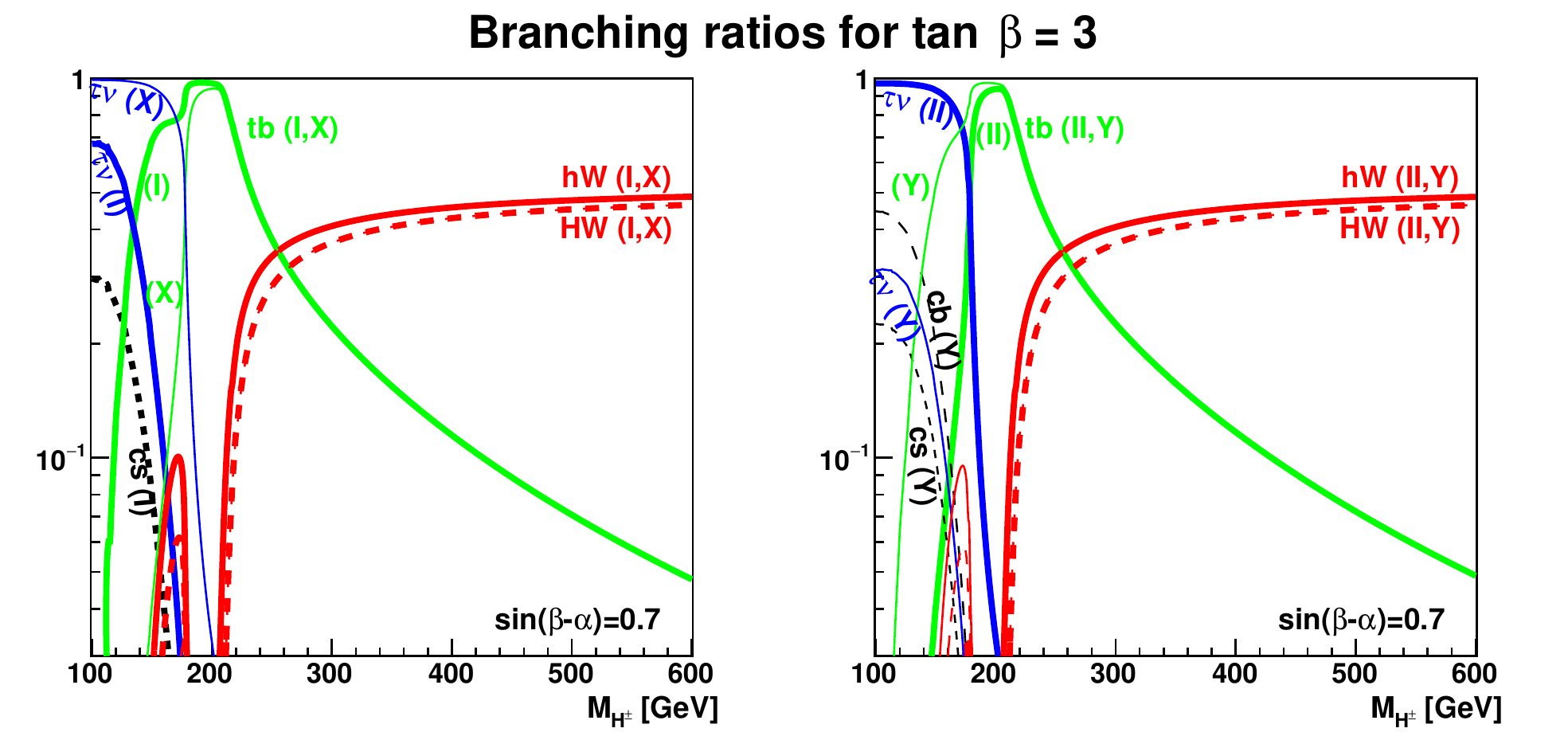}
\includegraphics[scale=0.60]{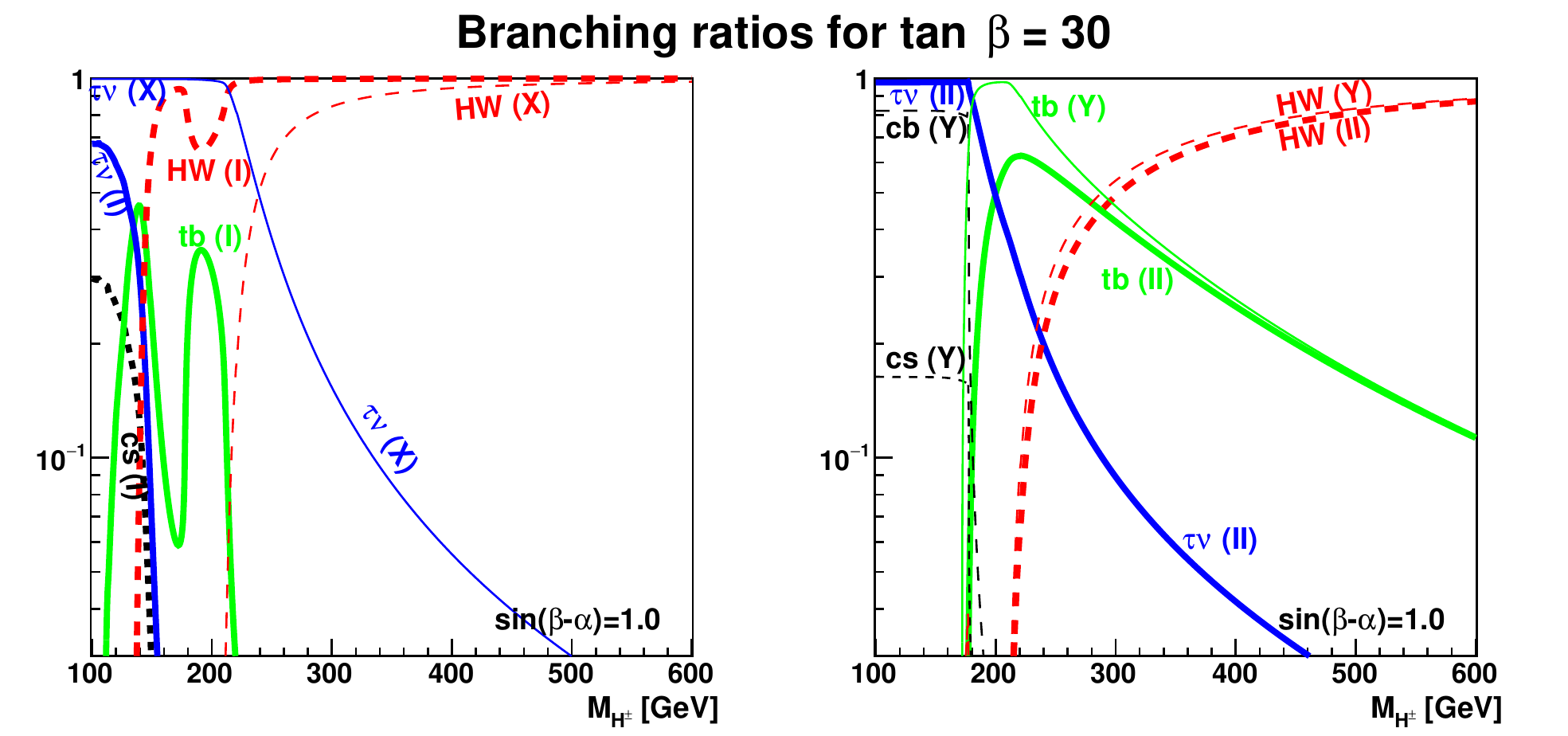}
\end{center}
\vspace*{-6mm}
\caption{Charged-Higgs branching ratios vs $M_{H^\pm}$, for $\tan\beta=3$ and 30, with two light neutral Higgs bosons $h$ and $H$ (125~GeV and 130~GeV). Left: Models~I and X, right: Models~II and Y.
Top: $\sin(\beta-\alpha)=0.7$, bottom: $\sin(\beta-\alpha)=1$}
\end{figure}

The picture from Figs.~\ref{Fig:cpc-br-ratios-low} and \ref{Fig:cpc-br-ratios-120} is confirmed: At low masses, the $\tau\nu$ channel dominates, whereas at higher masses, the $tb$ channel will compete against $hW$ and $HW$, if these channels are kinematically open, and not suppressed by some particular values of the mixing angles.

Of course, for $\tan\beta=1$ (Fig.~\ref{Fig:cpc-br-ratios-vs-mass-1}), all four Yukawa models give the same result. Qualitatively, the result is simple. At low masses, the $\tau\nu$ and $cs$ channels dominate, whereas above the $t$ threshold, the $tb$ channel dominates. There is however some competition with the $hW$ and $HW$ channels. 
Similar results hold for $\sin(\beta-\alpha)=1$, the only difference being that the $HW$ branching ratio rises faster with mass, and the $hW$ mode disappears completely in this limit.
Even below the $hW$ threshold, branching ratios for three-body decays via an off-shell $W$ can be significant \cite{Djouadi:1995gv}.
The strength of the $hW$ channel is proportional to $\cos^2(\beta-\alpha)$, and is therefore absent for $\sin(\beta-\alpha)=1$ (not shown). 

At higher values of $\tan\beta$ (Fig.~\ref{Fig:cpc-br-ratios-vs-mass-3-30}), the interplay with the $HW$ and $hW$ channels becomes more complicated. At 
high charged-Higgs masses, the $HW$ rate can be important (if kinematically open). On the other hand, the $hW$ channel can dominate over $HW$, because of the larger phase space.
Here, we present the case of $\sin(\beta-\alpha)=0.7$. The case of $\sin(\beta-\alpha)=1$ is similar, the main difference is a higher $HW$ branching ratio, while the $hW$ channel disappears.
It should be noted that three-body channels that proceed via $hW$ and $HW$ can be important also below threshold, if the $tb$ channel is closed.

\begin{figure}[htb]
\refstepcounter{figure}
\label{Fig:br-top}
\addtocounter{figure}{-1}
\begin{center}
\includegraphics[scale=0.60]{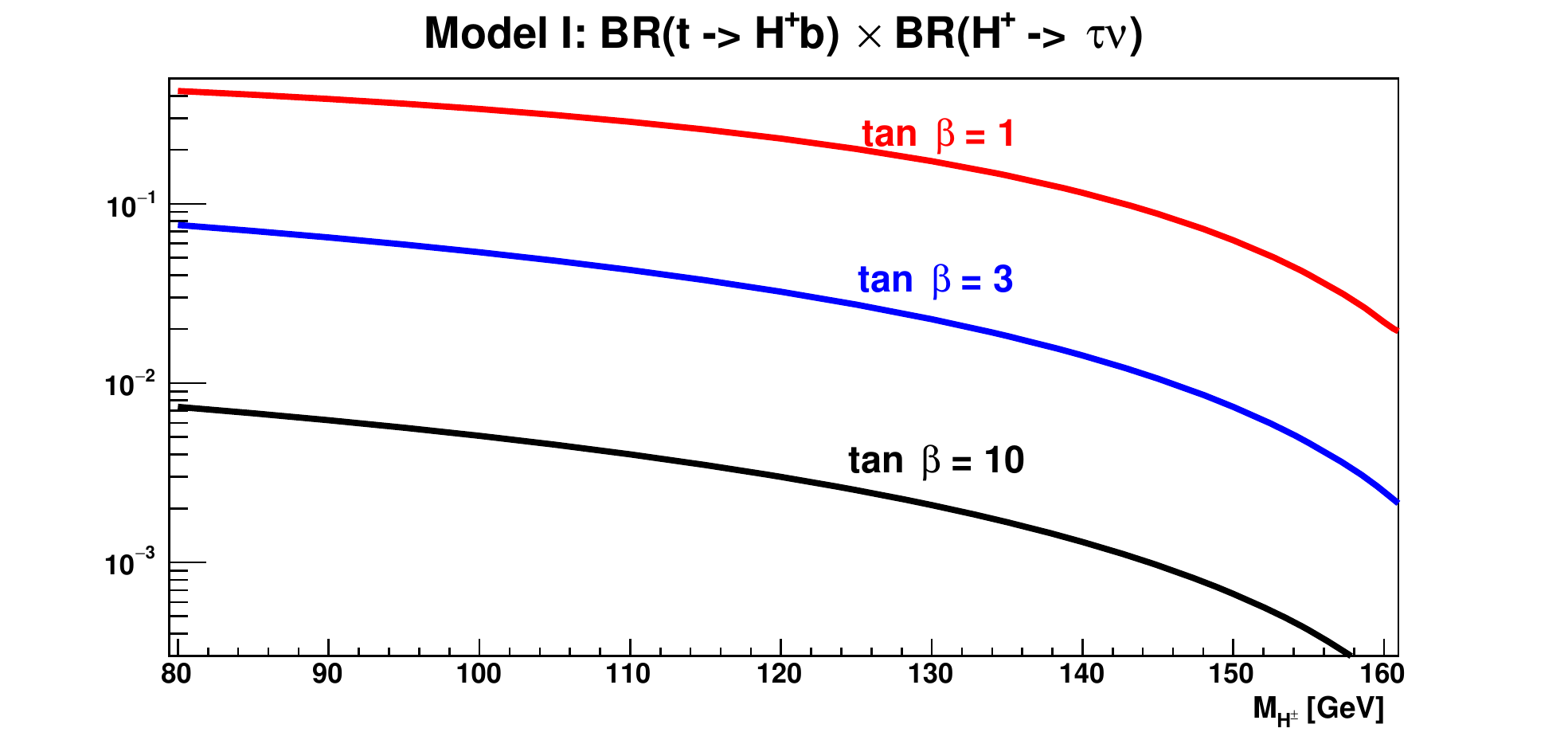}
\end{center}
\vspace*{-4mm}
\caption{Product of branching ratios, $\text{BR}(t\to H^+b)\times\text{BR}(H^+\to\tau^+\nu)$, for Model~I, and three values of $\tan\beta$, as indicated.}
\end{figure}

\subsection{Top decay to $H^+b$} 
\label{subsec:top-BR}

A light charged Higgs boson may emerge in the decay of the top quark
\begin{equation}
t\to H^+b,
\end{equation}
followed by a model-dependent $H^+$ decay. In Model~I possible channels are $H^+\to \tau^+\nu$ and $H^+\to c\bar s$, as shown in Fig.~\ref{Fig:cpc-br-ratios-low}. For the former case, the product $\text{BR}(t\to H^+b)\times\text{BR}(H^+\to\tau^+\nu)$ is shown in Fig.~\ref{Fig:br-top} for three values of $\tan\beta$. Note that recent LHC data have already excluded a substantial region of the low-$\tan\beta$ and low-$M_{H^\pm}$ parameter region in Model~I, see section~\ref{subsect:LHCsearches}.

\subsection{The $H^+\to H_1 W^+$ partial width} 
\label{subsec:br-toWH1}
In this section we consider the decay mode $H^+\to H_1 W^+$, allowing for the possibility that the lightest Higgs boson, $H_1$, is not an eigenstate of CP.

The $H^+\to H_1W^+$ coupling is given by Eq.~(\ref{Eq:H_chWH_1}). The partial width, relative to its maximum value, is given by the quantity
\begin{equation} \label{Eq:HWH1}
\cos^2\alpha_2\sin^2(\beta-\alpha_1)+\sin^2\alpha_2,
\end{equation}
which is shown in Fig.~\ref{Fig:br-toWH1}. We note that there is no dependence on the mixing angle $\alpha_3$. If $\alpha_3=0$ or $\pm\pi/2$, then CP is conserved along the axis $\alpha_2=0$ with $H_1=h$.

\begin{figure}[htb]
\refstepcounter{figure}
\label{Fig:br-toWH1}
\addtocounter{figure}{-1}
\begin{center}
\vspace*{-3mm}
\includegraphics[scale=0.32]{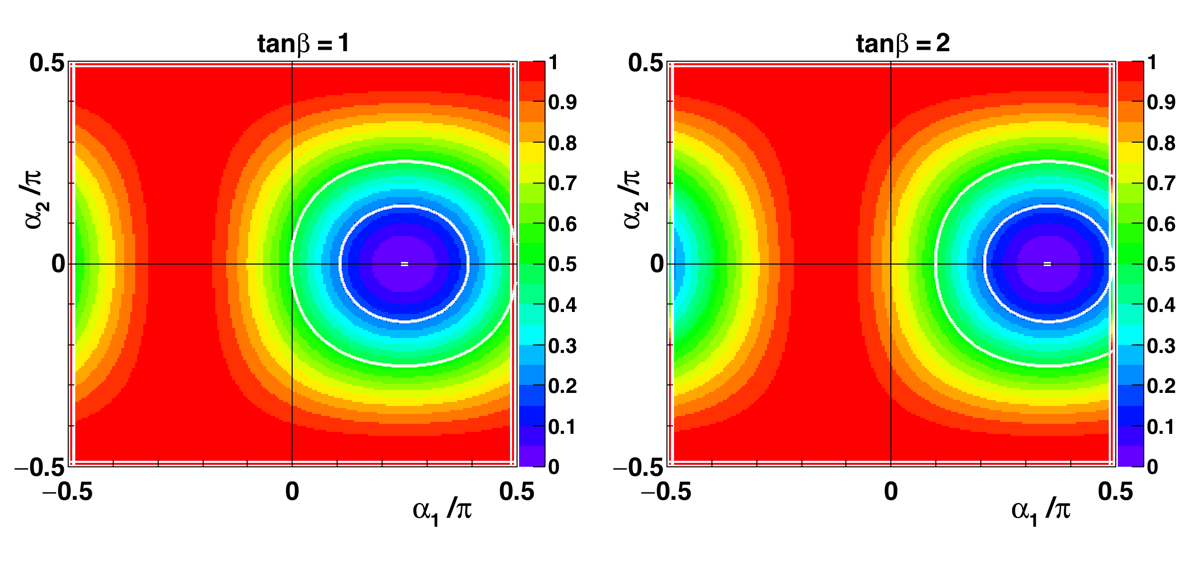}
\end{center}
\vspace*{-6mm}
\caption{Relative partial width for $H^+\to H_1 W^+$, given by Eq.~(\ref{Eq:HWH1}), vs $\alpha_1$ and $\alpha_2$, for $\tan\beta=1$ and 2. The white ``circles'' outline the region within which the $VVH_1$ coupling squared deviates by at most 10\% or 30\% from the SM value.}
\end{figure}

In the alignment limit,
\begin{equation}
\alpha_1=\beta, \quad \alpha_2=0,
\end{equation}
which is closely approached by the LHC data on the Higgs-gauge-boson coupling, the $H^+H_1W^+$ coupling actually vanishes. 

Hence, the $H^+\to H_1 W^+$ decay crucially depends on {\it some deviation from this limit}. 
We note that the $VVH_1$ coupling is proportional to $\cos\alpha_2\cos(\beta-\alpha_1)$. Thus, the deviation of the square of this coupling from unity (which represents the SM-limit), is given by the expression (\ref{Eq:HWH1}).
Note that the experimental constraint (on the deviation of the coupling squared from unity) is 15--20\% at the 95\% CL \cite{Khachatryan:2014jba,Aad:2015ona}.

For comparison, a recent study of decay modes that explicitly exhibit
CP violation in Model~II \cite{Fontes:2015xva}, compatible with all
experimental constraints, considers $\tan\beta$ values in the range 1.3 to 3.3, with parameter points displaced from the alignment limit by $\sqrt{(\Delta\alpha_1/\pi)^2+(\Delta\alpha_2/\pi)^2}$ ranging from 1.5\% to 83.2\% (the one furthest away has a negative value of $\alpha_1$).

This decay channel is also interesting for Model~I \cite{Keus:2015hva}.

\section{$H^+$ production mechanisms at the LHC} 
\label{sect:production}
\setcounter{equation}{0}
This section describes $H^+$ production and detection channels at
the LHC. Since a charged Higgs boson couples to mass, it will
predominantly be produced in connection with heavy fermions, $\tau$, $c$, $b$ and
$t$, or bosons, $W^\pm$ or $Z$, and likewise for the decays. 
The cross sections given here, are for illustration only. For the studies presented in sections~\ref{sect-benchmarks}--\ref{sect:inert-models} they are calculated independently.

We shall here split the
discussion of possible $H^+$ production mechanisms into two mass regimes,
according to whether the charged Higgs boson can be produced (in the on-shell approximation) in a top
decay or whether it could decay to a top and a bottom quark. 
These two mass regimes will be referred to as ``low'' and ``high'' $M_{H^\pm}$ mass, respectively.

While discussing such processes in hadron-hadron collisions
one should be aware that there are two approaches to the treatment of heavy quarks in the initial state.
One may take the heavy flavors as being generated from the gluons, then the
relevant number of active quarks is $N_f=4$ (or sometimes 3).
Alternatively, the $b$-quark can be  included as a constituent of the hadron, then an $N_f=5$ parton density should be used in the calculation of the corresponding cross section.
These two approaches are referred to as the 4-flavor and 5-flavor schemes, abbreviated 4FS and 5FS.
This should be kept in mind when referring to the lists of possible subprocesses initiated by heavy quarks and the corresponding figures in the following discussion.
Below, we will use the notation $q'$, $Q$ and $Q'$ to denote quarks which are not $b$-quarks. We only indicate $b$-quarks when they couple to Higgs bosons, thus enhancing the rate.

For some discussions it is useful to distinguish ``bosonic'' and ``fermionic'' production mechanisms, since the former, corresponding to final states involving only $H^+$ and $W^-$, may proceed via an intermediate neutral Higgs, and thus depend strongly on its mass, see e.g., Ref.~\cite{Basso:2015dka}.

\subsection{Production processes} 
Below, we list all important $H^+$ production processes represented in Figs.~\ref{Fig:feyn-figures-single-cs}-\ref{Fig:feyn-figures-pair-def} in the 5FS.\footnote{Charge-conjugated processes are not shown separately. Higgs radiation from initial-state quarks are not shown explicitly.}

\begin{figure}[htb]
\refstepcounter{figure}
\label{Fig:feyn-figures-hwetc}
\addtocounter{figure}{-1}
\begin{center}
\includegraphics[scale=0.55]{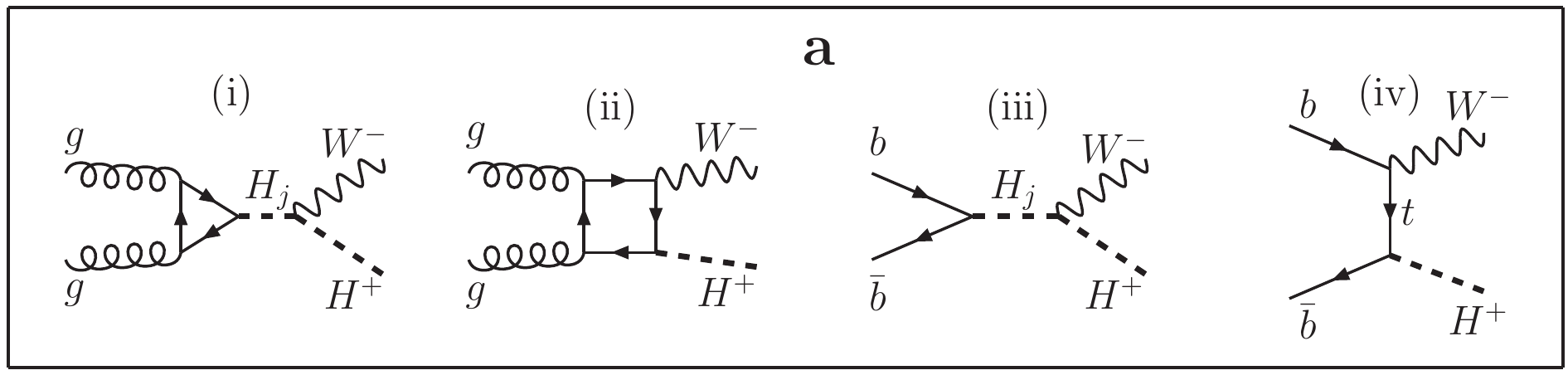} \\
\includegraphics[scale=0.55]{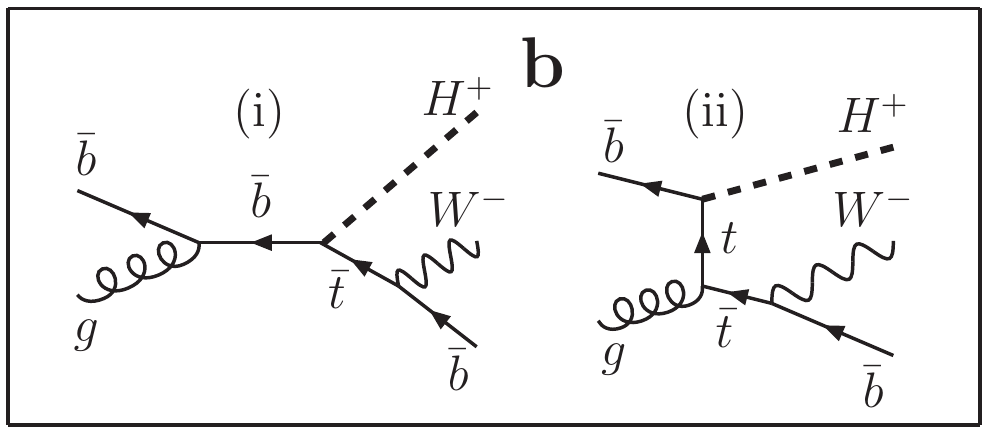}
\end{center}
\vspace*{-4mm}
\caption{Feynman diagrams for the production processes (\ref{Eq:WH}) and (\ref{Eq:WbH}).}
\end{figure}

\subsubsection{Single $H^+$ production} 
\label{sect:single-H+-production}

A single $H^+$ can be accompanied by a $W^-$ (Fig.~\ref{Fig:feyn-figures-hwetc}a, ``bosonic'') \cite{Dicus:1989vf,BarrientosBendezu:1998gd,Moretti:1998xq,BarrientosBendezu:1999vd,Brein:2000cv,Hollik:2001hy,Asakawa:2005nx,Eriksson:2006yt,Hashemi:2010ce}:
\begin{subequations} \label{Eq:WH}
\begin{align} 
gg&\to W^-H^+, \label{Eq:gg-WH}\\
b\bar b&\to W^-H^+,
\end{align}
\end{subequations}
or by a $W^-$ and a $b$ jet (Fig.~\ref{Fig:feyn-figures-hwetc}b, ``fermionic'') \cite{Gunion:1986pe,DiazCruz:1992gg,Moretti:1996ra,Miller:1999bm,Moretti:1999bw,Zhu:2001nt,Plehn:2002vy,Berger:2003sm,Kidonakis:2004ib,Weydert:2009vr,Kidonakis:2010ux,Flechl:2014wfa,Degrande:2015vpa,Kidonakis:2016eeu,Degrande:2016hyf}:\footnote{Note that in the 5FS (\ref{Eq:WbH}) can be a tree-level process, whereas (\ref{Eq:gg-WH}) can not.}
\begin{equation}
g\bar b \ (\to \bar t H^+) \to \bar b W^- H^+. \label{Eq:WbH} 
\end{equation}

The pioneering study \cite{Dicus:1989vf} of the bosonic process (\ref{Eq:WH}) already discussed both the triangle and box contributions to the one-loop $gg$-initiated production, but considered massless $b$-quarks, i.e., the $b$-quark Yukawa couplings were omitted. This was subsequently restored in a complete one-loop calculation of the $gg$-initiated process \cite{BarrientosBendezu:1998gd,BarrientosBendezu:1999vd}, and it was realized that there can be a strong cancellation between the triangle- and box diagrams. This interplay of triangle and box diagrams has also been explored in the MSSM \cite{Brein:2000cv}.

NLO QCD corrections to the $b\bar b$-initiated production process were found to reduce the cross section by ${\cal O}(10-30\%)$ \cite{Hollik:2001hy}. On the other hand, possible $s$-channel resonant production via heavier neutral Higgs bosons (see Fig.~\ref{Fig:feyn-figures-hwetc}a (i) and (iii)) was seen to provide possible enhancements of up to two orders of magnitude \cite{Asakawa:2005nx}. These authors also pointed out that one should use running-mass Yukawa couplings, an effect which significantly reduced the cross section at high mass \cite{Eriksson:2006yt}.

A first comparison of the $H^+\to t\bar b$ signal with the $t\bar t$ background \cite{Moretti:1998xq} (in the context of the MSSM) concluded that the signal could not be extracted from the
background. More optimistic conclusions were reached for the $H^+\to \tau^+\nu$ channel \cite{Eriksson:2006yt,Hashemi:2010ce}, again in the context of the MSSM.

The first study \cite{Gunion:1986pe} of the fermionic process (\ref{Eq:WbH}) pointed out that there is a double counting issue (see sect.~\ref{sect:double counting}).
Subsequently, it was realized \cite{DiazCruz:1992gg,Borzumati:1999th} that the $g\bar b\to H^+\bar t$ process could be described as $gg\to H^+\bar t b$, where a gluon splits into $b\bar b$ and one of these is not observed. As mentioned above, this approach is in recent literature referred to as the four-flavor scheme (4FS) whereas in the five-flavor scheme (5FS) one considers $b$-quarks as proton constituents.

NLO QCD corrections to the $g\bar b\to H^+\bar t$ cross section have been calculated \cite{Zhu:2001nt,Plehn:2002vy,Degrande:2016hyf}, and the resulting scale dependence studied \cite{Plehn:2002vy,Berger:2003sm}, both in the 5FS and the 4FS. In a series of papers by Kidonakis \cite{Kidonakis:2004ib,Kidonakis:2010ux,Kidonakis:2016eeu}, soft-gluon corrections have been included at the ``approximate NNLO'' order and found to be significant near threshold, i.e., for heavy $H^+$.
A recent study \cite{Degrande:2016hyf} is devoted to total cross sections in the intermediate-mass region, $M_{H^+}\sim m_t$, providing a reliable interpolation between low and high masses.

These fixed-order cross section calculations have been merged with parton showers \cite{Alwall:2004xw,Weydert:2009vr,Flechl:2014wfa,Degrande:2015vpa}, both at LO and NLO, in the 4FS and in the 5FS. The 5FS results are found to exhibit less scale dependence \cite{Degrande:2015vpa}.

Different background studies \cite{Moretti:1996ra,Miller:1999bm,Moretti:1999bw} compared triple $b$-tagging vs 4-$b$-tagging, identifying parameter regions where either is more efficient.

\begin{figure}[htb]
\refstepcounter{figure}
\label{Fig:feyn-figures-hqq}
\addtocounter{figure}{-1}
\begin{center}
\includegraphics[scale=0.55]{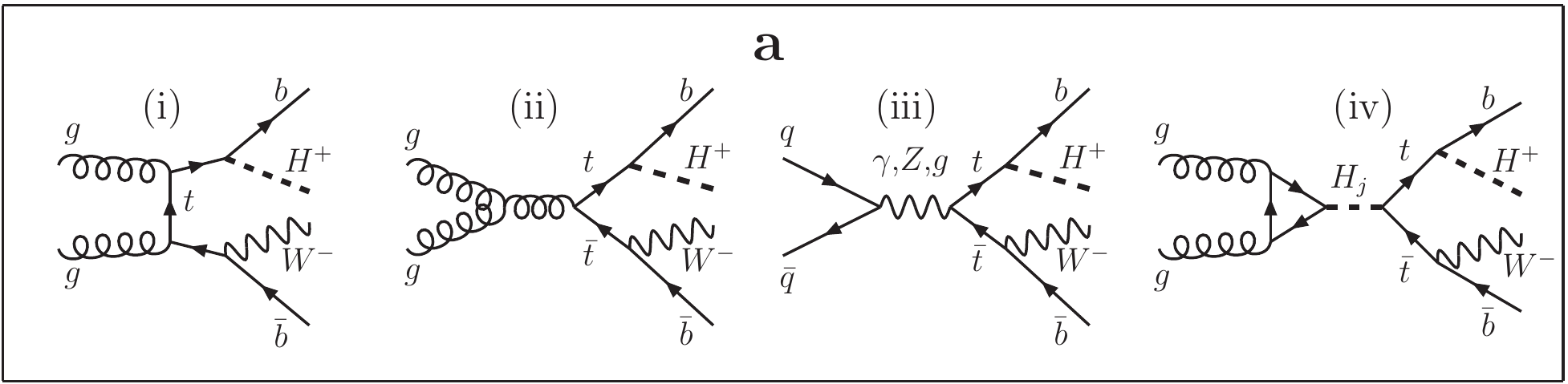}\\
\includegraphics[scale=0.55]{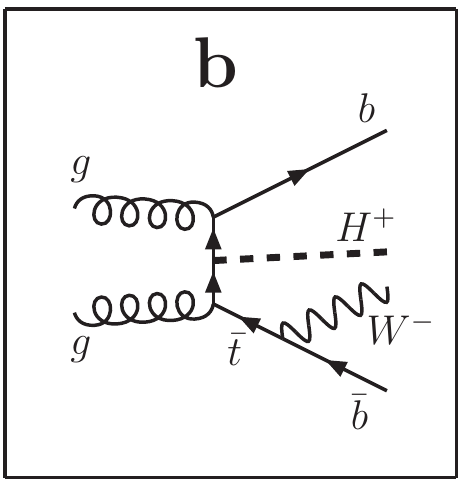}
\end{center}
\vspace*{-4mm}
\caption{Feynman diagrams for the production processes (\ref{Eq:WHbb}).  }
\end{figure}

\begin{figure}[htb]
\refstepcounter{figure}
\label{Fig:feyn-figures-hqq=bcd}
\addtocounter{figure}{-1}
\begin{center}
\includegraphics[scale=0.55]{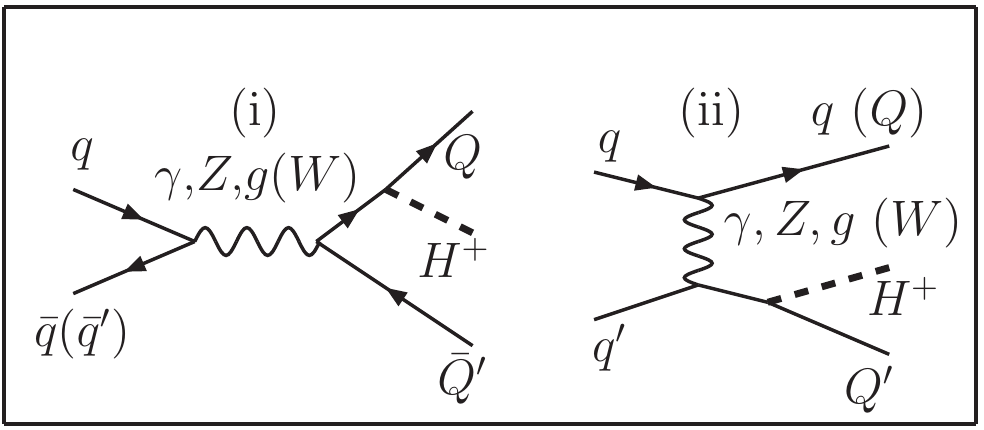}
\end{center}
\vspace*{-6mm}
\caption{Feynman diagrams for the production processes (\ref{Eq:Hqq}). If the line has no arrow, it represents either a quark or an antiquark.}
\end{figure}

\begin{figure}[htb]
\refstepcounter{figure}
\label{Fig:feyn-figures-arhrib}
\addtocounter{figure}{-1}
\begin{center}
\includegraphics[scale=0.55]{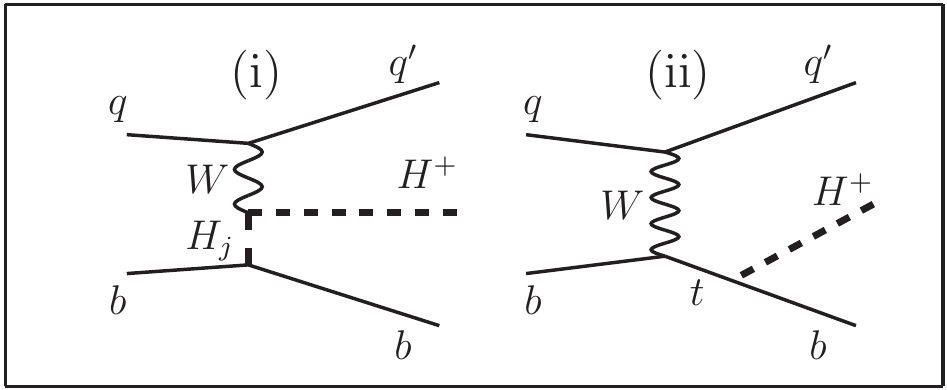}
\end{center}
\vspace*{-6mm}
\caption{Feynman diagrams for the production processes (\ref{Eq:arhrib}).}
\end{figure}

\begin{figure}[htb]
\refstepcounter{figure}
\label{Fig:feyn-figures-single-cs}
\addtocounter{figure}{-1}
\begin{center}
\includegraphics[scale=0.55]{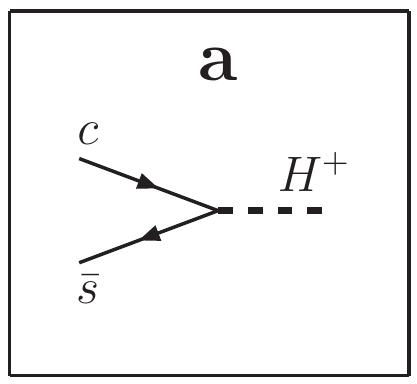}
\hspace{1mm}
\includegraphics[scale=0.55]{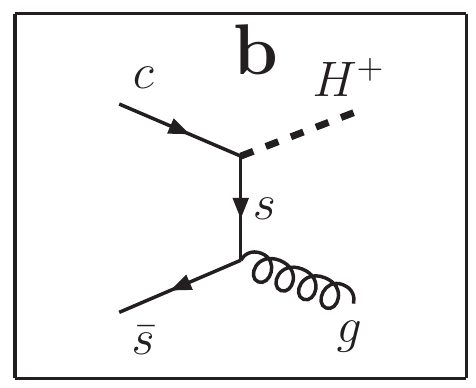}
\end{center}
\vspace*{-4mm}
\caption{Feynman diagrams for the production processes (\ref{Eq:single-production-cs}).}
\end{figure}

In addition to the importance of the $t\bar t$ channel at low mass, 
the following processes containing two accompanying $b$ jets (see Fig.~\ref{Fig:feyn-figures-hqq}) are important at high charged-Higgs mass:
\begin{subequations}\label{Eq:WHbb}
\begin{gather} 
gg,q\bar q,b\bar b\ (\to t\bar t \to b \bar t H^+) \to b\bar b W^- H^+,  \\
gg, q\bar q \ (\to b\bar t H^+) \to b\bar b W^- H^+.  \label{Eq:btH}
\end{gather}
\end{subequations}
There are also processes with a single $H^+$ and two jets (see Fig.~\ref{Fig:feyn-figures-hqq=bcd}):
\begin{equation}
\text{(i): }q\bar q (\bar q')\to Q\bar Q' H^+, \quad \text{(ii): }qq'\to q (Q)Q'H^+. \label{Eq:Hqq}
\end{equation}
In this particular case, with many possible gauge boson couplings, one of the final-state jets could be a $b$.

In addition, single $H^+$ production can be initiated by a $b$-quark,
\begin{equation} \label{Eq:arhrib}
qb\to q^\prime H^+b,
\end{equation}
as illustrated in Fig.~\ref{Fig:feyn-figures-arhrib}.

In the 5FS, single $H^+$ production can also take place from $c$ and $s$ quarks, typically accompanied by a gluon jet \cite{He:1998ie,DiazCruz:2001gf,Slabospitsky:2002gw,Dittmaier:2007uw} (Fig.~\ref{Fig:feyn-figures-single-cs}):
\begin{subequations} \label{Eq:single-production-cs}
\begin{align}
c\bar s&\to H^+, \\
c\bar s&\to H^+ g.
\end{align}
\end{subequations}
Similarly, one can consider $c\bar b$ initial states.

At infinite order the 4FS and the 5FS should only differ by terms of ${\cal O}(m_b)$, but the perturbation series of the two schemes are organized differently. Some authors (see, e.g., Ref.~\cite{Flechl:2014wfa}) advocate combining the two schemes according to the ``Santander matching'' \cite{Harlander:2011aa}:
\begin{equation}
\sigma=\frac{\sigma(4\text{FS})+w\sigma(5\text{FS})}{1+w},
\end{equation}
with the relative weight factor
\begin{equation}
w=\log\frac{M_{H^\pm}}{m_b}-2,
\end{equation}
since the difference between the two schemes is logarithmic, and in the limit of $M_{H^\pm}\gg m_b$ the 5FS should be exact.

\subsubsection{The double counting and NWA issues}
\label{sect:double counting}
A $b$-quark in the initial state may be seen as a constituent of the proton (5FS), or as resulting from the gluon splitting into $b\bar b$ (4FS). Adding $gg\to b\bar b g \to b H^+\bar t$ (with one $b$ possibly not detected) and $g\bar b\to H^+\bar t$ in the 5FS one may therefore commit double counting \cite {Barnett:1987jw,Olness:1987ep}. 
The resolution lies in subtracting a suitably defined infrared-divergent part of the gluon-initiated amplitude \cite{Alwall:2004xw}.\footnote{For a complete discussion on the flavour scheme choice in inclusive charged Higgs production associated with fermions see IV.3.2 of \cite{deFlorian:2016spz} and references therein.}
The problem can largely be circumvented by choosing either the 5FS or the 4FS.
For a more pragmatic approach, see Refs.~\cite{Belyaev:2001qm,Belyaev:2002eq}.

\begin{figure}[htb]
\refstepcounter{figure}
\label{Fig:feyn-figures-single-associated}
\addtocounter{figure}{-1}
\begin{center}
\includegraphics[scale=0.55]{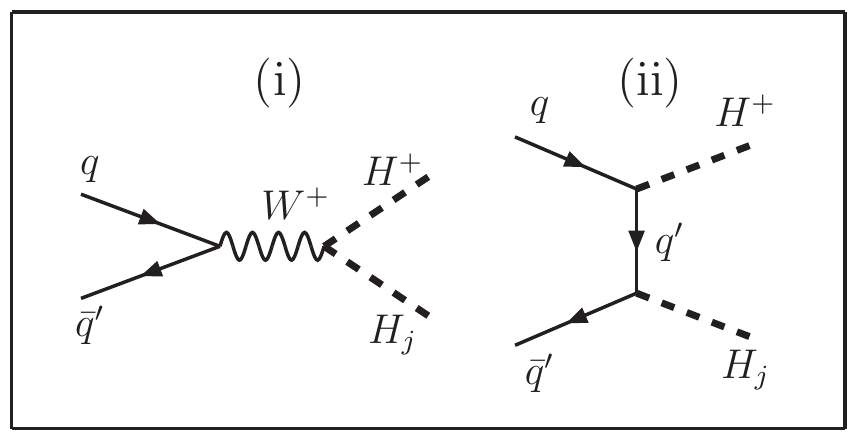}
\end{center}
\vspace*{-6mm}
\caption{Feynman diagrams for the production processes (\ref{Eq:single-associated}).}
\end{figure}

\begin{figure}[htb]
\refstepcounter{figure}
\label{Fig:feyn-figures-pair}
\addtocounter{figure}{-1}
\begin{center}
\includegraphics[scale=0.55]{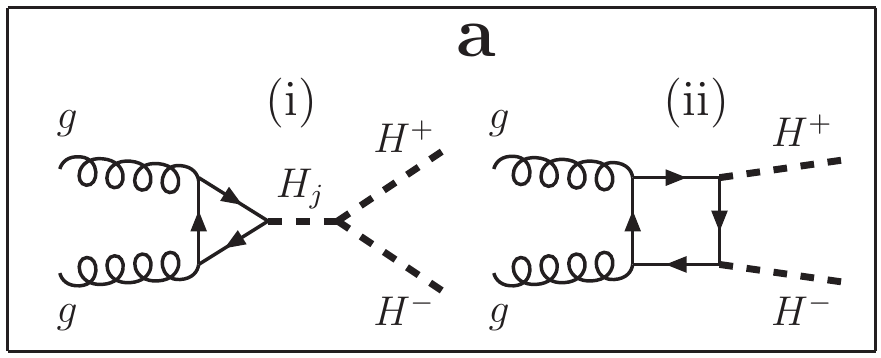}
\hspace*{-2mm}
\includegraphics[scale=0.55]{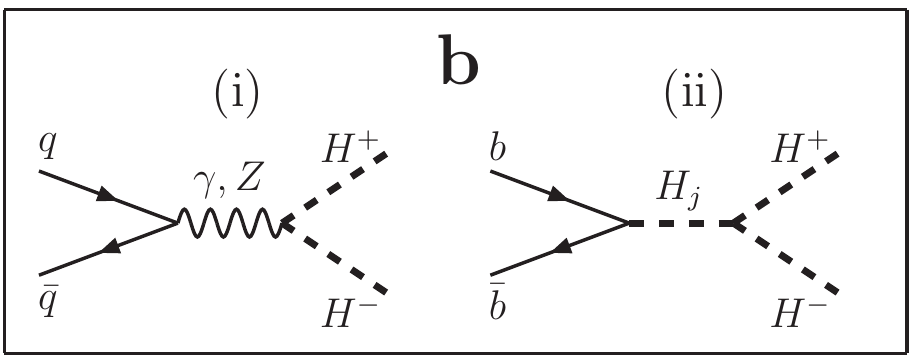} 
\end{center}
\vspace*{-6mm}
\caption{Feynman diagrams for the pair production processes (\ref{Eq:gg_HH}).}
\end{figure}

\begin{figure}[htb]
\refstepcounter{figure}
\label{Fig:feyn-figures-pair-def}
\addtocounter{figure}{-1}
\begin{center}
\includegraphics[scale=0.55]{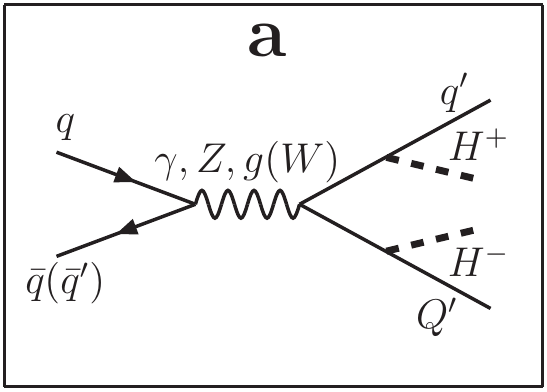}
\hspace*{-2mm}
\includegraphics[scale=0.55]{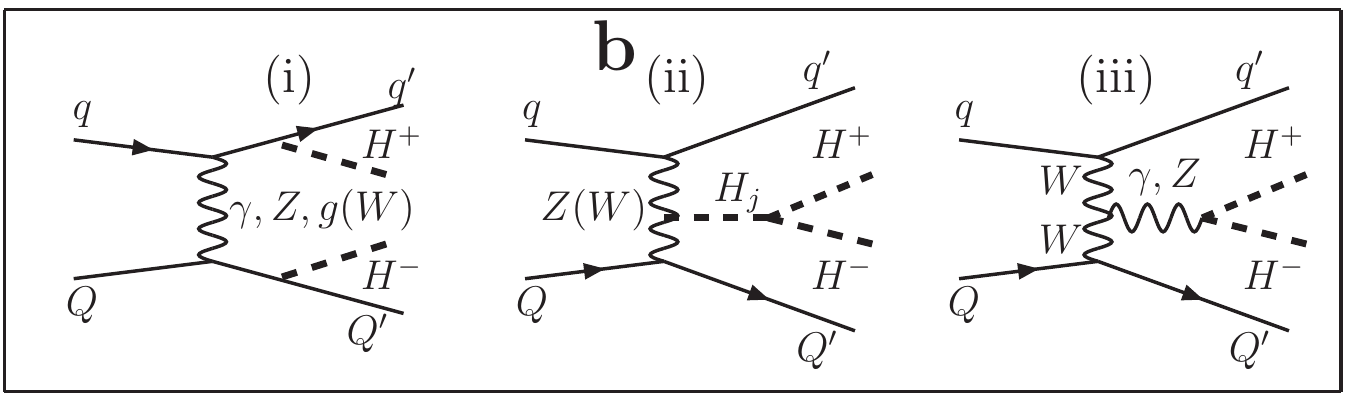}
\end{center}
\vspace*{-6mm}
\caption{Feynman diagrams for the pair production processes (\ref{Eq:VBF}).}
\end{figure}

A related issue is the one of low-mass $H^+$ production via $t$-quark decay, $gg,q\bar q\to t\bar t$ followed by $t\to H^+ b$ (with $\bar t$ a spectator), usually treated in the Narrow Width Approximation (NWA). 
The NWA however fails the closer the top and charged Higgs masses are, in which case the finite top width needs to be accounted for, which in turn implies that the full gauge invariant set of diagrams yielding $gg,q\bar q\to H^+b \bar t$ has to be computed. 
Considerable effort has been devoted to understanding this implementation, see also Refs.~\cite{Guchait:2001pi,Alwall:2003tc,Assamagan:2004gv}.

\subsubsection{$H^+H_j$ and $H^+H^-$ production} 

We can have a single $H^+$ production in association with a neutral Higgs boson $H_j$ \cite{Kanemura:2001hz,Akeroyd:2003bt,Akeroyd:2003jp,Cao:2003tr,Belyaev:2006rf,Miao:2010rg}:
\begin{equation} 
\label{Eq:single-associated}
q\bar q' \to H^+ H_j,
\end{equation}
as shown in Fig.~\ref{Fig:feyn-figures-single-associated}.

For $H^+H^-$ pair production we have \cite{Eichten:1984eu,Willenbrock:1986ry,Glover:1987nx,Dicus:1987ic,Jiang:1997cg,Krause:1997rc,BarrientosBendezu:1999gp,Brein:1999sy,Moretti:2001pp,Moretti:2003px,Alves:2005kr}:
\begin{subequations}\label{Eq:pair-production}
\begin{gather} 
gg, q\bar q, b \bar b \to H^+H^-, \label{Eq:gg_HH} \\
q\bar q(\bar q'), qQ \to q'Q' H^+H^-, \label{Eq:VBF}
\end{gather}
\end{subequations}
as illustrated in Figs.~\ref{Fig:feyn-figures-pair} and \ref{Fig:feyn-figures-pair-def}, respectively.
These mechanisms would be important for light charged Higgs bosons, as allowed in Models~I and X.

\subsection{Production cross sections} 

In this section, predictions for single Higgs production at 14~TeV for the CP-conserving 2HDM, Models~I and II (valid also for X and Y) are discussed.

In Fig.~\ref{Fig:sigma-production}, $pp\to H^+ X$ cross sections for the main production channels are shown at leading order, sorted by the parton-level mechanism \cite{Basso:2015dka}\footnote{In the Feynman diagrams $t$ is represented by its dominant decay products $W^+ b$.}. The relevant partonic channels can be categorized as:
\begin{itemize}
\item ``fermionic'':\quad
$g\bar{b}\to H^+ \bar{t}$, Fig.~\ref{Fig:feyn-figures-hwetc} {\bf b} (solid),
\item ``fermionic'':\quad
$gg\to H^+ b\bar{t}$, Fig.~\ref{Fig:feyn-figures-hqq} {\bf a}, {\bf b} (dotted),
\item ``bosonic'':\quad
$gg\to H_j\to H^+ W^-$, Fig.~\ref{Fig:feyn-figures-hwetc} {\bf a} (i) (dash-dotted).
\end{itemize}
The charge-conjugated channels are understood to be added unless specified otherwise. No constraints are imposed here, neither from theory (like positivity, unitarity), nor from experiments.

The CTEQ6L (5FS) parton distribution functions \cite{Pumplin:2002vw} are adopted here, with the scale $\mu=M_H$.
Three values of $\tan\beta$ are considered, and $M_H$ and $M_A$ are held fixed at $(M_H,M_A)=(500,600)~\text{GeV}$. Furthermore, we consider the CP-conserving alignment limit, with $\sin(\beta-\alpha)=1$. 
The bosonic cross section is accompanied by a next-to-leading order QCD $K$-factor enhancement \cite{Spira:1995rr}. 

\begin{figure}[htb]
\refstepcounter{figure}
\label{Fig:sigma-production}
\addtocounter{figure}{-1}
\begin{center}
\includegraphics[scale=0.80]{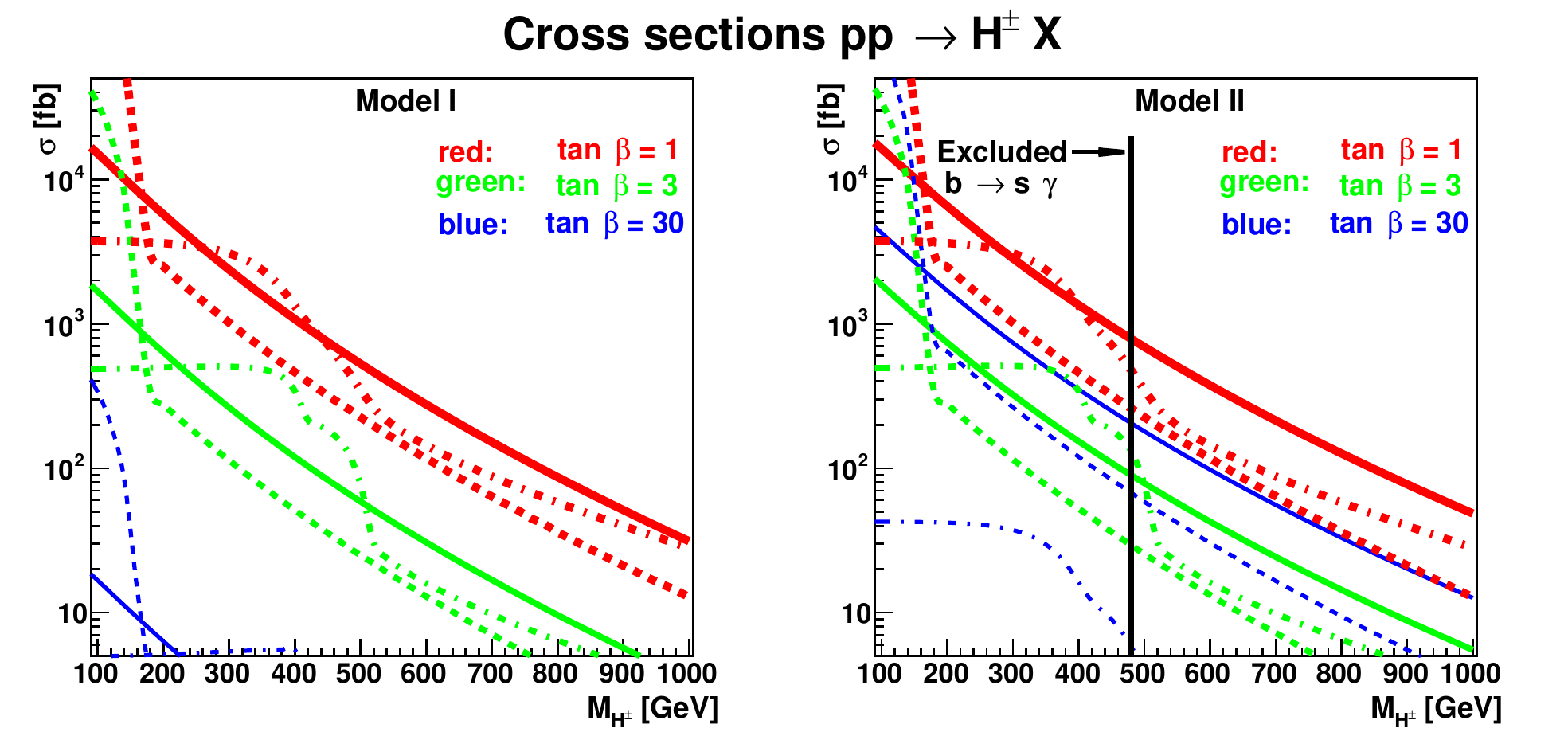}
\end{center}
\vspace*{-4mm}
\caption{Charged Higgs production cross sections in the 2HDM, at 14 TeV.
Left: Model~I (or X). Right: Model~II (or Y). Solid and dotted curves refer to ``fermionic'' channels, whereas dash-dotted refer to ``bosonic'' ones (see text).}
\end{figure}

Several points are worth mentioning:
\begin{itemize}
\item
To any contribution at fixed order in the perturbative expansion of the gauge coupling, the three cross sections are to be merged with regards to the interpretation in different flavour schemes, as discussed above. In the following, we focus on the first fermionic channel in the 5FS at the tree level.
\item
The enhancement exhibited by the dotted curve at low masses is due to resonant production of $t$-quarks which decay to $H^+b$. However, in Model~I this mode is essentially excluded by LHC data (see section~\ref{sect:LHC-search-summary}), and in Model~II it is excluded by the $B\to X_s\gamma$-constraint (see section~\ref{sect:constraint-h_loops}).
\item
Model~I differs from Model~II also for $\tan\beta=1$, because of a different relative sign between the Yukawa couplings proportional to $m_t$ and those proportional to $m_b$, see Table~\ref{tab:couplings}.
\item
Models~X and Y will have the same production cross sections as Models~I and II, respectively, but the sensitivity in the $\tau\nu$-channel would be different.
\item
The bumpy structure seen for the bosonic mode is due to resonant production of neutral Higgs bosons, and depends on the values of $M_H$ and $M_A$.
Note that in the MSSM the masses of the heavier neutral Higgs bosons are close to that of the charged one, and this resonant behavior is absent.
\end{itemize}

While recent studies (see section~\ref{sect:single-H+-production}) provide a more accurate calculation of the $g\bar b \to H^+\bar t$ cross section than what is given here, they typically leave out the 2HDM model-specific $s$-channel (possibly resonant) contribution to the cross section.

\begin{figure}[htb]
\refstepcounter{figure}
\label{Fig:sigma-production-b}
\addtocounter{figure}{-1}
\begin{center}
\includegraphics[scale=0.50]{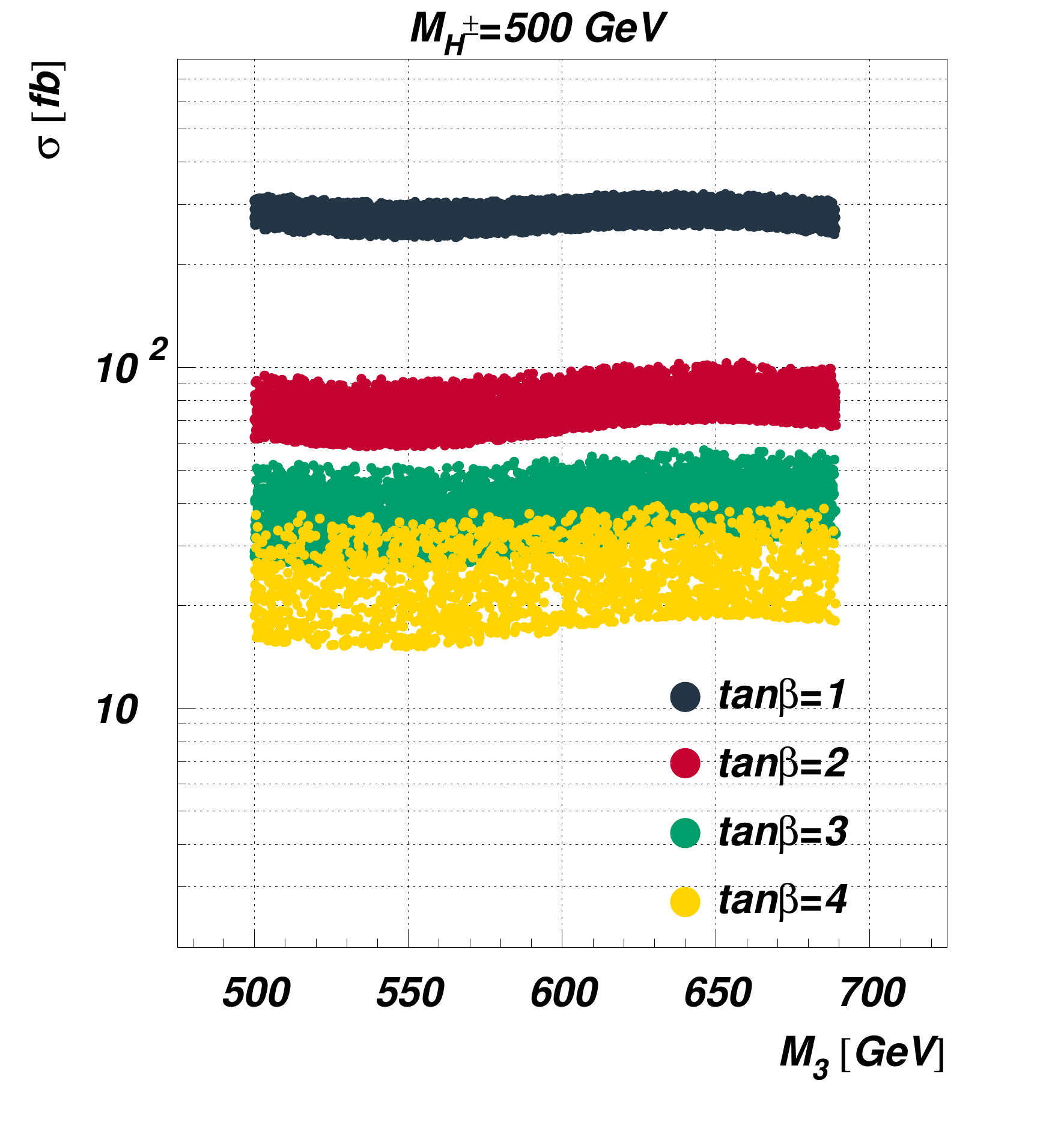}
\end{center}
\vspace*{-4mm}
\caption{Charged Higgs bosonic production cross sections in the 2HDM, Model~II, for 14~TeV,
and a fixed value of $M_{H^\pm}=500~\text{GeV}$,
plotted vs $M_3\equiv M_A$ for $\tan\beta=1$, 2, 3 and 4.}
\end{figure}

In Fig.~\ref{Fig:sigma-production-b}, the bosonic charged-Higgs
production cross section vs $M_3\equiv M_A$ for a set of CP-conserving
parameter points that satisfy the theoretical and experimental
constraints \cite{Basso:2015dka} (see also
\cite{Basso:2012st,Basso:2013wna}) are presented. These are shown in different
colors for different values of $\tan\beta$. The spread in cross
section values for each value of $\tan\beta$ and $M_A$ reflects the
range of allowed values of the other parameters scanned over, namely $\mu$, $M_H$ and $\alpha$.

Low values of $\tan\beta$ are enhanced for the bosonic mode due to the contribution of the $t$-quark in the loop, whereas the modulation is due to resonant $A$ production. In the CP-violating case, this modulation is more pronounced \cite{Basso:2015dka}.

As summarized by the LHC Top Physics Working Group 
the $pp \to t \bar t$ cross section has been calculated at next-to-next-to leading order (NNLO) in QCD including resummation of next-to-next-to-leading logarithmic (NNLL) soft gluon terms with the software {\tt Top++2.0}~\cite{Beneke:2011mq, Cacciari:2011hy, Czakon:2011xx, Baernreuther:2012ws, Czakon:2012zr, Czakon:2012pz,Czakon:2013goa}. The decay
width $\Gamma (t \to b W^+)$ is available at NNLO~\cite{Czarnecki:1998qc, Chetyrkin:1999ju, Blokland:2004ye, Blokland:2005vq, Czarnecki:2010gb, Gao:2012ja, Brucherseifer:2013iv},
while the decay width $\Gamma (t \to b H^+)$  is available at NLO~\cite{Czarnecki:1992zm}.

\section{Experimental constraints} 
\label{sect:ex-constraints}
\setcounter{equation}{0}

Here we review various experimental constraints for charged Higgs bosons derived from different low (mainly $B$-physics) and high (mainly LEP, Tevatron and LHC) energy processes. Also some relevant information on the neutral Higgs sector is presented.
Some observables depend solely on $H^+$ exchange, and are thus independent of CP violation in the potential, whereas other constraints depend on the exchange of neutral Higgs bosons, and are sensitive to the CP violation introduced via the mixing discussed in subsection~\ref{subsect:masses}.
Due to the possibility of $H^+$, in addition to $W^+$
exchange, we are getting constraints from a variety of processes, some at tree and some at the loop level.
In addition, we present general constraints coming from electroweak precision measurements, $S$, $T$, the muon magnetic moment and the electric dipole moment of the electron.
The experimental constraints listed below are valid only for Model II, if not stated otherwise.\footnote{Analyses with general Yukawa couplings can be found in Refs.~\cite{Mahmoudi:2009zx} and \cite{Crivellin:2013wna}.}
Also, some of the constraints are updated, with respect to those used in the studies presented in later sections.

The charged-Higgs contribution may substantially modify the branching ratios for $\tau\nu_\tau$-production in $B$-decays \cite{Krawczyk:1987zj}.  An attempt to describe various $\tau$ and $B$ anomalies (also $W\to \tau \nu$) in the 2HDM, Model III, with a novel ansatz relating up- and down-type Yukawa couplings, can be found in \cite{Cline:2015lqp}. 
This analysis points towards an $H^+$ mass around 100~GeV, with masses of other neutral Higgs bosons in the range 100--125 GeV.
A similar approach to describe various low energy anomalies by introducing additional scalars
can be found in \cite{Crivellin:2015hha}. Here, a lepton-specific 2HDM (i.e., of type X)
with non-standard Yukawa couplings has been analysed with
the second neutral CP-even Higgs boson light (below 100~GeV) and a
relatively light $H^+$, with a mass of the order of 200~GeV.

\subsection{Low-energy constraints}

As mentioned above, several decays involving heavy-flavor quarks could be affected by $H^+$ in addition to $W^+$-exchange. Data on such processes provide constraints on the coupling (represented by $\tan\beta$) and the mass, $M_{H^\pm}$. Below, we discuss the most important ones.

\subsubsection{Constraints from $H^+$ tree-level exchange}

\paragraph{{\boldmath $B\to \tau \nu_\tau (X)$}:}
The measurement of the branching ratio of the inclusive process
$B\to \tau \nu_\tau X$ \cite{Abbiendi:2001fi} leads to the
following constraint, at the $95\%$ CL,
\begin{equation}
\frac{\tan\beta}{M_{H^\pm}} <  0.53 \gev^{-1}.
\end{equation}
This is in fact a very weak constraint. (A similar result can be
obtained from the leptonic tau decays at the tree level
\cite{Krawczyk:2004na}.)  A more recent measurement for the exclusive case gives 
$\text{BR}(B\to\tau\nu_\tau) =(1.14\pm0.27)\times10^{-4}$ 
\cite{Agashe:2014kda}\footnote{The error of the $B\to \tau \nu$ measurement, given by HFAG \cite{Amhis:2014hma} and released after the PDG 2014 \cite{Agashe:2014kda}, is slightly lower: ($1.14 \pm 0.22)\times10^{-4}$.}. 
With a Standard Model prediction of $(0.733\pm0.141)\times10^{-4}$
\cite{Charles:2004jd}\footnote{We have added in quadrature symmetrized statistical and systematic errors.} , we obtain
\begin{equation}
r_{H\,\text{exp}}=\frac{\text{BR}(B\to\tau\nu_\tau)}
{\text{BR}(B\to\tau\nu_\tau)_\text{SM}}
=1.56\pm0.47.
\end{equation}
Interpreted in the framework of the 2HDM at the tree level, one finds
\cite{Hou:1992sy,Grossman:1994ax,Grossman:1995yp}
\begin{equation}
r_{H\,\text{2HDM}}=\biggl[1-\frac{m_B^2}{M_{H^\pm}^2}\,\tan^2\beta\biggr]^2.
\end{equation}
Two sectors of the ratio $\tan\beta/M_{H^\pm}$ are excluded.
Note that this exclusion is relevant for high values of $\tan\beta$.

\paragraph{{\boldmath $B\to D\tau \nu_\tau$}:}
The ratios \cite{Aubert:2007dsa}
\begin{equation}
R^\text{exp}(D^{(*)})
=\frac{\text{BR}(B\to D^{(*)}\tau\nu_\tau)}{\text{BR}(B\to D^{(*)}\ell\nu_\ell)},
\quad \ell=e,\mu,
\end{equation}
are sensitive to $H^+$-exchange, and
lead to constraints similar to the one following from $B\to \tau \nu_\tau X$ \cite{Nierste:2008qe}.
In fact, there has been some tension between BaBar results \cite{Aubert:2007dsa,Lees:2012xj,Lees:2013uzd} and both the 2HDM (II) and the SM. These ratios have also been measured by Belle \cite{Huschle:2015rga,Abdesselam:2016cgx} and LHCb \cite{Aaij:2015yra}. Recent averages \cite{Freytsis:2015qca,Cline:2015lqp} are summarized in Table~\ref{table:BDtaunu}, together with the SM predictions \cite{Fajfer:2012vx,Lattice:2015rga,Na:2015kha}.
They are compatible at the $2\sigma$--$3\sigma$ level. A comparison with the 2HDM (II) concludes \cite{Huschle:2015rga} that the results are compatible for $\tan\beta/M_{H^\pm}=0.5/\text{GeV}$. 
However, in view of the high values for $M_{H^\pm}$ required by the $B
\to X_s \gamma$ constraint, uncomfortably high values of $\tan\beta$ would be required.
The studies given for Model~II in section~\ref{sect:benchmarks-high} do not take this constraint into account.

\begin{table}[ht]
\begin{center}
\begin{tabular}{|c|c|c|c|}
\hline
Ratio & Experiment & SM  \\
\hline
 $R(D^{*})$ & $0.321\pm 0.021$ & $0.252\pm 0.005$ \\
 $R(D)$ \cite{Freytsis:2015qca} & $0.388\pm 0.047$ & $0.300\pm 0.010$ \\
$R(D)$ \cite{Cline:2015lqp} & $0.408\pm 0.050$&$0.297\pm0.017$\\ 
\hline
\end{tabular}
\end{center}
\caption{Average experimental values \cite{Freytsis:2015qca,Cline:2015lqp} and SM predictions \cite{Fajfer:2012vx,Lattice:2015rga,Na:2015kha,Cline:2015lqp,Crivellin:2015hha }.
\label{table:BDtaunu}}
\end{table}

\paragraph{{\boldmath $D_s\to \tau \nu_\tau$}:}
Severe constraints can be obtained, which are competitive with those from $B\to\tau \nu_\tau$ \cite{Akeroyd:2009tn}.

\subsubsection{Constraints from $H^+$ loop-level exchange}
\label{sect:constraint-h_loops}

\begin{figure}[htb]
\refstepcounter{figure}
\label{Fig:misiak-br-models}
\addtocounter{figure}{-1}
\begin{center}
\includegraphics[scale=0.38]{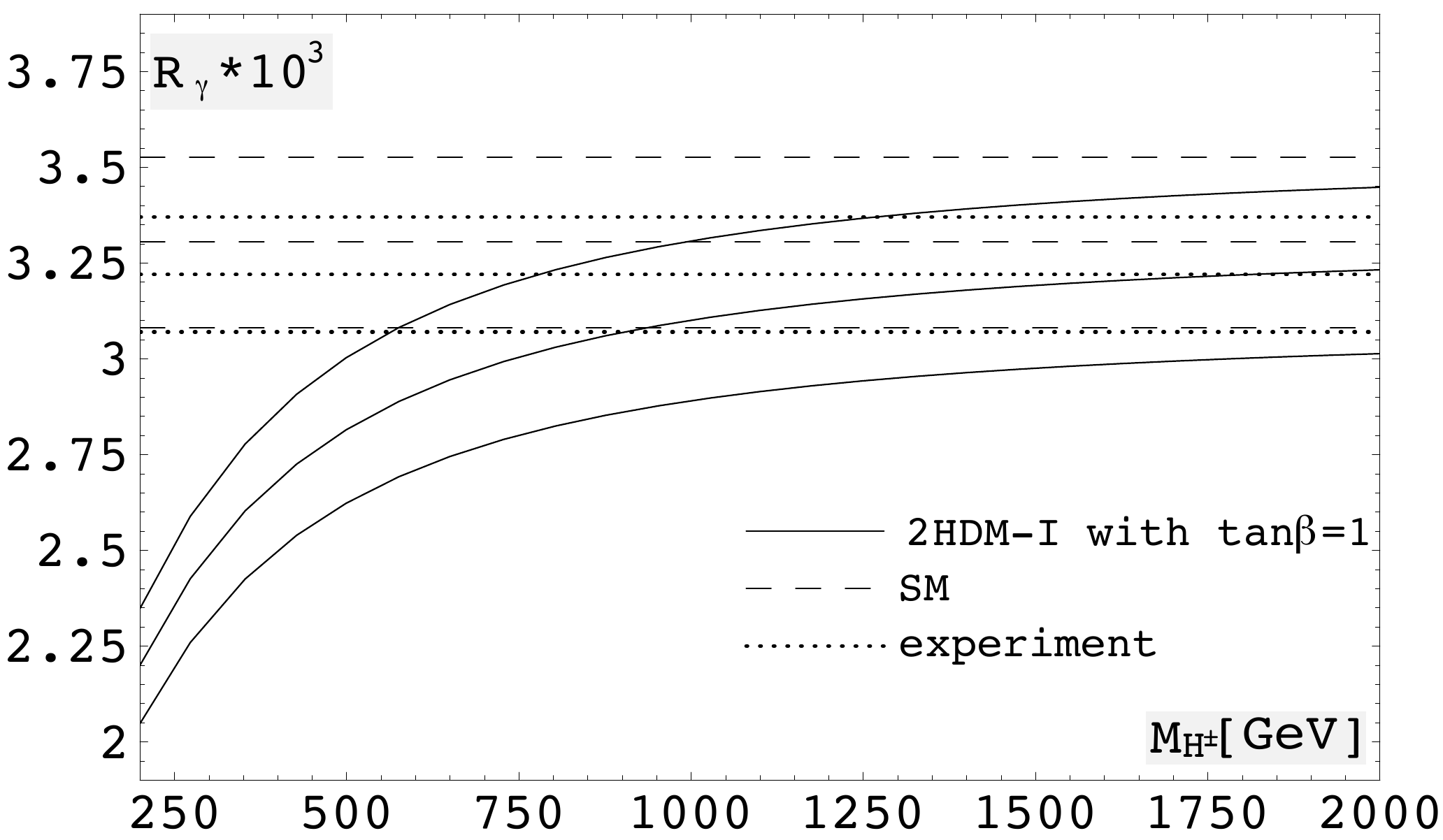}
\includegraphics[scale=0.38]{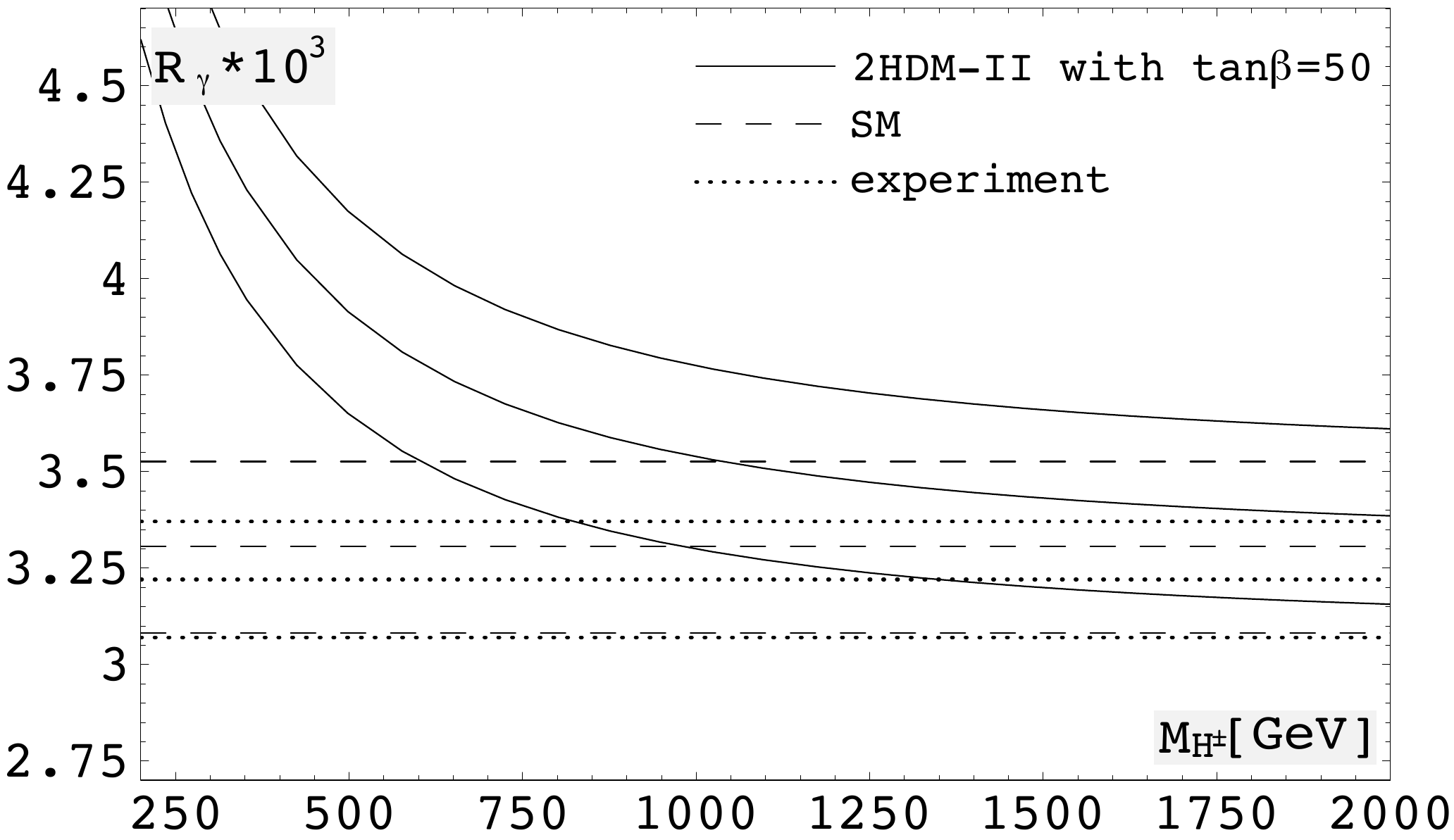}
\end{center}
\vspace*{-4mm}
\caption{BR$(B^-\to X_s\gamma)$ as a function of $M_{H^\pm}$ for Model~I (left) and Model~II (right), at two values of $\tan\beta$. Solid and dashed lines correspond to the NNLO 2HDM and SM predictions, respectively. (Shown are central values with $\pm1\sigma$ shifts.) Dotted curves represent the experimental average.
[Reprinted with kind permission from the authors, Fig.~2 of \cite{Misiak:2017bgg}].}
\end{figure}

\paragraph{{\boldmath $B\to X_s \gamma$}:}
The $B \to X_s\gamma$ transition may also proceed via charged Higgs boson exchange, which is sensitive
to the values of $\tan\beta$ and $M_{H^\pm}$. The allowed region depends on higher-order QCD effects. A huge effort has been devoted to the calculation of these corrections, the bulk of which are the same as in the SM \cite{Chetyrkin:1996vx,Buras:1997bk,Bauer:1997fe,Bobeth:1999mk,Buras:2002tp,Misiak:2004ew,Neubert:2004dd,Melnikov:2005bx,Misiak:2006zs,Misiak:2006ab,Asatrian:2006rq,Czakon:2006ss,Boughezal:2007ny,Ewerth:2008nv,Misiak:2010sk,Asatrian:2010rq,Ferroglia:2010xe,Misiak:2010tk,Kaminski:2012eb,Czakon:2015exa}. They are now complete up to NNLO order. On top of these, there are 2HDM-specific contributions \cite{Ciafaloni:1997un,Ciuchini:1997xe,Borzumati:1998tg,Bobeth:1999ww,Gambino:2001ew,Misiak:2015xwa} that depend on $M_{H^\pm}$ and $\tan\beta$. The result is that mass roughly up to $M_{H^\pm}=480~\text{GeV}$ is excluded for high values of $\tan\beta$ \cite{Misiak:2015xwa},
with even stronger constraints for very low values of $\tan\beta$. Recently, a new analysis \cite{Trabelsi:2015} of Belle results \cite{Saito:2014das} concludes that the lower limit is 540~GeV.  Also note the new result of Misiak and Steinhauser \cite{Misiak:2017bgg}  with lower limit in the range 570--800 GeV, see  Fig.~\ref{Fig:misiak-br-models} (right) for high $\tan \beta$ and high $H^+$ masses. 
We have here adopted the more conservative value of 480~GeV, however our results can easily be re-interpreted for this new limit.  Constraints from $B \to X_s \gamma$ decay for lower $H^+$ masses are presented in Fig.~\ref{Fig:gfitter} together with other constraints.

For low values of $\tan\beta$, the constraint is even more
severe. This comes about from the charged-Higgs coupling to $b$ and
$t$ quarks ($s$ and $t$) containing terms proportional to
$m_t/\tan\beta$ and $m_b\tan\beta$ ($m_s\tan\beta$). The product of
these two couplings determine the loop contribution, where there is an
intermediate $tH^-$ state, and leads to terms proportional to
$m_t^2/\tan^2\beta$ (responsible for the constraint at low
$\tan\beta$) and $m_tm_b$ (responsible for the constraint that is
independent of $\tan\beta$). For Models~I and X, on the other hand,
both these couplings are proportional to $\cot\beta$. Thus, the $B \to
X_s \gamma$ constraint is in these models only effective at low values of $\tan\beta$.\footnote{For early studies, see \cite{Grossman:1994jb,Akeroyd:1994ga}.}
This can be seen in Fig.~\ref{Fig:misiak-br-models} (left) and Fig.~\ref{Fig:misiak-models}, where the new results from the $B \to X_s \gamma$ analysis applied to Model~I of the 2HDM are shown.
We stress that Model I can avoid the $B \to X_s \gamma$ constraints and hence it can accommodate a light $H^+$.

\begin{figure}[htb]
\refstepcounter{figure}
\label{Fig:misiak-models}
\addtocounter{figure}{-1}
\begin{center}
\includegraphics[scale=0.5]{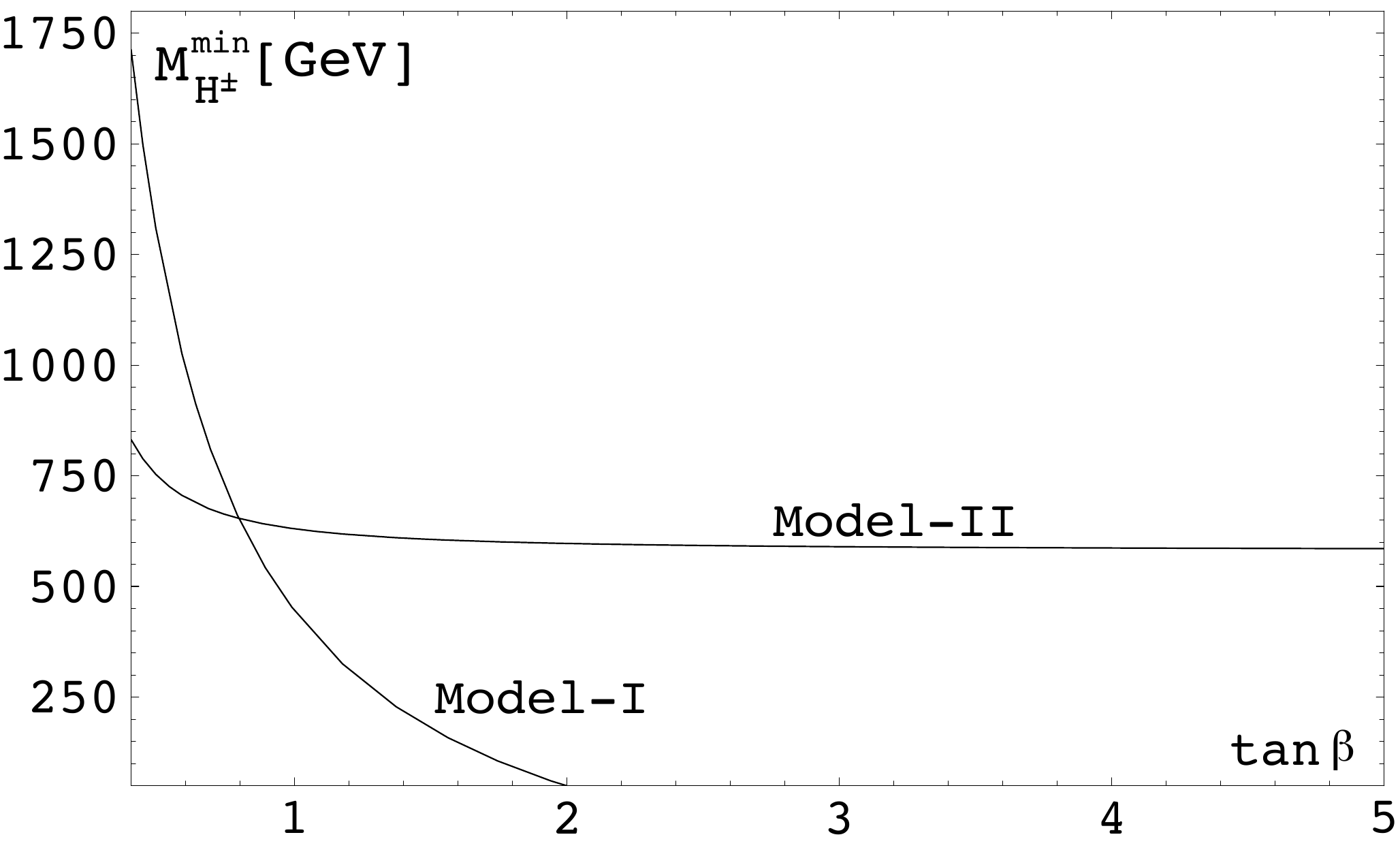}
\end{center}
\vspace*{-4mm}
\caption{2HDM 95\%C.L. $B\to X_s\gamma$ exclusion (lower part) in the plane of $\tan\beta$ and $M_{H^\pm}$.
[Reprinted with kind permission from the authors, Fig.~4 of \cite{Misiak:2017bgg}].}
\end{figure}

\paragraph{{\boldmath $B_0-\bar{B}_0$} mixing:}
Due to the possibility of charged-Higgs exchange, in addition to $W^+$
exchange, the $B_0-\bar{B}_0$ mixing constraint excludes low values of
$\tan\beta$ (for $\tan\beta<{\cal O}(1)$) and low values of $M_{H^\pm}$
\cite{Abbott:1979dt,Inami:1980fz,Athanasiu:1985ie,Glashow:1987qe,Geng:1988bq,Urban:1997gw}.
Recent values for the oscillation parameters $\Delta m_d$ and $\Delta m_s$ are given in Ref.~\cite{Deschamps:2009rh}, only at very low values of $\tan\beta$ do they add to the constraints coming from $B\to X_s\gamma$.

\subsubsection{Other precision constraints}

\paragraph{\boldmath $T$ and $S$:}
The precisely measured electroweak (oblique) parameters $T$ and $S$ correspond to radiative corrections, and are (especially $T$) sensitive to the mass splitting of the additional scalars of the theory.  In papers \cite{Grimus:2007if,Grimus:2008nb}  general expressions for these quantities are derived for the MHDMs and by confronting them with experimental results, in particular $T$, strong constraints are obtained on the masses of scalars.
In general, $T$ imposes a constraint on the splitting in the scalar sector, a mass splitting among the neutral scalars gives a negative contribution to $T$, whereas a splitting between the charged and neutral scalars gives a positive contribution.
A recent study \cite{Gorbahn:2015gxa} also demonstrates how RGE running may induce contributions to $T$ and $S$.
Current data on $T$ and $S$ are given in \cite{Agashe:2014kda}.

\paragraph{The muon anomalous magnetic moment:}
We are here considering heavy Higgs bosons ($M_1, M_{H^\pm} \gsim 100~\text{GeV}$), with a focus on the Model~II,  therefore, according to
\cite{Cheung:2003pw,Chang:2000ii,WahabElKaffas:2007xd}, the 2HDM contribution to the muon
anomalous magnetic moment is negligible even for $\tan\beta$ as high as $\sim
40$ (see, however, \cite{Krawczyk:2002df}).

\paragraph{The electron electric dipole moment:}
The bounds on electric dipole moments constrain the allowed amount of CP
violation of the model. For the study of the CP-non-conserving Model~II presented in section~\ref{sect:benchmarks-high}, the bound \cite{Regan:2002ta} (see also \cite{Pilaftsis:2002fe}):
\begin{equation}
|d_e|\lsim1\times10^{-27} [e\,\text{cm}],
\end{equation}
was adopted at the $1\sigma$ level. (More recently, an order-of-magnitude stronger bound has been established \cite{Baron:2013eja}.)
The contribution due to neutral Higgs exchange, via the
two-loop Barr--Zee effect \cite{Barr:1990vd}, is given by Eq.~(3.2) of
\cite{Pilaftsis:2002fe}.

\subsubsection{Summary of low-energy constraints}

A summary of constraints of the 2HDM Model~II coming  from low-energy physics  performed by the ``Gfitter'' group \cite{Flacher:2008zq} is presented on Fig.~\ref{Fig:gfitter}. The more recent inclusion of higher-order effects pushes the $B\to X_s\gamma$ constraint up to around 480~GeV \cite{Misiak:2015xwa} or even higher, as discussed above. See also Refs.~\cite{Deschamps:2009rh,Bona:2009cj,Enomoto:2015wbn}.
\begin{figure}[htb]
\refstepcounter{figure}
\label{Fig:gfitter}
\addtocounter{figure}{-1}
\begin{center}
\includegraphics[scale=1.2]{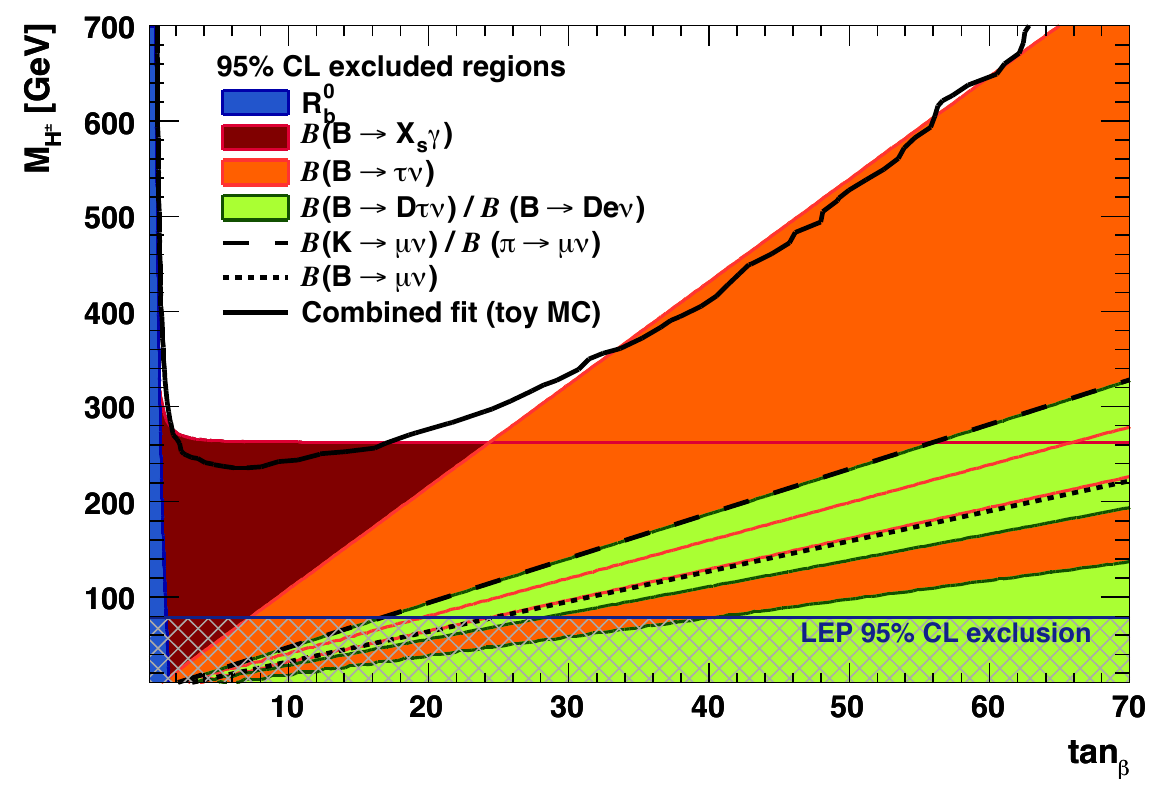}
\end{center}
\vspace*{-4mm}
\caption{Model~II 95\% CL exclusion regions in the ($\tan\beta$, $M_{H^\pm}$) plane. [Reprinted with kind permission from EPJC and the authors of ``Gfitter''  \cite{Flacher:2008zq}]. A new analysis, including the updated bound from $B\to X_s\gamma$, is being prepared by the ``Gfitter'' group.}
\end{figure}
\goodbreak

\subsection{High-energy constraints}
\label{subsect:HEconstraints}
Most bounds on charged Higgs bosons are obtained in the low-mass region, where a charged Higgs might be produced in the decay of a top quark, $t\to H^+ b$,
with the $H^+$ subsequently decaying according to Eqs.~(\ref{Eq:fermion-decay-channels}a-c), (\ref{Eq:gauge-decay-channels}) or  (\ref{Eq:WHj-light}).
Of special interest are the decays $H^+\to \tau^+\nu$ and $H^+\to c\bar s$. For comparison with data, products like $\text{BR}(t\to H^+ b)\times\text{BR}(H^+\to \tau^+\nu)$ are relevant, as presented in section~\ref{subsec:top-BR}.
At high charged-Higgs masses, the $HW$ rate can be important (if kinematically open). On the other hand, the $hW$ channel can dominate over $HW$, because of the larger phase space. However, as illustrated in Fig.~\ref{Fig:cpc-br-ratios-vs-mass-3-30}, it vanishes in the alignment limit.

\subsubsection{Charged-Higgs constraints from  LEP}

The branching ratio $R_b \equiv \Gamma_{Z\to b\bar b} /\Gamma_{Z\to {\rm had}}$ would be affected by Higgs exchange.
Experimentally $R_b= 0.21629 \pm 0.00066$ \cite{Agashe:2014kda}.
The contributions from neutral Higgs
bosons to $R_b$ are negligible \cite{ElKaffas:2006nt}, however, charged Higgs boson contributions, as
given by \cite{Denner:1991ie}, Eq.~(4.2), exclude low values of $\tan\beta$
and low $M_{H^\pm}$.  See also Fig.~\ref{Fig:gfitter}.

LEP and the Tevatron have given limits on the mass and couplings,
for charged Higgs bosons in the 2HDM. At LEP a lower mass limit of 80 GeV that refers to the Model~II scenario for $\text{BR}(H^+ \to\tau^+ \nu)+\text{BR}(H^+ \to  c \bar s)=100\%$  was derived.  The mass limit for $\text{BR}(H^+\to \tau^+ \nu) = 100\%$  is 94~GeV (95\% CL), and for $\text{BR}(H^+\to c \bar s) = 100\%$ the region below 80.5 as well as the region 83--88~GeV are excluded (95\% CL). 
Search for the decay mode $H^+ \to A W^+$ with $A\to b\bar b$, which is not negligible in Model~I, leads  to the corresponding $M_{H^\pm}$ limit of 72.5 GeV (95\% CL) if $M_A > 12~\text{GeV}$ \cite{Abbiendi:2013hk}.

\subsubsection{Search for charged Higgs at the Tevatron}

A D0 analysis \cite{:2009zh} with an integrated luminosity $1~\text{fb}^{-1}$ has been performed for $t\to H^+ b$, with $H^+\to c\bar s$ and $H^+\to \tau^+ \nu$.
In the SM one has BR($t\to W^+b$)=100\% with $W\to l\nu/q'\overline q$.
The presence of a sizeable BR($t\to H^+ b$) would change these ratios.
For the optimum case of BR($H^+ \to q'\overline q)=100\%$, upper bounds on
BR$(t\to H^+ b)$ between 19\% and 22\% were obtained for
$80 \,{\rm GeV}< M_{H^\pm} < 155$ GeV. In \cite{:2009zh} the decay $H^+ \to q'\overline q$ was assumed to be
entirely $H^+ \to c\bar s$. But these limits on BR$(t\to H^+ b)$ also apply to the case
of both $H^+ \to c\bar s$ and $H^+ \to c\bar b$ having sizeable BRs, as discussed in \cite{Logan:2010ag}.
This is because the search strategy merely requires that  $H^+$ decays to quark jets.

An alternative strategy was adopted in the CDF analysis \cite{Aaltonen:2009ke}
with an integrated luminosity 2.2 fb$^{-1}$. A direct search for the decay $H^+ \to q'\overline q$ was performed by looking for a peak
centered at $M_{H^\pm}$ in the dijet invariant mass distribution, which would be distinct from the peak at $M_W$
arising from the SM decay $t\to W^+b$ with $W\to q'\overline q$.  
For the optimum case of BR($H^+ \to q'\overline q)=100\%$, upper bounds on
BR$(t\to H^+ b)$ between 32\% and 8\% were obtained for
$90~\text{GeV}< M_{H^\pm} < 150~\text{GeV}$.
No limits on BR$(t\to H^+ b)$ were given for the region $70~\text{GeV}< M_{H^\pm} < 90~\text{GeV}$ due to the
large background from $W\to q'\overline q$ decays. For the region $60~\text{GeV}< M_{H^\pm} < 70~\text{GeV}$,
limits on BR$(t\to H^+ b)$ between 9\% and 12\% were derived.

A search for charged-Higgs production has also been carried out by D0 \cite{Abazov:2008rn} at higher masses, where
$H^+\to t\bar b$.
Bounds on cross section times branching ratio have been obtained for Models~I and III, in the range $180~\text{GeV}\leq M_{H^\pm}\leq300~\text{GeV}$, for $\tan\beta=1$ and $\tan\beta>10$.

\begin{figure}[htb]
\refstepcounter{figure}
\label{Fig:lhc-cms-low}
\addtocounter{figure}{-1}
\vspace*{-4mm}
\begin{center}
\includegraphics[scale=0.35]{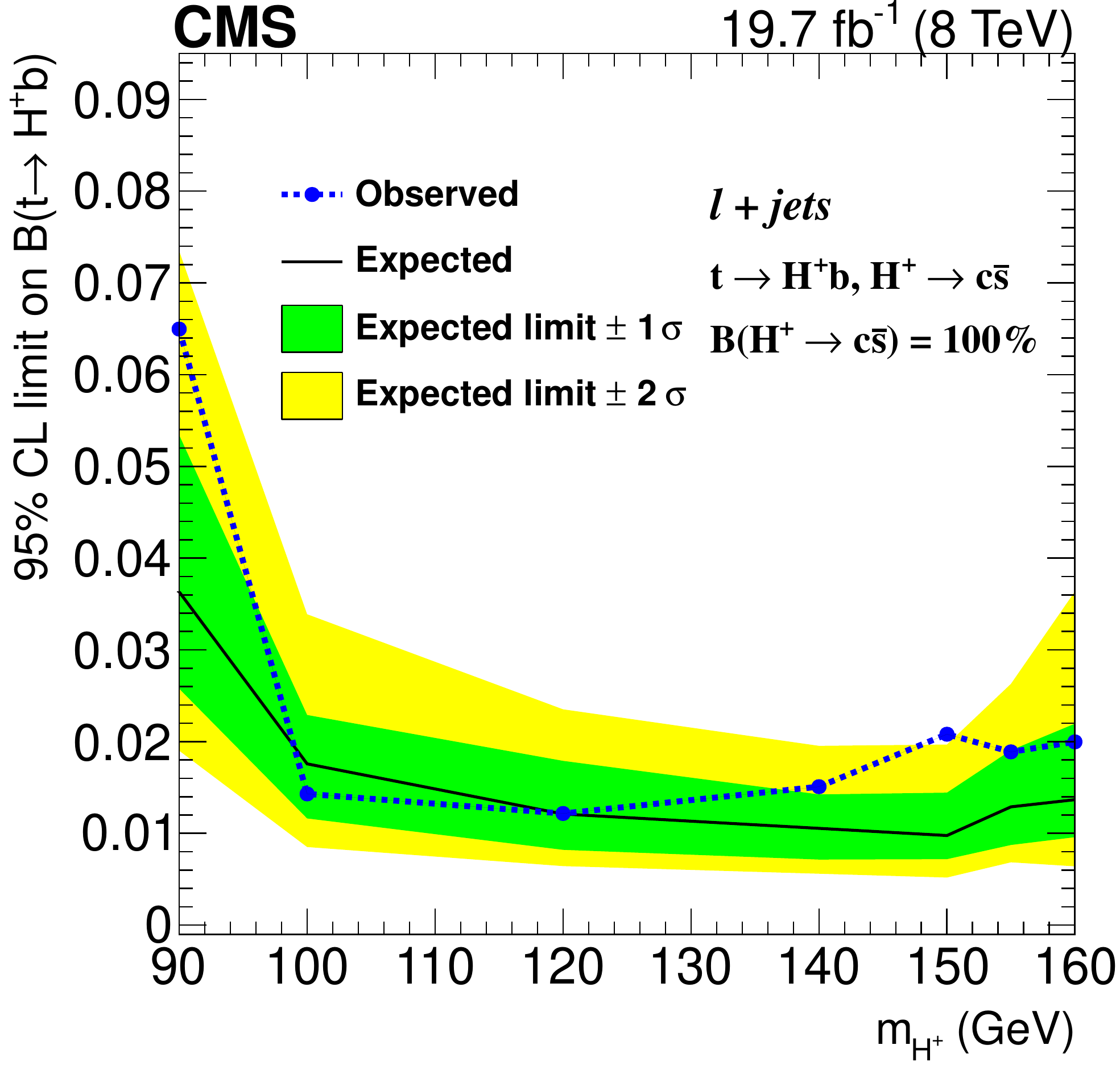}
\end{center}
\vspace*{-6mm}
\caption{CMS exclusion limit on the branching fraction BR$(t\to H^+b)$ as a function of $M_{H^\pm}$ assuming BR$(H^+\to c\bar s) = 100\%$. [Reprinted with kind permission from JHEP and the authors, Fig.~6 of \cite{Khachatryan:2015uua}].}
\end{figure}

\begin{figure}[htb]
\refstepcounter{figure}
\label{Fig:lhc-cms-fig8}
\addtocounter{figure}{-1}
\vspace*{-2mm}
\begin{center}
\includegraphics[scale=0.3]{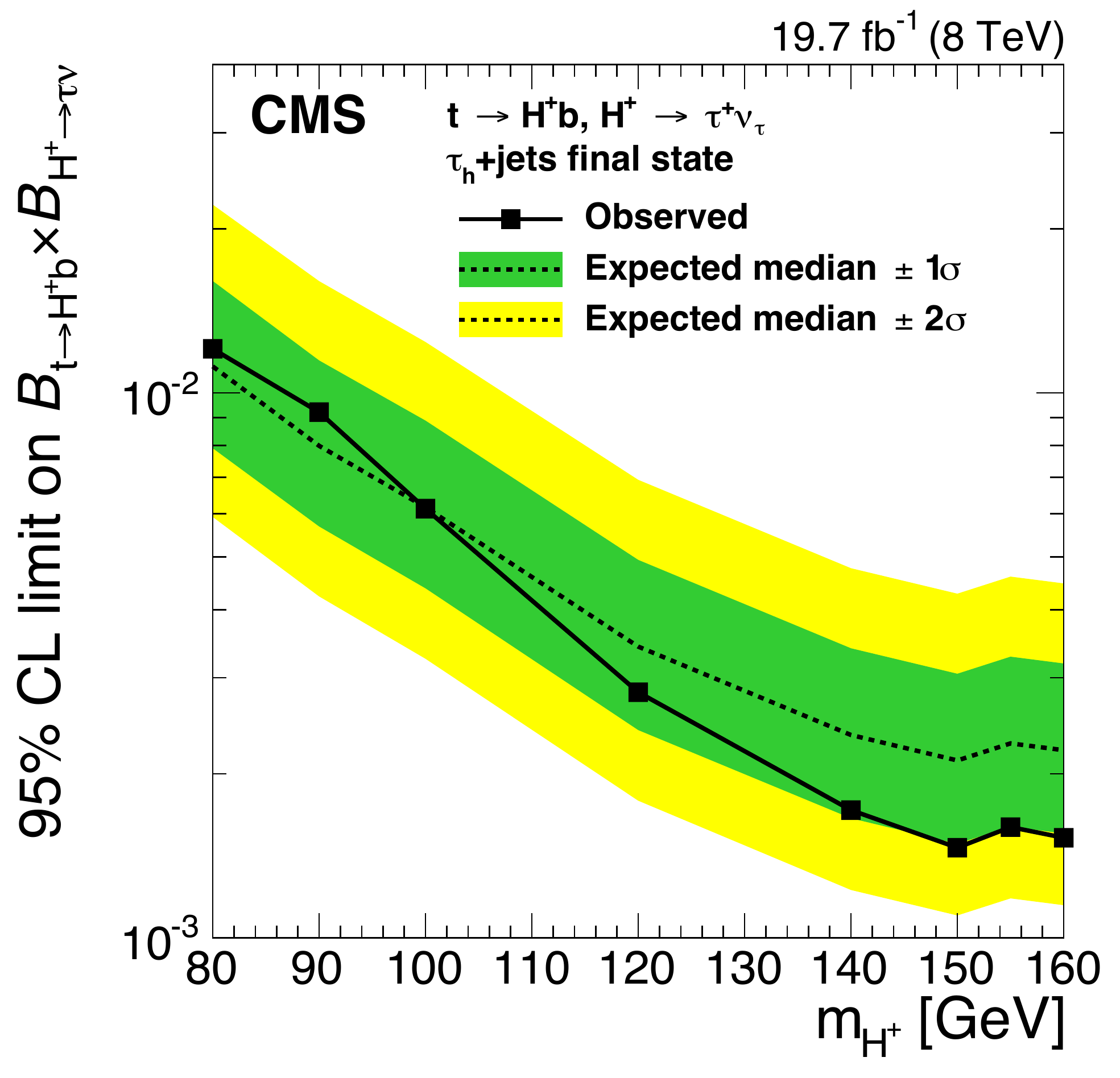}
\includegraphics[scale=0.3]{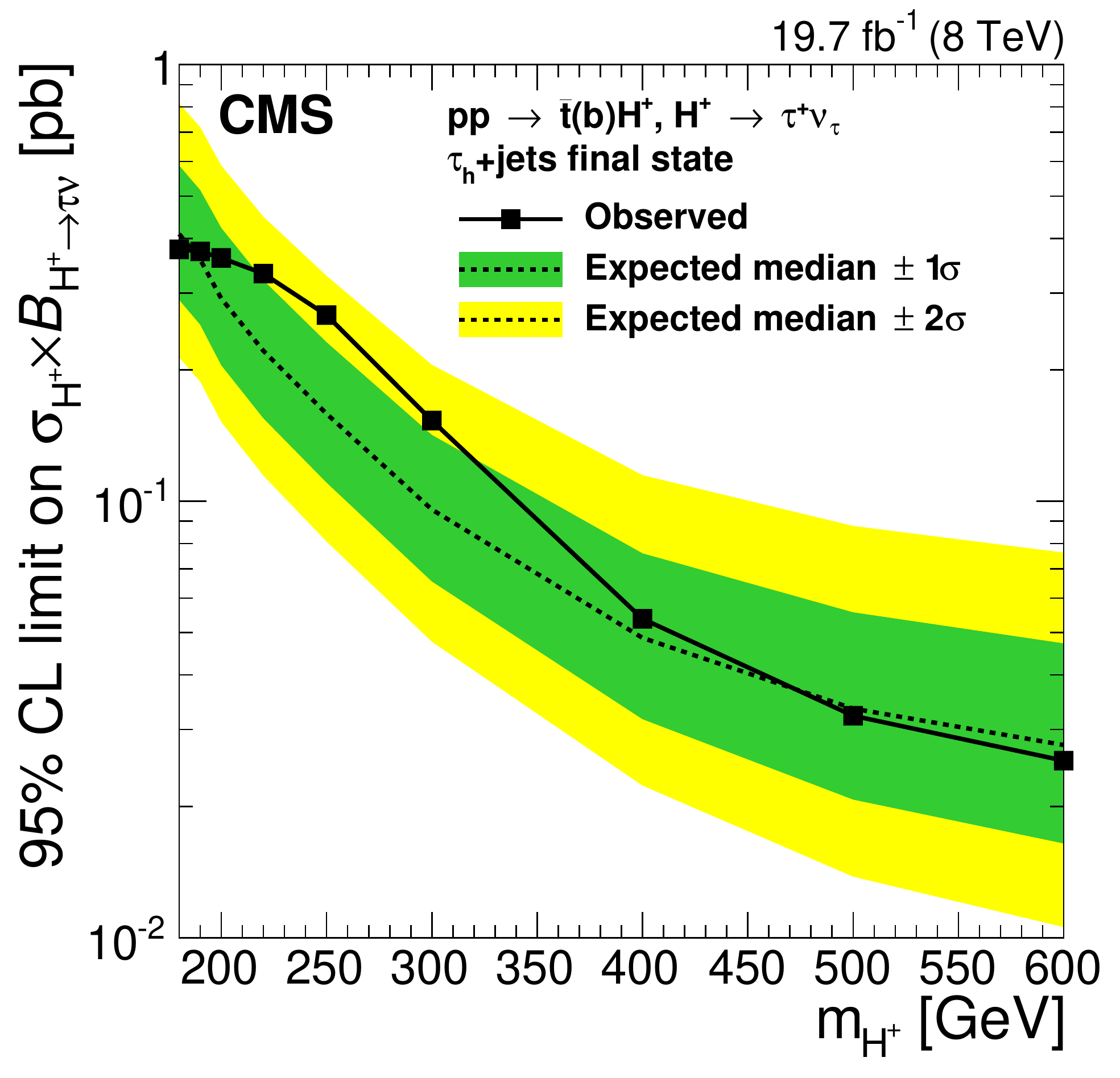}
\end{center}
\vspace*{-6mm}
\caption{CMS model-independent upper limits on BR$(t\to H^+b)\times\text{BR}(H^+ \to \tau^+\nu_\tau)$ (left) and on $\sigma(pp\to t(b)H^+)\times\text{BR}(H^+ \to \tau^+\nu_\tau)$ (right). [Reprinted with kind permission from JHEP and the authors, Fig.~8 of \cite{Khachatryan:2015qxa}].}
\end{figure}

\subsubsection{LHC searches for charged Higgs}
\label{subsect:LHCsearches}

A search for $t\to H^+ b$ followed by the decay $H^+ \to c\bar s$ at the LHC (7~TeV) has
been performed by the ATLAS collaboration with 4.7 fb$^{-1}$ \cite{Aad:2013hla}.
Assuming BR$(H^+ \to c\bar s)=100\%$, the derived upper limits on BR$(t\to H^+ b)$
are 5.1\%, 2.5\% and 1.4\%  for $M_{H^\pm}=90 \, {\rm GeV}, 110 \,{\rm GeV}$ and $130$ GeV,
respectively.  These limits are superior to those from the Tevatron search \cite{Aaltonen:2009ke},
and exclude a sizeable region of the Yukawa-coupling
plane\footnote{See section \ref{sect:multi-Higgs-doublet-models} for
  details.}, not excluded by $B \to X_s \gamma$. The recent data from
CMS \cite{Khachatryan:2015uua} on the production in the $t\bar t$
channel of light charged Higgs bosons decaying to $c \bar s$  at the
collision energy of 8~TeV and with an integrated luminosity $19.7~\text{fb}^{-1}$ show no deviation from the SM.
Assuming BR$(H^+ \to c\bar s)=100\%$, the derived upper limits on BR$(t\to H^+ b)$
are 1.2\% to 6.5\% for $M_{H^\pm}$ in the range (90--160 GeV), see Fig.~\ref{Fig:lhc-cms-low}. The data
points are found to be consistent with the signal-plus-background hypothesis for a charged Higgs boson
mass of $150~\text{GeV}$ for a best-fit branching fraction value of $(1.2\pm0.2)\%$
including both statistical and systematic errors. The local observed significance is
$2.4\sigma$ ($1.5\sigma$ including the look-elsewhere effect).

\begin{figure}[htb]
\refstepcounter{figure}
\label{Fig:lhc-atlas-fig7}
\addtocounter{figure}{-1}
\begin{center}
\includegraphics[scale=0.35]{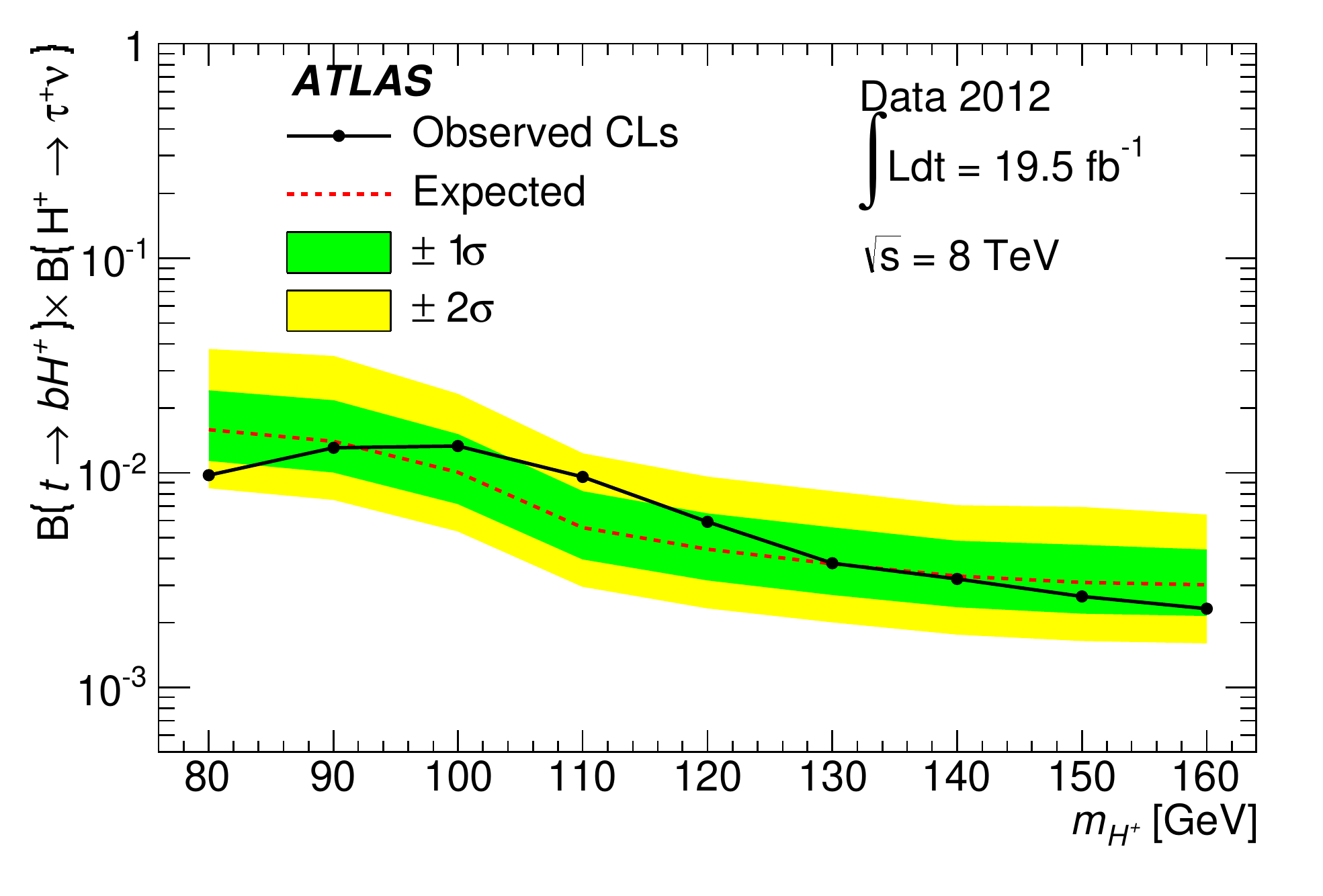}
\includegraphics[scale=0.35]{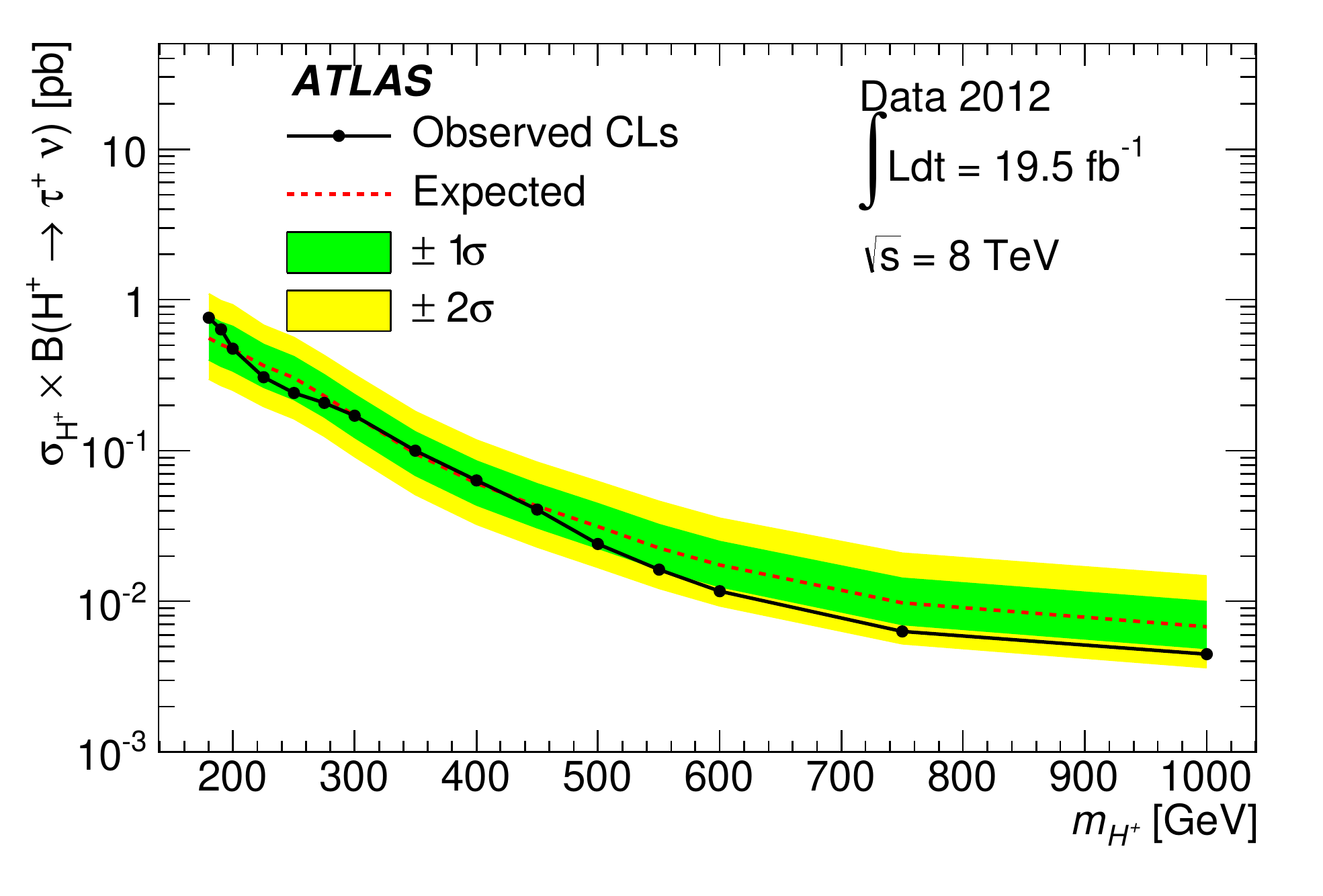}
\end{center}
\vspace*{-4mm}
\caption{ATLAS upper limits on the production and decay of low-mass (left) and high-mass (right) charged Higgs bosons. [Reprinted with kind permission from JHEP and the authors, Fig.~7 of \cite{Aad:2014kga}].}
\end{figure}

\begin{figure}[htb]
\refstepcounter{figure}
\label{Fig:lhc-cms-fig10}
\addtocounter{figure}{-1}
\begin{center}
\includegraphics[scale=0.3]{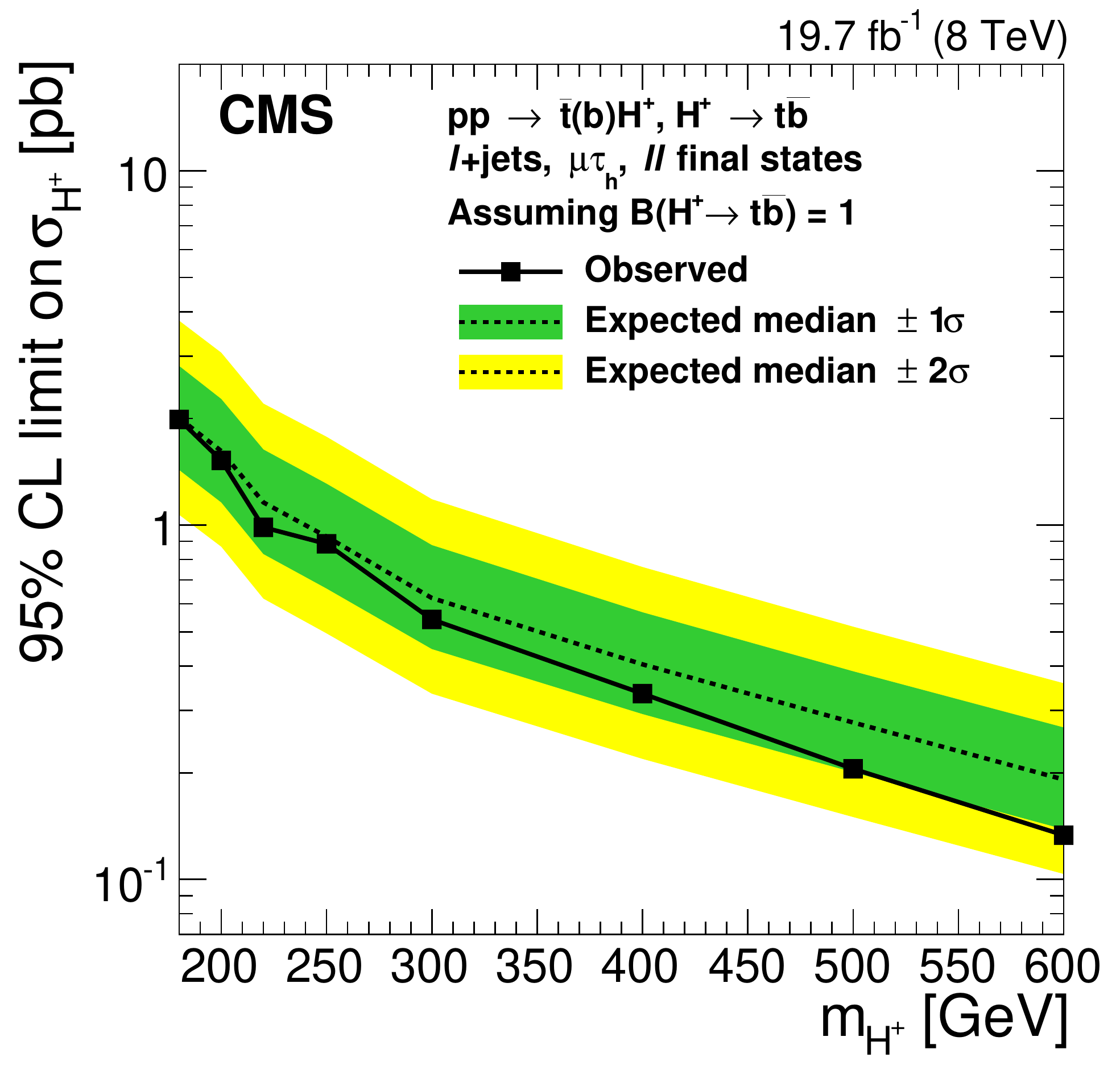}
\includegraphics[scale=0.38]{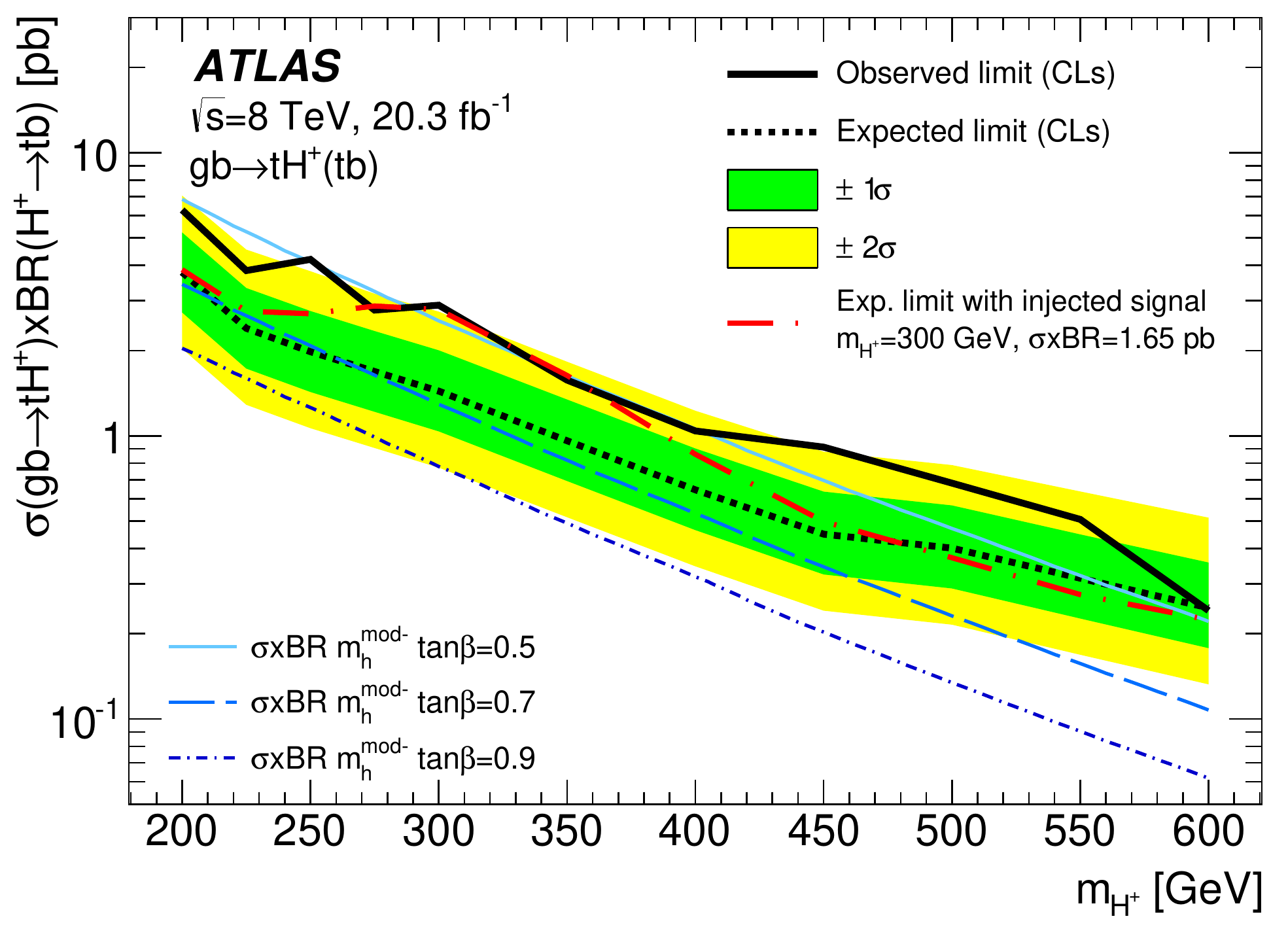}
\end{center}
\vspace*{-4mm}
\caption{Left: CMS upper limits on $\sigma(pp\to t(b)H^+)$ for the
  combination of the $\mu\tau_\text{h}$, $\ell+\text{jets}$, and
  $\ell\ell^\prime$ final states assuming BR$(H^+\to t\bar b)=100\%$. [Reprinted with kind permission from JHEP and the authors, Fig.~10 of \cite{Khachatryan:2015qxa}].
Right: ATLAS upper limits for the production of $H^+ \to t\bar b$ in association with a top quark. The red dash-dotted line shows the expected limit obtained for a simulated signal injected at $M_{H^\pm}= 300~\text{GeV}$.
[Reprinted with kind permission from JHEP and the authors, Fig.~6 of \cite{Aad:2015typ}].}
\end{figure}

\begin{figure}[htb]
\refstepcounter{figure}
\label{Fig:lhc-atlas-high}
\addtocounter{figure}{-1}
\begin{center}
\includegraphics[scale=0.35]{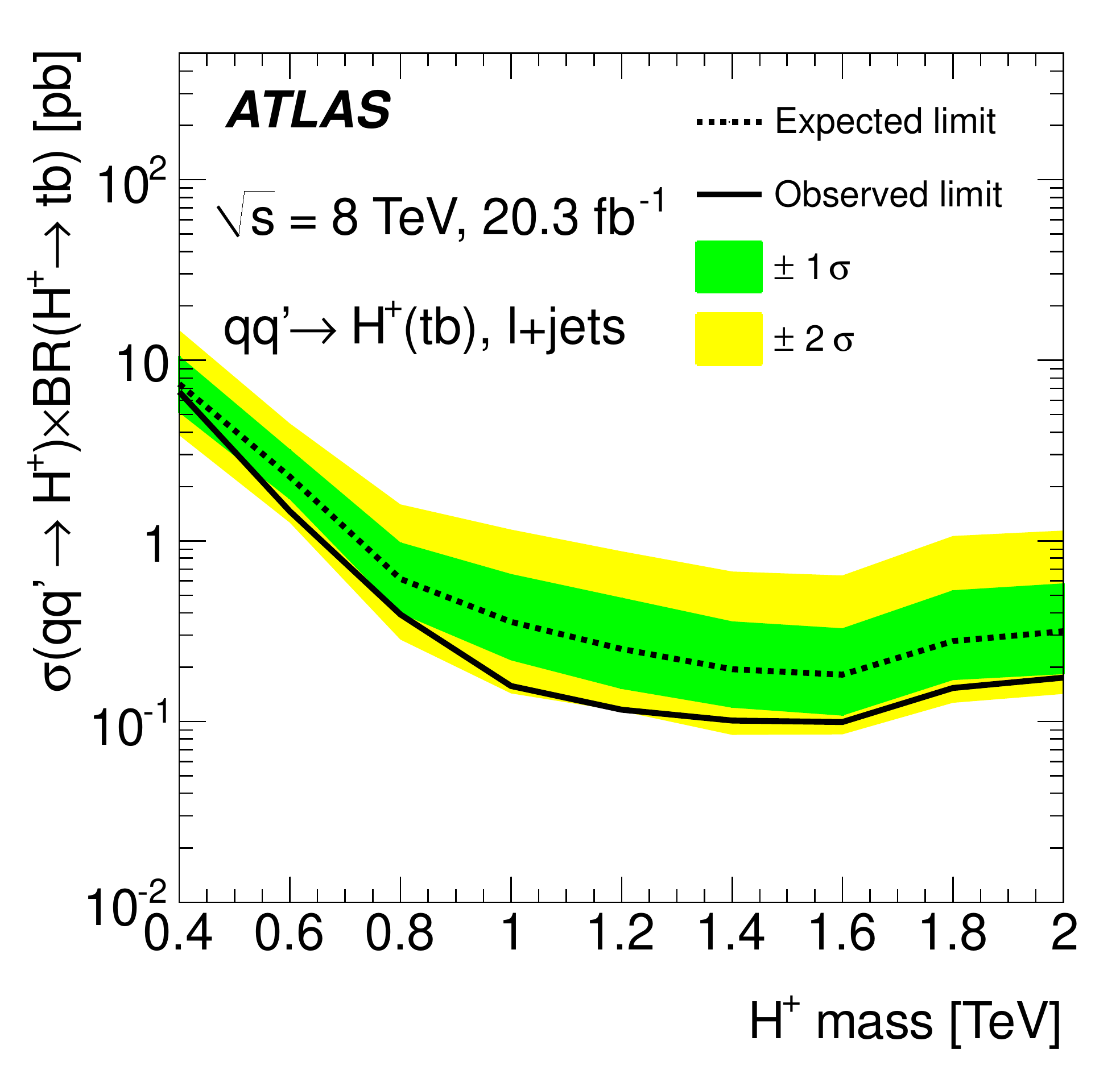}
\includegraphics[scale=0.35]{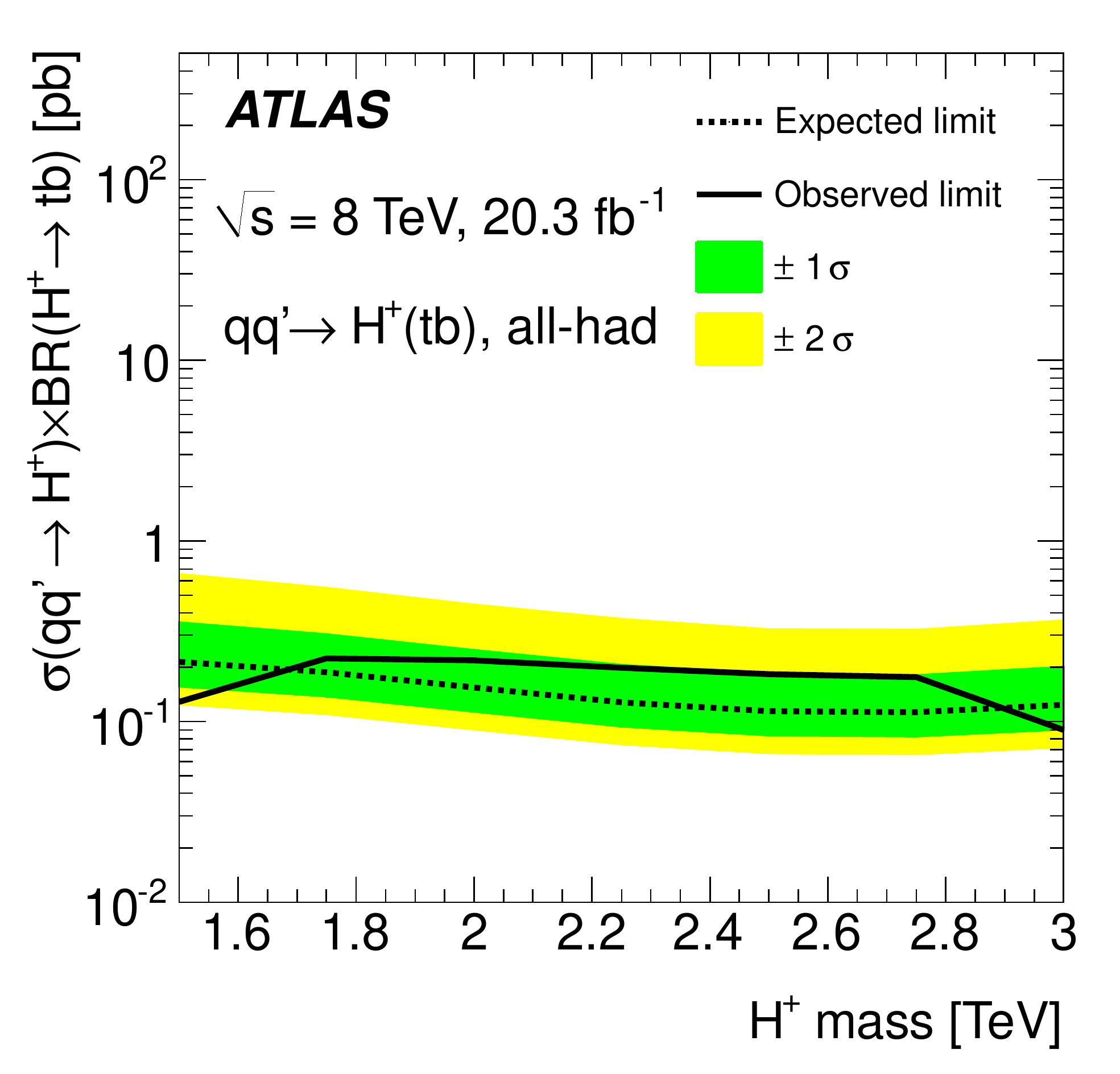}
\end{center}
\vspace*{-4mm}
\caption{ATLAS limits on the $s$-channel production cross section
  times branching fraction for $H^+\to t\bar b$ as a function of the charged Higgs boson mass, for particular final states, using the narrow-width approximation. [Reprinted with kind permission from JHEP and the authors, Fig.~10 of \cite{Aad:2015typ}].}
\end{figure}

Likewise, a search for a light charged Higgs boson produced in the decay $t\to H^+b$ and decaying to $\tau^+ \nu$ has been performed by CMS \cite{Chatrchyan:2012vca,Khachatryan:2015qxa}, see Fig.~\ref{Fig:lhc-cms-fig8}. For charged Higgs boson mass between 80 and 160 GeV, they obtain upper limits on the product of branching fractions $\text{BR}(t\to H^+b)\times\text{BR}(H^+\to\tau^+ \nu)$ in the range 0.23\% to 1.3\%.

Similarly, constraints are obtained by ATLAS \cite{Aad:2014kga} from the 8 TeV measurements at the LHC, with luminosity $19.5~\text{fb}^{-1}$. Results for low and high mass $H^+$ are shown in Fig.~\ref{Fig:lhc-atlas-fig7}, for BR$(t \to H^+ b)\times\text{BR}(H^+\to \tau^+\nu)$ (left) and  for $\sigma(pp \to H^+t +X)\times \text{BR}(H^+\to \tau^+ \nu)$ (right), respectively. 

In Fig.~\ref{Fig:lhc-cms-fig10} (left) CMS results
\cite{Khachatryan:2015qxa} for the case BR$(H^+\to t\bar b)=100\%$ are presented.
Results of a recent ATLAS analysis, performed using a multi-jet final state  for the process $gb \to t H^-$ are presented in Fig.~\ref{Fig:lhc-cms-fig10} (right). An excess of events above the background-only hypothesis is observed across a wide mass range, amounting to up to $2.4\sigma$.

In addition, ATLAS provides limits on the $s$-channel production cross section, via the decay mode $H^+\to t\bar b$ for heavy charged Higgs bosons (masses from 0.4~TeV to 3~TeV), for two categories of final states, see Fig.~\ref{Fig:lhc-atlas-high}.

It should be noted that in all these figures, ``expectations'' are a measure of the instrumental capabilities, and the amount of data. In fact, theoretical (model-dependent) expectations can be significantly lower.
In particular, in Model~I and Model~II, the branching ratio for $H^+\to\tau^+\nu$ is at high masses very low, see Fig.~\ref{Fig:cpc-br-ratios-vs-mass-3-30}. Thus, these models are not yet constrained by the high-mass results shown in Figs.~\ref{Fig:lhc-cms-fig8} and \ref{Fig:lhc-atlas-fig7} \cite{Dorsch:2016tab}. However, for Model~X the $\tau^+\nu$ branching ratio is sufficiently high for these searches to be already relevant.

\subsubsection{Summary of search for charged scalars at high energies}
\label{sect:LHC-search-summary}
The LEP lower limits on the mass for light $H^+$ are 80.5 GeV -- 94 GeV, depending on the assumption on the $H^+$ decaying 100\% into $c\bar s$, $b\bar s$ or $c\bar s+b\bar s$ channels.

For  low mass $H^+$, $\sim80$ (90) -- 160~GeV, limits for the top decay to $H^+b$ were derived at the Tevatron and the LHC (ATLAS and CMS) at the level of a few per cent ($5.1\%-1.2\%$) for the assmption of 100\% decay to $c\bar s$.  CMS results on $\text{BR}(t\to H^+ b)\times\text{BR}(H^+\to \tau^+ \nu)$ reached down to $1.3\% - 0.23\%$.

For heavy $H^+$ the region between 200 and 600~GeV was studied at LHC for $\sigma(pp\to t(b)H^+)\times\text{BR}(H^+\to \tau^+ \nu)$. A special search for an $s$-channel resonance with mass of $H^+$ up  to 3~TeV with the decay mode to $t\bar b$ was performed by ATLAS.

Some excesses at $2.4\sigma$ for $H^+$ mass equal to 150~GeV, as well as for masses between $220-320~\text{GeV}$,  are reported by CMS \cite{Khachatryan:2015uua,Khachatryan:2015qxa} and for a very wide $H^+$ mass range $200-600~\text{GeV}$ by ATLAS \cite{Aad:2015typ}.

\subsubsection{LHC constraints from the neutral Higgs sector}

After the discovery in 2012 of the SM-like Higgs particle with a mass of 125~GeV, measurements of its properties lead to  serious constraints on the parameters space of the 2HDM, among others on the mass of the $H^+$.  

Constraints on the gauge coupling of the lightest neutral 2HDM Higgs boson were recently obtained by ATLAS \cite{Aad:2015pla} for four Yukawa models. The results support the SM-like scenario for $h$ with $\sin (\beta-\alpha) \approx 1$, the  allowed ($95\%$ CL) small value of $\cos (\beta-\alpha)$, e.g.\ for Model~II is up to 0.2 for $\tan \beta=1$ while it extends up to $\pm$ 0.4 for large $\tan \beta$ in Model~I, see Fig.~\ref{Fig:lhc-atlas-dedicated}. There are also ``wrong sign'' regions allowed for Yukawa couplings for larger values of $\cos (\beta-\alpha)$ for Model~II, as mentioned in section~\ref{sect:Yukawa}.

Two further aspects of the recent model-independent neutral-Higgs studies at the LHC \cite{ATLAS:2012ae,Chatrchyan:2012tx} are important: 
\begin{itemize}
\item[(i)]
The production and subsequent decay of a neutral Higgs $H_1$ to $\gamma\gamma$, at $M=125~\text{GeV}$ should be close to the SM result. Assuming the dominant production to be via gluon fusion (and adopting the narrow-width approximation), this can be approximated as a constraint on
\begin{equation}
R_{\gamma\gamma}=\frac{\Gamma(H_1\to gg)\text{BR}(H_1\to\gamma\gamma)}
{\Gamma(H_\text{SM}\to gg)\text{BR}(H_\text{SM}\to\gamma\gamma)}.
\end{equation}
For Model~II channels discussed in section~\ref{sect:benchmarks-high} \cite{Basso:2012st,Basso:2015dka}, a generous range $0.5\leq R_{\gamma\gamma}\leq2$ was adopted, whereas recent ATLAS and CMS results ($\pm 2\sigma$ regions) are 
$0.63 \leq R_{\gamma\gamma}\leq 1.71$ \cite{Aad:2015gba} and $0.64\leq R_{\gamma\gamma}\leq 1.6$ \cite{Khachatryan:2014jba}, respectively.
Note, that this quantity is sensitive to the $H^+$, since its loop
contribution proportional to the $H_1H^+H^-$ coupling can have a
constructive or destructive interference with the SM contribution. The non-decoupling property of the $H^+$ contribution to the $H_1\to\gamma\gamma$  effective coupling may lead to sensitivity to even a very heavy $H^+$ boson. 
\item[(ii)]
The production and subsequent decay, dominantly via $ZZ$ and $WW$ are constrained in the mass ranges of heavier neutral Higgs bosons $H_2$ and $H_3$ from $130~\text{GeV}$ to $500~\text{GeV}$. We consider the quantity
\begin{equation}
R_{ZZ}=\frac{\Gamma(H_j\to gg)\text{BR}(H_j\to ZZ)}
{\Gamma(H_\text{SM}\to gg)\text{BR}(H_\text{SM}\to ZZ)},
\end{equation}
for $j=2,3$ and require it to be below the stronger 95\% CL obtained
by ATLAS or CMS in the scans described in section~\ref{sect:benchmarks-high}.
\end{itemize}

\begin{figure}[htb]
\refstepcounter{figure}
\label{Fig:lhc-atlas-dedicated}
\addtocounter{figure}{-1}
\begin{center}
\includegraphics[scale=0.35]{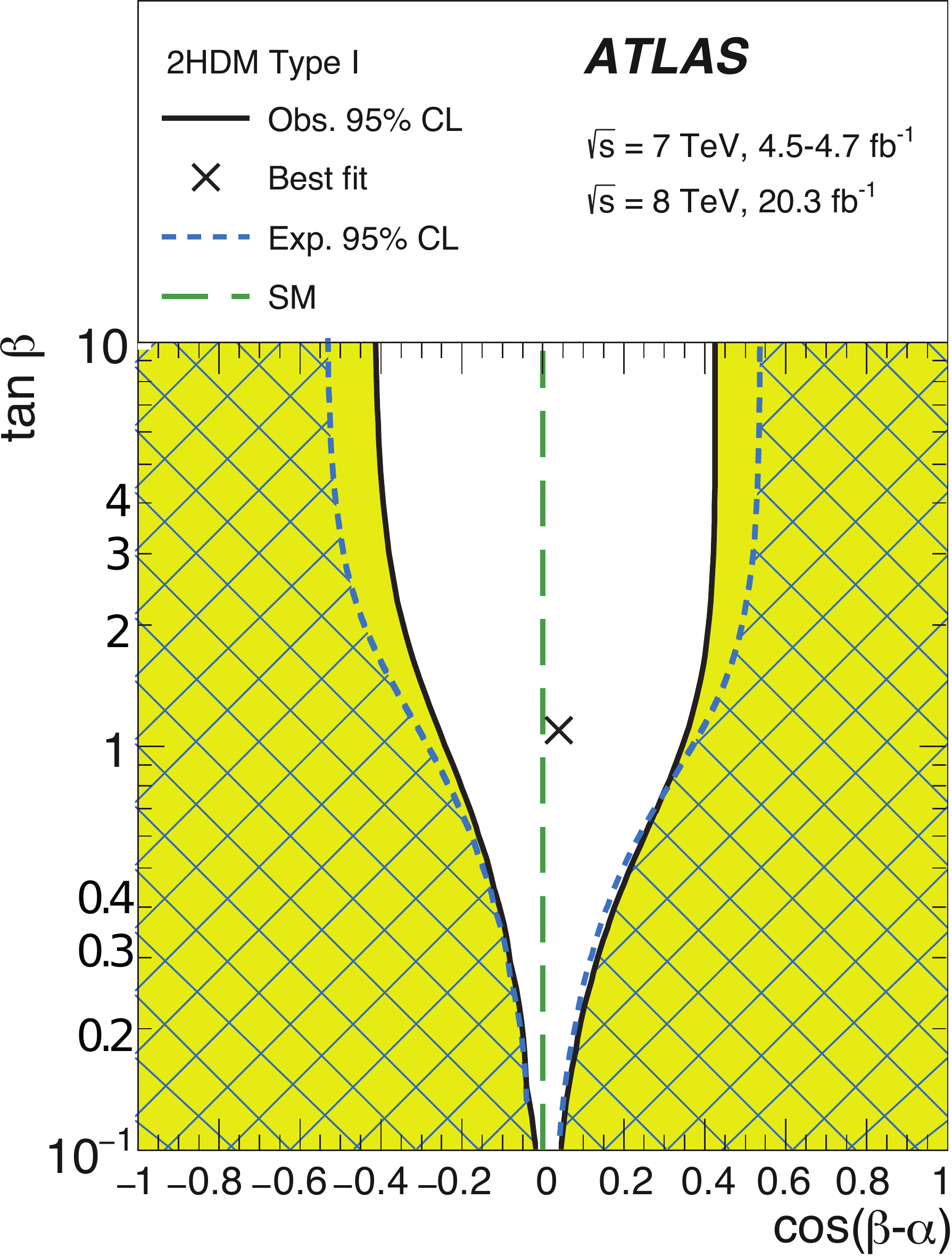}
\includegraphics[scale=0.35]{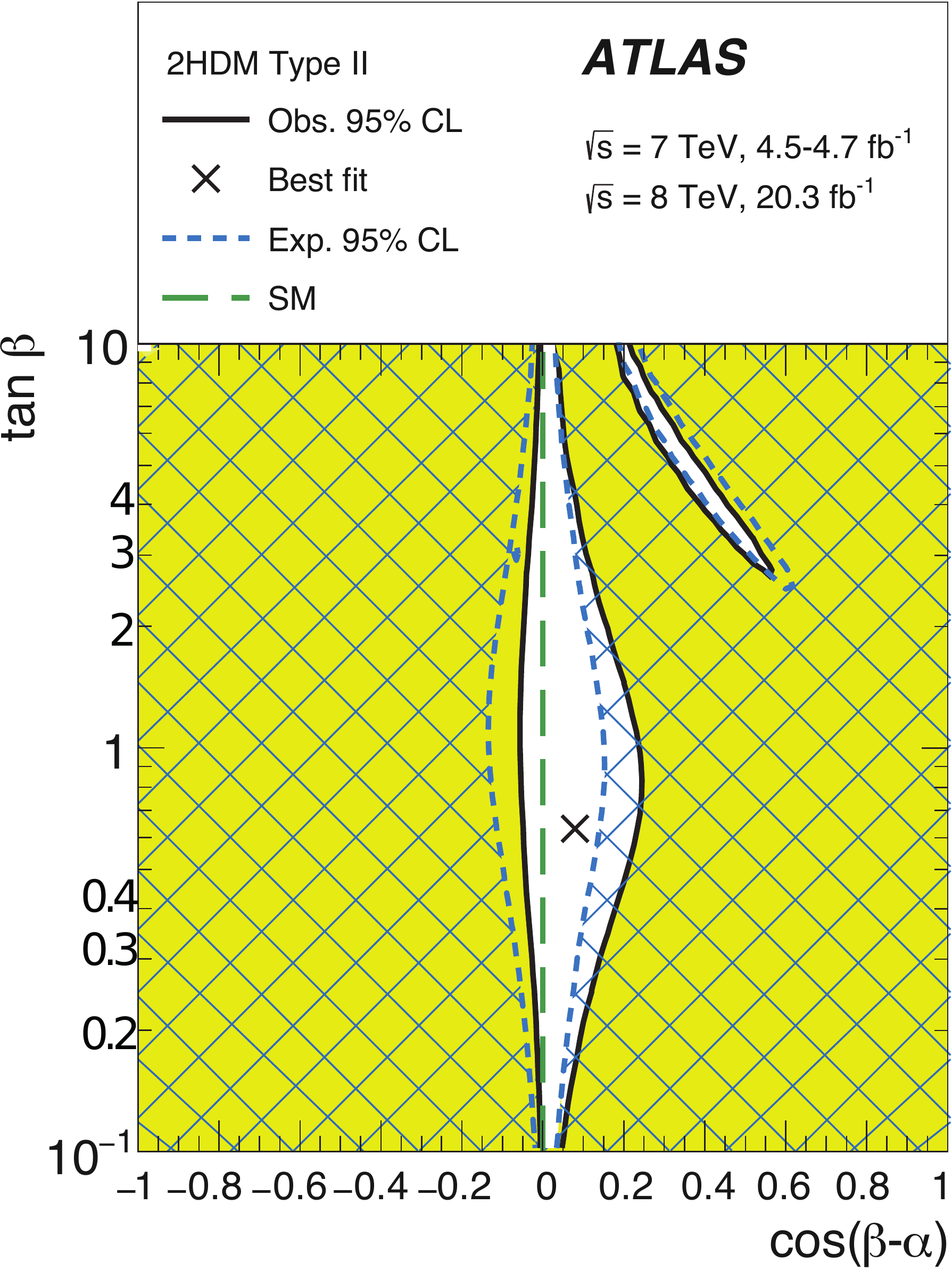}
\end{center}
\vspace*{-4mm}
\caption{ATLAS regions of the [$\cos(\beta-\alpha)$, $\tan\beta$] plane for the 2HDM Model~I (left) and Model~II (right) excluded by fits to the measured rates of Higgs boson production and decays. [Reprinted with kind permission from JHEP and the authors, Fig.~5 of \cite{Aad:2015pla}].}
\end{figure}

\section{Further search for $H^+$ at  the LHC}
\label{sect-benchmarks}
\setcounter{equation}{0}

Here, a discussion of possible search strategies for charged scalars at the LHC is presented. 
The stakes of a possible discovery from an extended scalar sector are very high,
 these searches should be pursued in all conceivable channels. Some propositions are described below, separately 
for low and high masses of the $H^+$ boson.

As discussed in previous sections, a light charged Higgs boson is only viable in Models~I and X. 
In the more familiar Model~II (and also Y), the $B \to X_s \gamma$ constraint enforces $M_{H^\pm}\gsim480~\text{GeV}$ or even higher for all values of $\tan\beta$ \cite{Misiak:2015xwa}.

\subsection{Channels for $M_{H^\pm}\lsim m_t$}
\label{Sec:LightBenchmarks}
For low $M_{H^\pm} $ mass, the proposed searches can be divided into two categories, based on single $H^+$ production or $H^+H^-$ pair production. For all channels presented here,
$\tau\nu$ decays of charged Higgs bosons are the recommended ones.

\subsubsection{Single $H^+$ production}
In Ref.~\cite{Aoki:2011wd} processes with a single $H^+$ were studied for Models~I and X. Here, the production mechanism depends on the $H^+b\bar t$ Yukawa coupling, proportional to $1/\tan\beta$, thus falling off sharply at high $\tan\beta$. Concentrating on processes without neutral-Higgs-boson intermediate states\footnote{This process, involving neutral scalars $H_j$, depends on the Yukawa model and needs a dedicated analysis \cite{Arhrib:2009hc, Kanemura:2009mk}.} (eqs.~(\ref{Eq:Hqq})--(\ref{Eq:single-production-cs})), it was found that for 30 $\text{fb}^{-1}$ of integrated luminosity the reach at the 95\% CL allows exploring low values of $\tan\beta$, up to about 10.
At higher values of $\tan\beta$, the Model~I branching ratio for $t\to H^+b$ becomes too small (see Fig.~\ref{Fig:br-top}) for the search to be efficient.
In table~\ref{tab:bench1} we present promising parameters for two proposed channels from this analysis.

\begin{table}[htb]
\begin{center}
\begin{tabular}{|c c | c c | c c | c c c c c c c c c c c c|} \hline \hline
\multicolumn{2}{|c|}{$M_{H^{\pm}}$}   &\multicolumn{2}{c|}{100 GeV}& \multicolumn{2}{c|}{150 GeV}  \\ \hline 
\multicolumn{2}{|c|}{$\tan \beta$}     & 3 &    10      & 3 &    10 \\ \hline \hline
$H^+W^- b \bar{b} $ &(\ref{Eq:btH}) \qquad \qquad  & $\surd$  & $\surd$  & $\surd$  &  
\\ \hline
$H^+ b q$ &(\ref{Eq:Hqq}) \qquad \qquad  & $\surd$   & ($\surd$)   & $\surd$   & 
\\ \hline \hline
\end{tabular}
\caption{Proposed channels, denoted by $\surd$, for Models~I and X at 30 $\text{fb}^{-1}$. The case denoted by ($\surd$) requires higher luminosity.}
\label{tab:bench1}
\end{center}
\vskip -0.2cm
\end{table}

\subsubsection{$H^+H^-$ pair production}
Charged Higgs boson pair production, see Eq.~(\ref{Eq:pair-production}) and Figs.~\ref{Fig:feyn-figures-pair} and \ref{Fig:feyn-figures-pair-def}, can be sensitive also to higher values of $\tan\beta$ \cite{Aoki:2011wd}.
This will require resonant production via $H_j$ decaying to $H^+H^-$, and assuming an enhancement of the coupling 
between charged and neutral Higgs bosons.
In table~\ref{tab:bench2} we present channels which would be viable in the case of resonant intermediate $H_j$ states,
as represented by the mechanism of Fig.~\ref{Fig:feyn-figures-pair} a (i).
\begin{table}[htb]
\begin{center}
\begin{tabular}{|c c | c c c | c c c | c c c c c c c c c c c c|} \hline \hline
\multicolumn{2}{|c|}{$M_{H^{\pm}}$}   &\multicolumn{3}{c|}{100 GeV}& \multicolumn{3}{c|}{150 GeV}  \\ \hline 
\multicolumn{2}{|c|}{$\tan \beta$}     & 3 &    10      & 30   & 3 &    10      & 30 \\ \hline \hline
$H^+ W^-$ &(\ref{Eq:WH}) \qquad \qquad  & $\surd$   & ($\surd$)   &    & $\surd$   & ($\surd$)   & 
\\ \hline 
$H^+H^-$ &(\ref{Eq:gg_HH})
\qquad \qquad  & $\surd$  & $\surd$   & $\surd$  & $\surd$  & $\surd$   & $\surd$  
\\ \hline
$H^+H^- q'Q' $ &(\ref{Eq:VBF}) \qquad \qquad  & $\surd$  & $\surd$  & $\surd$  & $\surd$  & $\surd$  & $\surd$ 
\\ \hline
\hline
\end{tabular}
\caption{Proposed channels, denoted by $\surd$, for Models~I and X, requiring resonant production, at 30 $\text{fb}^{-1}$. The cases denoted by ($\surd$) would need higher luminosity.}
\label{tab:bench2}
\end{center}
\vskip -0.2cm
\end{table}

\subsection{Channels for $m_t<M_{H^\pm} < 480~\text{GeV}$}
\label{sect:benchmarks-intermediate}
The intermediate mass region requires a dedicated discussion, since only Models~I and X are allowed. However, in contrast to the $M_{H^\pm} <m_t$-region, the $H^+\to t\bar b$ channel is now open. Also the channel $H^+\to W^+h$ is open in the higher mass range. These channels may thus compete with the $\tau\nu$ channel discussed for the low-mass case. Whereas the cross section becomes very small at high $\tan\beta$, where the $\tau\nu$ channel is interesting (see Fig.~\ref{Fig:sigma-production}, left panel), these other channels could be interesting at lower values of $\tan\beta$.

\subsection{Channels for $480~\text{GeV} < M_{H^\pm}$}
\label{sect:benchmarks-high}
For high masses, all four Yukawa models are permitted, and there are three classes of decay channels, 
$H^+\to H_1W^+$ (or $AW^+$, $HW^+$, $hW^+$), $H^+\to\tau^+\nu$ and $H^+\to t\bar b$. We shall here present studies of the first two, which only compete with a moderate QCD background.
Within the 2HDM~II, like in the MSSM \cite{Moretti:2000yg,Ghosh:2004wr}, the decay channel
(\ref{Eq:WHj-light}), $H^+\to H_1W^+$, can be used, with $H_1\to b\bar{b}$.
The $t\bar b$ channel competes with an enormous QCD background, but recent progress in $t$ and $b$ tagging have yielded the first results, as reported in section~\ref{subsect:LHCsearches}.

\subsubsection{The channel $H^+\to W^+ H_j\to W^+ b\bar b$}

A study \cite{Enberg:2014pua} of the process $pp\to H^+\bar t$ in Model~II, where the charged Higgs boson decays to a $W$ and the observed Higgs boson at 125~GeV, which in turn decays to $b\bar b$, for charged-Higgs mass up to about 500~GeV, concludes that an integrated luminosity of the order of $3000~\text{fb}^{-1}$ is required for a viable signal.

This search channel has recently been re-examined for Model~II, for high charged-Higgs mass and neutral-Higgs masses all low \cite{Moretti:2016jkp}. The discovered 125~GeV Higgs boson is taken to be $H$, the heavier CP-even one. Thus, the charged one could decay to $WH$, $WA$ and $Wh$. The dominant production mode at high charged-Higgs mass is from the channel (\ref{Eq:WbH}), $\bar b g \to H^+\bar t \to H^+\bar b W^-$. There will thus be at least three $b$ quarks in the final state, two of which will typically come from one of the neutral Higgs bosons:
\begin{equation}
pp (\bar b g) \to H^+ \bar t  X \to H^+ \bar b W^- X \to b \bar b \bar b W^+ W^- X 
\to b \bar b \bar b \ell \nu jj X.
\end{equation}
Interesting parameter regions are identified for $\tan\beta={\cal O}(1)$, and $\sin(\beta-\alpha)$ close to 0 (the discovered Higgs boson is the $H$), where a signal can be extracted over the background. The fact that the $H^+$ is heavy, means that the $W^+$ from its decay will be highly boosted. This fact is exploited to isolate the signal.
Since the interesting region has $\tan\beta={\cal O}(1)$, this channel remains relevant also for other Yukawa types.

Relaxing CP conservation, this option has been explored in \cite{Basso:2012st}. Imposing the theoretical and experimental constraints discussed in sections \ref{sect:th-constraints} and \ref{sect:ex-constraints}, one finds a surviving parameter space that basically falls into two regions: (i) low $\tan\beta$, with non-negligible CP violation and a considerable branching ratio $H^+\to H_1W^+$ (see section~\ref{subsec:br-toWH1}), and (ii) high $\tan\beta$, with little CP violation and only a modest decay rate $H^+\to H_1W^+$.

For the region (i), the channel
\begin{equation}
pp\to H^+ W^- X \to H_1W^+W^-X 
\to b\bar b \ell\nu jj X
\end{equation}
has been studied (see also Ref.~\cite{Basso:2015dka}).
A priori, there is a considerable $t\bar t$ background. However, imposing a series of kinematical cuts, it is found that this background can be reduced to a manageable level, yielding sensitivities of the order of 2--5 for a number of events of the order of 10--20, with an integrated luminosity of $100~\text{fb}^{-1}$ at 14~TeV. A more sophisticated experimental analysis could presumably improve on this. The more promising parameter points are presented in Table~\ref{table:points}. No point was found at higher values of $\tan\beta$ ($\gsim2$), within the now allowed range of $M_{H^\pm}$.

\begin{table}[ht]
\begin{center}
\begin{tabular}{|c|c|c|c|c|c|c|}
\hline
$\alpha_1/\pi$ & $\alpha_2/\pi$ & $\alpha_3/\pi$ & $\tan\beta$ & $M_2$ [GeV] & $M_3$ [GeV]\\
\hline
 $0.35$ & $-0.056$ & $0.43$ & $1$ & $400$  & 446\\
 $0.33$ & $-0.21$ & $0.23$ & $1$ & $450$  & 524 \\
\hline
\end{tabular}
\end{center}
\caption{Suggested points selected from the allowed parameter space \cite{Basso:2012st}. 
Note that $M_3$ is not an independent parameter.
Furthermore, $\mu=200~\text{GeV}$.  \label{table:points}}
\end{table}

While the above analysis focused on the bosonic production mode, where resonant production via $H_2$ or $H_3$ is possible, a study of the fermionic mode,
\begin{equation}
pp (gg) \to H^+ \bar t b X \to t \bar t b \bar b X,
\end{equation}
has been performed for the maximally symmetric 2HDM, which is based on the SO(5) group, and has natural SM alignment \cite{Dev:2014yca}. In this analysis, the ``stransverse'' mass, $M_{T2}$ \cite{Lester:1999tx}, is exploited, and it is found that by reconstructing at least one top quark, a signal can be isolated above the SM background. 

\subsubsection{The channel $H^+\to \tau^+ \nu$}

The $H^+\to\tau^+\nu$ channel is traditionally believed to have little background. However, a recent study of Model~II finds \cite{Basso:2015dka} that this channel can only be efficiently searched for at some future facility at a higher energy. This is due to a combination of many effects. At high mass ($M_{H^\pm}\gsim480~\text{GeV}$) the production rate goes down, whereas a variety of multi-jet processes also give events with an isolated $\tau$ and missing momentum. 

\subsubsection{The channel $H^+\to t\bar b$}
As discussed in section~\ref{sect:decays}, except for particular parameter regions allowing the $H^+\to W^+H_j$ modes, at high values of $M_{H^\pm}$ the $t\bar b$ channel is the dominant one. This channel has long been ignored because of the enormous QCD background, but methods are being developed to suppress this, as exemplified in Ref.~\cite{Aad:2015typ}.

\subsubsection{Exploiting top polarization}
At high masses the $b g\to H^- t$ production mechanism is dominant. If the $H^+$ decays fully hadronically and its mass is known, then semileptonic decays of the top quark can be analyzed in terms of its polarization.
Such studies can yield information on $\tan\beta$, since this parameter determines the chirality of the $H^+tb$ coupling \cite{Allanach:2006fy,Godbole:2006tq,Huitu:2010ad,Godbole:2011vw,Rindani:2013mqa}.

\subsection{Other scenarios}
\label{Sec:Scenarios}

Various scenarios for additional Higgs bosons have been discussed in the literature. These typically assume CP conservation. 
Several scenarios \cite{Haber:2015pua,Kling:2016opi} and channels have recently been presented, mostly focussing on the neutral sector, in particular the phenomenology of the heavier CP-even state, $H$.
In the ``Scenario D (Short cascade)'' of Ref.~\cite{Haber:2015pua}, it is pointed out that if $H$ is sufficiently heavy, it may decay as $H\to H^+W^-$, or even as $H\to H^+ H^-$. A version of the former is discussed above, in section~\ref{sect:benchmarks-high}, for Model~II.
In ``Scenario E (Long cascade)'', it is pointed out that for heavy
$H^+$, one may have the chain $H^+\to AW^+\to HZW^+$ or $H^+\to HW^+$,
whereas a heavy $A$ may allow $A\to H^+ W^-\to H W^+W^-$. The modes
$H^+\to HW^+$ and $H^+\to AW^+$ have also recently been discussed in Refs.~\cite{Coleppa:2014cca,Kling:2015uba}.

The class of $bW$ production mechanisms $qb\to q^\prime H^+b$ depicted in Fig.~\ref{Fig:feyn-figures-arhrib} has been explored in Ref.~\cite{Arhrib:2015gra}, where it is pointed out that in the alignment limit, with neutral Higgs masses close, $M_A\simeq M_H$, there is a strong cancellation among different diagrams. Thus, if $M_A$ should be light, this mechanism would be numerically important. It is also suggested that the $p_T$-distribution of the $b$-jet may be used for diagnostics of the production mechanism.

\section{Models with several charged scalars} 
\label{sect:other-models}
\setcounter{equation}{0}
\subsection{Multi-Higgs-Doublet models}
\label{sect:multi-Higgs-doublet-models}

Multi-Higgs Doublet Models (MHDM) are models 
with $n$ scalar SU(2) doublets, where $n \geq 3$ \cite{Grossman:1994jb}. 
The $n=1$ case corresponds to the Standard Model, the $n=2$ case corresponds to the 2HDM, the main topic of this paper. New phenomena will appear for $n\geq3$, for which we below often use the abbreviation MHDM.
The MHDM has the virtue of predicting $\rho=1$ at tree level,
as does the 2HDM. In the MHDM there are $n-1$ charged scalar pairs, $H_i^+$.
We shall discuss only the phenomenology of the lightest $H^+$ ($\equiv H_1^+$), assuming that the other $H^+$ are heavier. 

The Yukawa interaction of an $H_i^+$, $i=1,\ldots,n-1$ is described by the Lagrangian:
\begin{equation} \label{Eq:cal-F-MHDM}
{\cal L}_\text{ch}
=\frac{g}{\sqrt{2}\,m_W}
\bigl\{\bigl[\overline u(
m_dP_R{\cal F}_i^D
+m_uP_L{\cal F}_i^U)
d+ \overline{\nu}m_\ell P_R{\cal F}_i^L\ell\bigr]H_i^+ +\text{h.c.}
\bigr\}.
\end{equation}
It applies to the 2HDM ($n=2$); then the ${\cal F}$s given in Table~\ref{tab:couplings} of Appendix~A coincide with the ${\cal F}_1$ in the above equation.
In general, the ${\cal F}_i^D$,  ${\cal F}_i^U$ and ${\cal F}_i^L$ are complex numbers, which are defined in terms of an $n\times n$ matrix $U$,  diagonalizing the mass matrix of the charged scalars\footnote{For details, see Ref.~\cite{Grossman:1994jb}, where ${\cal F}_i^D$,  ${\cal F}_i^U$ and ${\cal F}_i^L$ are denoted $X_i$, $Y_i$ and $Z_i$, respectively.}.

It is evident that the branching ratios of the charged Higgs bosons,
$H_i^+$ depend on the parameters ${\cal F}_i^D$,  ${\cal F}_i^U$ and ${\cal F}_i^L$.
In the case of the 2HDM this shrinks to a single parameter, $\tan\beta$, which determines these three couplings.
This implies that certain combinations are constrained, for example, in Model~II  we have for each $i$, $|{\cal F}_i^D{\cal F}_i^U|=1$. 
  
As in the 2HDM (Models II and Y), an important constraint on the mass
and couplings of $H^+$ in the MHDM is provided by the decay $B \to X_s \gamma$. 
However, here, even 
a light $H^+$ (i.e., $M_{H^\pm} \lsim m_t$) is still a
possibility, because of a cancellation between the loop contributions from the different scalars. Recently, $2\sigma$ intervals in the ${\cal F}_1^D-{\cal
  F}_1^U$ parameter space for $M_{H^\pm}=100$~GeV were derived from $B
\to X_s \gamma$ \cite{Cree:2011uy,Jung:2010ik,Trott:2010iz}, 
assuming $|{\cal F}_1^U|<1$, in order to comply with constraints from $Z\to b\bar b$.

\begin{figure}[htb]
\begin{center}
\includegraphics[scale=0.33]{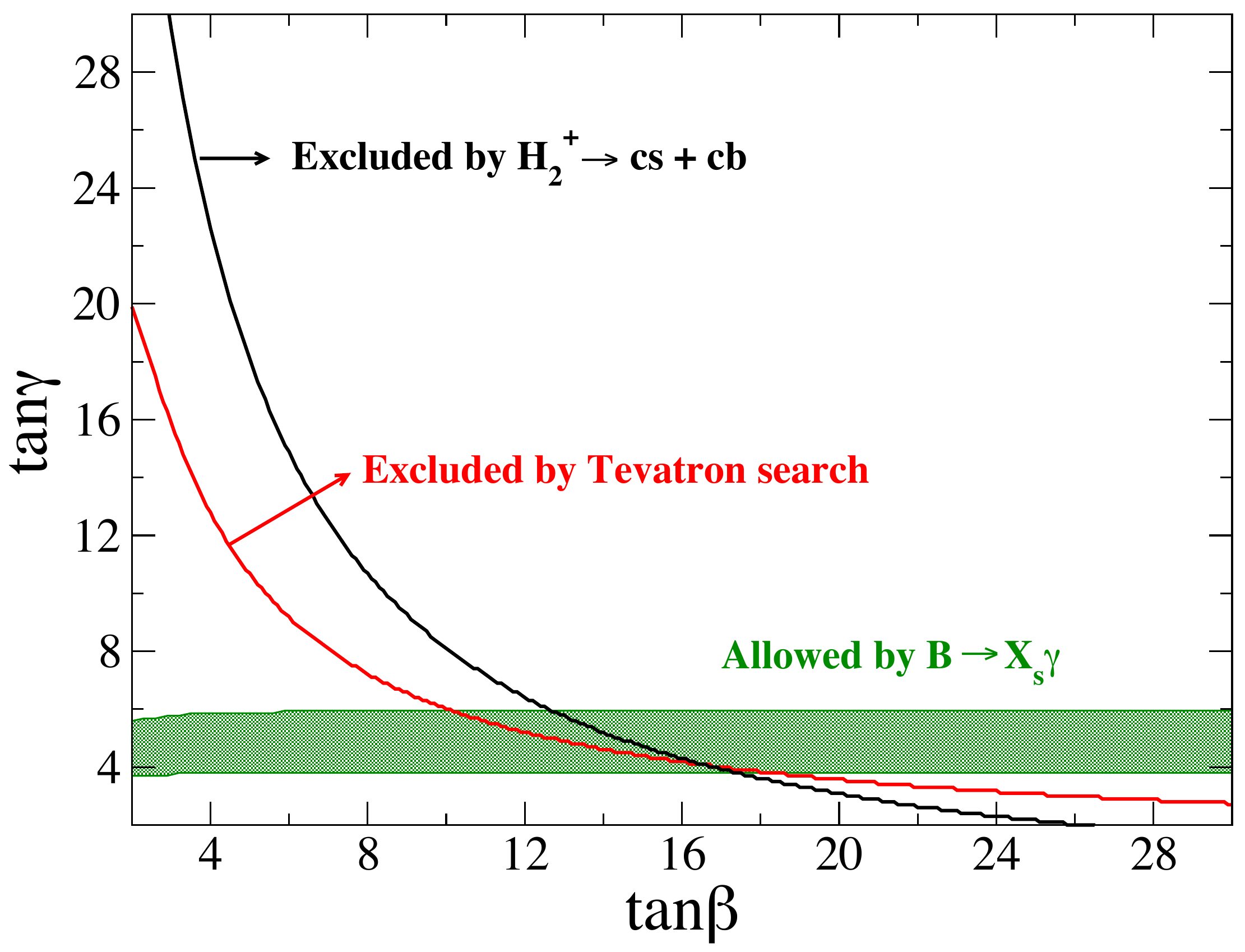}
\includegraphics[scale=0.36]{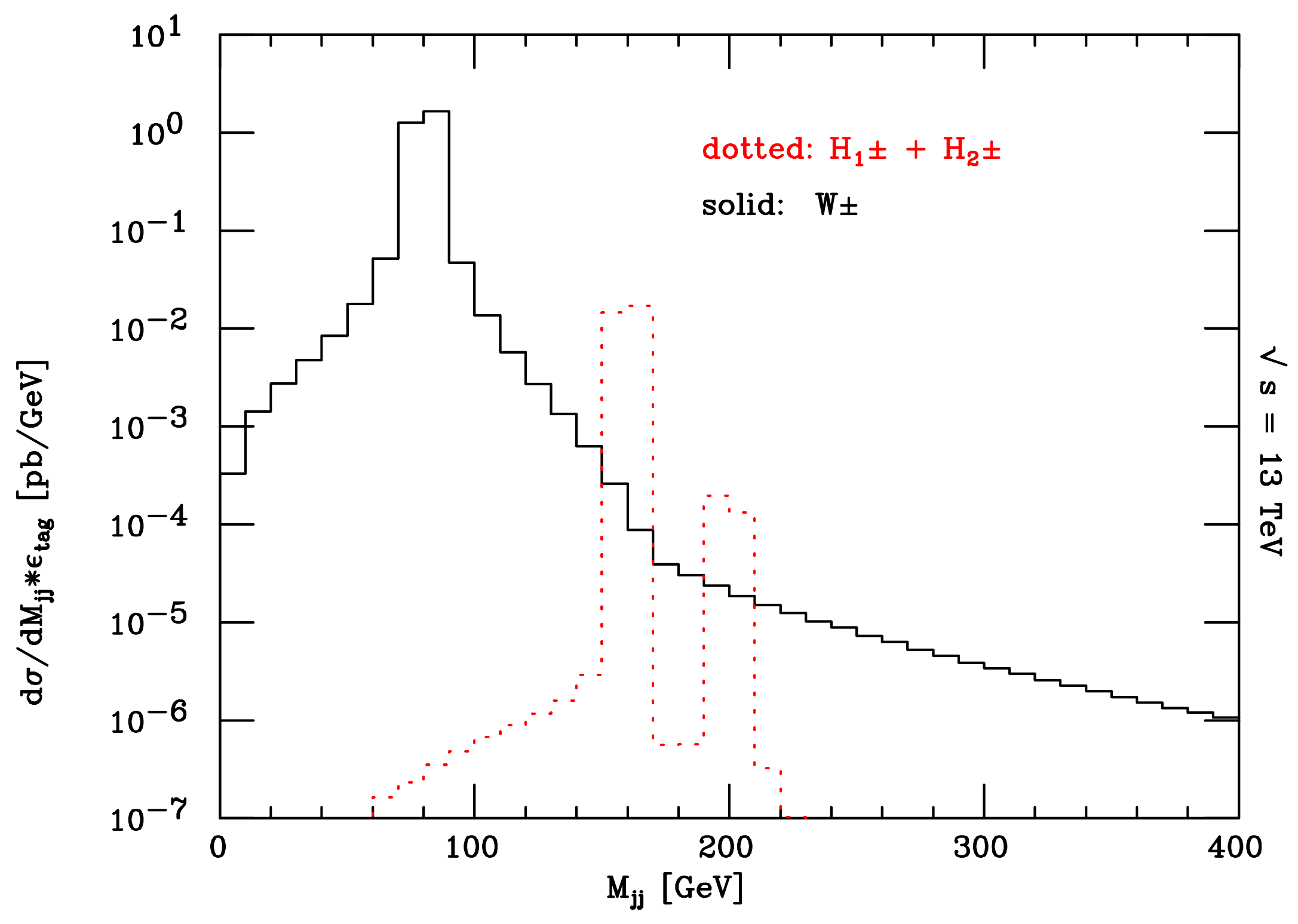}
\caption{Left: Allowed parameter space for the 3HDM (Model~Y)  with $M_{H_1^\pm}=83$ GeV, $M_{H_2^\pm}=160$ GeV and the mixing angle $\theta_C=-\pi/4$.
Only the green shaded region is allowed by the $B\to X_s\gamma$ constraint.
Right: Di-jet invariant mass distributions for signal (red) and background (black)
for a particular benchmark point (3HDM, Model~Y) at the 13~TeV LHC.}
\label{const3}
\end{center}
\end{figure}

The fully active 3HDMs with two softly-broken discrete $Z_2$ symmetries have 
two pairs of charged Higgs bosons, $H_1^\pm$ and $H_2^\pm$,
studied in~\cite{Akeroyd:2016ssd}. 
Depending on the $Z_2$ parity assignment, 
there are different Yukawa interactions. 
In each of these, the phenomenology of the charged Higgs
bosons is in the CP-conserving case described by five parameters:
the masses of the charged Higgs bosons, 
two ratios of the Higgs vev's $\tan\beta$ and $\tan\gamma$ and 
a mixing angle $\theta_C$ between $H_1^+$ and $H_2^+$. 
The BR($B\to X_s\gamma$) is determined by $W^+$, $H_1^+$ and $H_2^+$ loop contributions. 
The scenario with masses of ${\cal O}(100~\text{GeV})$  for the charged Higgs bosons is allowed.
Therefore, the search for a light charged Higgs boson, which in some Yukawa models dominantly decays into $c\bar{b}$,
may allow to distinguish 3HDMs from 2HDMs. Some results are presented in Fig.~\ref{const3} for the 3HDM Model~Y.

Experimental constraints on $t\to H^+ b$ followed by $H^+\to \tau^+\nu$ and 
$H^+\to c\bar{s}+c\bar{b}$ are relevant here. 
Scenarios with both $M_{H_1^\pm},M_{H_2^\pm}\lsim m_t$ are highly constrained 
from $B\to X_s\gamma$ and the LHC direct searches. 
The particular case $M_{H_1^\pm}\simeq m_W$ with 90 GeV$<M_{H_2^\pm}< m_t$ is allowed
(also by the Tevatron and LEP2). 
The region of 80 GeV$<M_{H_1^\pm}<90$ GeV is not constrained by current LHC searches
for $t\to H^+ b$ followed by dominant decay $H^+ \to c\bar{s}/c\bar{b}$, and 
this parameter space is only weakly constrained from LEP2 and Tevatron searches, see Fig.~\ref{const3} (left). 
Any future signal in this region could readily be accommodated by $H_1^\pm$ from 
a 3HDM. 

A Monte Carlo simulation of the $H^+_{1,2}$ signals and $W^+$ background 
via the processes $gg,q\bar q\to t\bar b H^-_{1,2}$ and
$gg,q\bar q\to t\bar b W^-$, respectively, followed by the corresponding 
di-jet decays is shown in Fig.~\ref{const3} (right). The charged Higgs boson signals should be accessible
at the LHC, provided that $b$-tagging is enforced so as to single out the $c\bar b$ component above the $c\bar s$ one (see the following subsection).
Therefore, these (multiple) charged Higgs boson  signatures can be used not only to distinguish between 2HDMs and 3HDMs but also to identify the particular Yukawa model realising the latter. Some benchmark points are provided in \cite{Akeroyd:2016ssd}.

\subsection{Enhanced $H^+\to c\bar b$ branching ratio}

In the 2HDM, the magnitude of BR($H^+ \to c\bar b$) is always less than a few 
percent, with the exception of Model~Y (see Fig.~\ref{Fig:cpc-br-ratios-low}), since the
decay rate is suppressed by the small CKM element $V_{cb}\, (\ll V_{cs})$.

A distinctive signal of $H^+$ from a 3HDM for $M_{H^+}\lsim m_t$ could be a 
sizeable branching ratio for $H^+ \to c\bar b$ \cite{Grossman:1994jb,Akeroyd:1994ga,Akeroyd:1998dt}.
The scenario with $|{\cal F}_1^D|\gg |{\cal F}_1^U|,|{\cal F}_1^L|$ corresponds to a ``leptophobic'' $H^+$ with\footnote{A similar situation arises in the 2HDM (Y), for $\tan\beta \gg 3$.} 
$\text{BR}(H^+\to c\bar s)+\text{BR}(H^+\to c\bar b) \sim 100\%. $

\begin{figure}[htb]
\begin{center}
\includegraphics[origin=c, angle=0, scale=0.5]{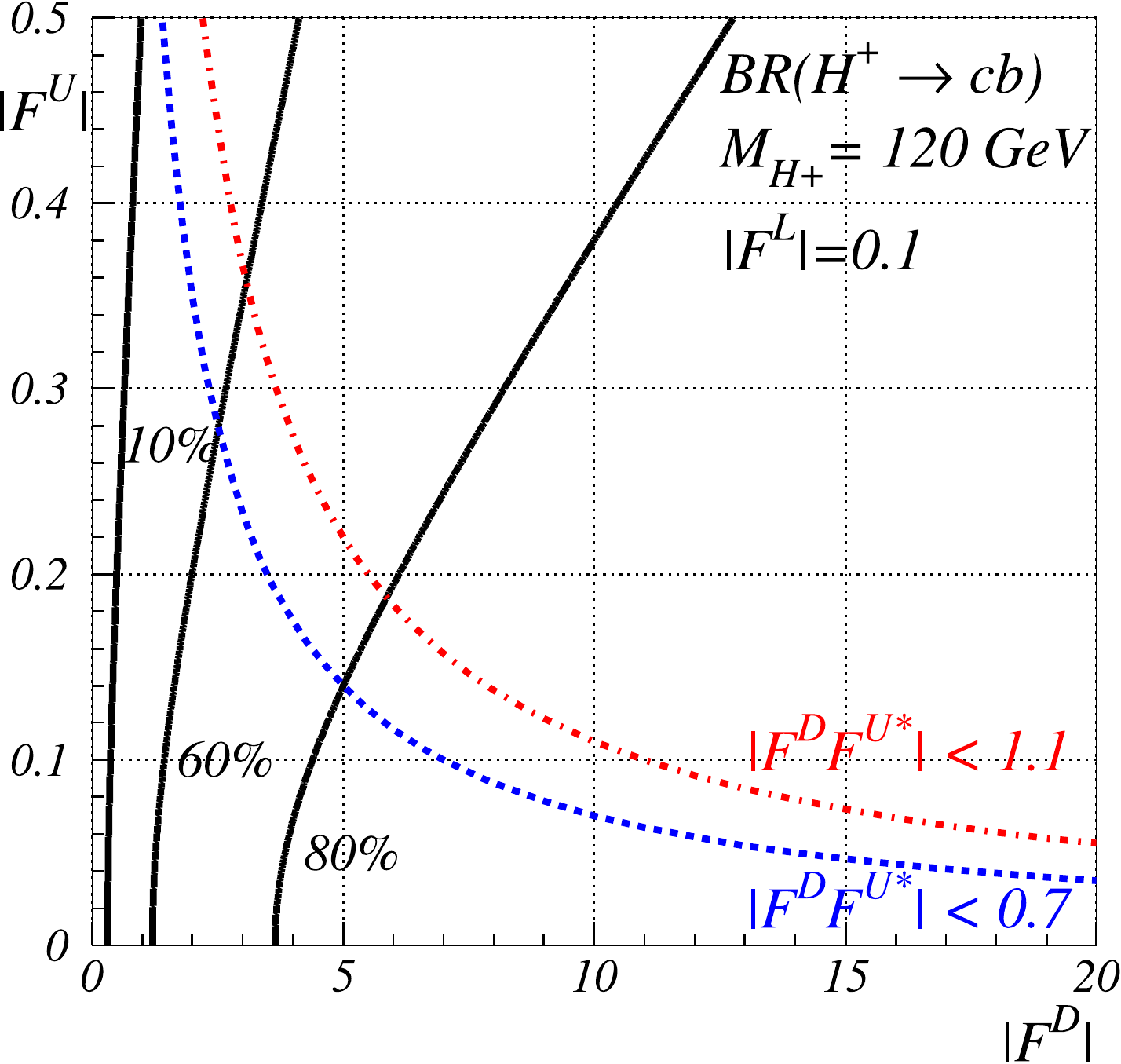}
\includegraphics[scale=0.5]{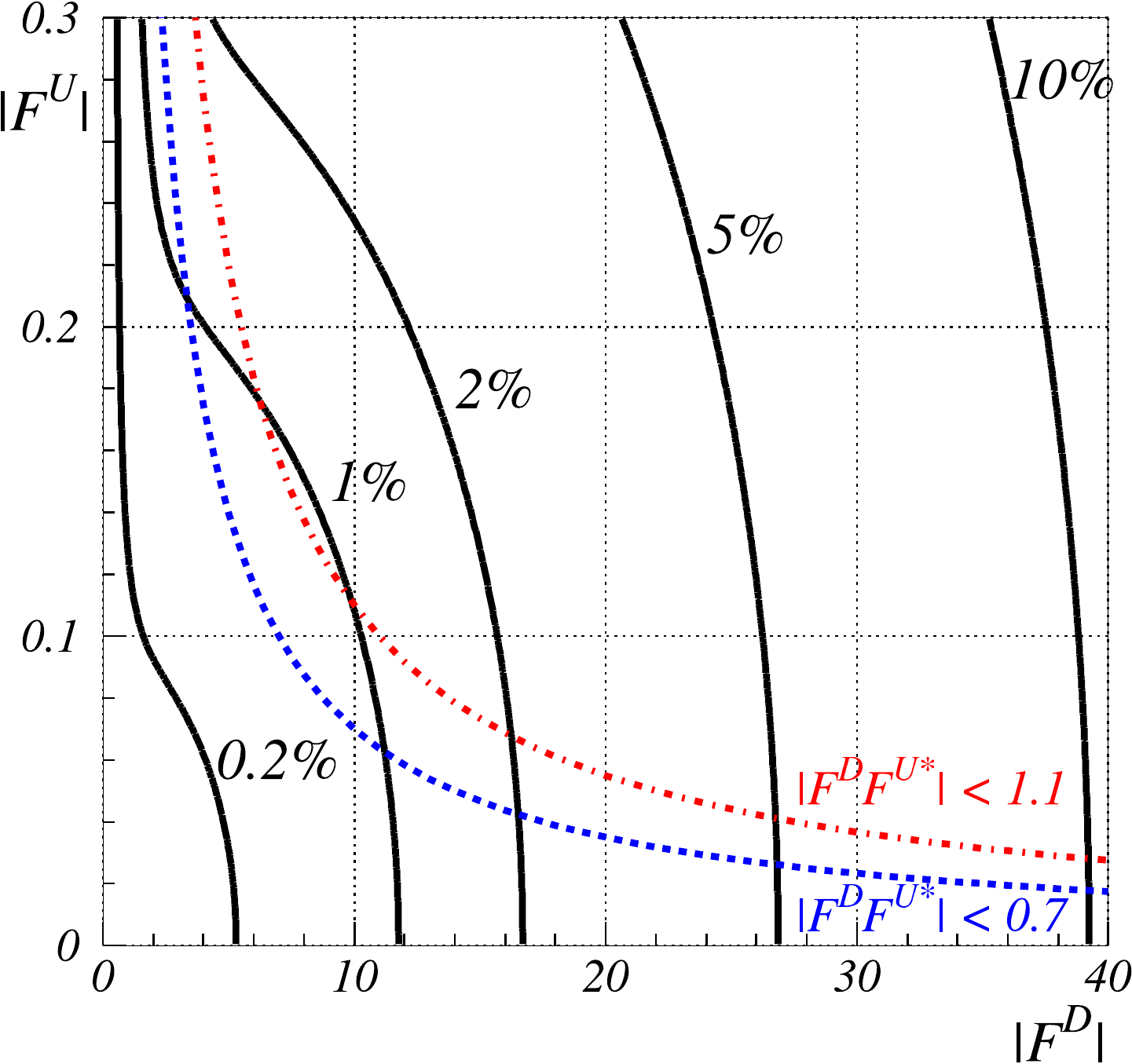}
\caption{Left: Contours of BR$(H^+\to cb$) in the $|{\cal F}^D_1|$--$|{\cal F}^U_1|$-plane with $|{\cal F}^L_1|=0.1$.
The $B \to X_s \gamma$ constraint removes the region above the red or blue hyperbolas, see the text.
Note that the two scales on the axes are very different.
Right: Contours of BR$(t\to H^+ b)\times {\rm BR}(H^+\to cb$).}
\label{fig:brcb}
\end{center}
\end{figure}

In this limit, the ratio of BR$(H^+\to c\bar b)$ and BR$(H^+\to c\bar s)$ can be expressed as follows:
\begin{equation}
\frac{{\rm BR}(H^+\to c\bar b)}{{\rm BR}(H^+\to c\bar s)}
\equiv R_{bs}
\sim \frac{|V_{cb}|^2} {|V_{cs}|^2} \,\frac{m_b^2}{m_s^2}.
\label{cbcsratio}
\end{equation}

In Ref.~\cite{Akeroyd:2012yg} the magnitude of BR($H^+ \to c\bar b$) 
as a function of the couplings ${\cal F}^U_1$, ${\cal F}^D_1$ and ${\cal F}^L_1$
was studied, updating the numerical study of \cite{Akeroyd:1994ga}.
As an example,
in Fig.~\ref{fig:brcb} (left), BR$(H^+\to c\bar b)$ in a 3HDM is displayed in 
the $|{\cal F}^D_1|$--$|{\cal F}^U_1|$-plane for $M_{H^\pm}=120$ GeV,
with $|{\cal F}^L_1|=0.1$.
The maximum value is BR$(H^+\to c\bar b)\sim 81\%$.
The bound from $B \to X_s \gamma$ is also shown, 
which is $|{\cal F}^D_1{\cal F}^U_1| < 1.1$  (0.7) for ${\cal F}^D_1{\cal F}^{U*}_1$ being real and negative (positive).

Increased sensitivity can be achieved by requiring also a $b$-tag on the jets 
from the decay of $H^+$.
In Fig.~\ref{fig:brcb} (right) for $M_{H^\pm}=120$ GeV we show contours of
${\rm BR}(t\to H^+ b)\times {\rm BR}(H^+\to c\bar b)$,
starting from 0.2\%, accessible at the LHC.
In this case, a large part of the region of $|{\cal F}^D_1|<5$ 
could be probed, even for $|{\cal F}^U_1|<0.2$. 

In summary, a distinctive signal of $H^+$ from a 3HDM for $M_{H^+}\lsim m_t$ could be a 
sizeable branching ratio for $H^+ \to c\bar b$.
A dedicated search for $t\to H^+ b$ and $H^+ \to c\bar b$, in which
the additional $b$-jet originating from $H^+$ is tagged,
would be a well-motivated and (possibly) straightforward extension of the ongoing searches 
with the decay $H^+ \to c\bar s$.

\section{Models with charged scalars and DM candidates} 
\label{sect:inert-models}
\setcounter{equation}{0}
It is possible that the issues of dark matter (DM) and mass generation
are actually related.\footnote{The SM Higgs boson mass term, 
$\sim \Phi^\dagger \Phi$,
may allow for a connection (Higgs portal) to a hypothetical hidden sector \cite{Patt:2006fw}.}
Such models must of course contain a Standard-Model-like neutral Higgs particle, with  mass at $M_h=125~\text{GeV}$.
Additionally, there appear charged Higgs particles and other charged (as well as neutral) scalars.
In these models, the DM relic density provides a constraint on the charged scalars.

In order to have a stable DM candidate some $Z_2$ symmetry is typically introduced,
under which an SU(2) singlet or doublet involving the DM particle, is odd. The $Z_2$-odd scalars are often called dark scalars. Among them, the lightest neutral one is a DM candidate. Below, we will denote the charged ones $S^+$, and the neutral ones $A$ and $S$, with $S$ being the lightest. In some models, there may be several scalars, then referred to as $S_i^+$ and $S_i$.

Typically, the charged scalars of these models have some features in
common with the charged Higgs of Model~I (and X). They do not couple
to the $b$ and $s$ quarks (in fact, they does not couple to {\it any}
fermion), and thus are not affected by the $B \to X_s \gamma$ constraint. Hence, they can be rather light. LEP searches for charginos can be used to establish a lower mass bound of about 70~GeV \cite{Pierce:2007ut} for such charged scalars. 

The first model which allows this relationship between the Higgs and DM sectors was introduced many years ago \cite{Deshpande:1977rw}, and will here be referred to as the ``Inert Doublet Model'', or IDM \cite{Ma:2006km,Barbieri:2006dq,LopezHonorez:2006gr}.\footnote{The initial motivation was to provide a mechanism for neutrino mass generation.} Here, one SU(2) doublet ($\Phi_1$) plays the same role as the SM scalar doublet, the other one with zero vacuum expectation value does not couple to fermions. 
An extension of this model with an extra doublet \cite{Grzadkowski:2009bt,Grzadkowski:2010au,Keus:2014jha,Keus:2015xya} or a singlet \cite{Bonilla:2014xba} allows also for CP violation. This improves the prospects for describing baryogenesis \cite{Riotto:1999yt}.

Alternatively, a $Z_2$-odd scalar SU(2) singlet ${\cal S}$ \cite{McDonald:1993ex,Burgess:2000yq,Barger:2007im,Barger:2008jx} mixed with a $Z_2$-odd scalar SU(2) doublet $\Phi_2$ may provide a framework for dark matter.\footnote{This is in contrast to the mechanism discussed in \cite{Bonilla:2014xba}, where the singlet has a non-zero vev and is not related to the dark matter.}
It was shown in Refs.~\cite{Kadastik:2009dj,Kadastik:2009cu}  that 
the high-energy theory leading to electroweak-scale scalar DM models
can be a non-SUSY SO(10) Grand Unified Theory (GUT) \cite{Fritzsch:1974nn}.
Indeed,  the discrete $Z_2$ symmetry which makes DM stable, could be
an unbroken discrete remnant of some underlying U(1) gauge subgroup \cite{Krauss:1988zc,Martin:1992mq,DeMontigny:1993gy}.
Unlike in the IDM, in the GUT-induced scalar DM scenario the lightest dark scalar is predicted by RGEs to be dominantly singlet.

A corresponding charged dark scalar $S^+$ can be searched for in the decays
 \begin{equation} \label{Eq:darkHplusdecay}
S^+ \to S_i  f\bar{f^\prime},
\end{equation}
mediated by a virtual or real $W^\pm$,
where $S_i$ is a neutral dark scalar, and $f$ and $f^\prime$ denote SM fermions. 
 
The production at the LHC of $S^+S^-$ pairs, $pp\to S^+S^-$, and of $S^+$ together with neutral dark scalars $pp\to S^+S_i$ were investigated \cite{Cao:2007rm,Dolle:2009ft,Miao:2010rg,Huitu:2010uc}.
Because of relic density and electroweak precision measurement constraints, $S^+$ and $S$ tend to be close in mass,
$M_{S^\pm}-M_S\ll m_W$.
In such regions of parameter space, and in the limit of massless fermions, the $S^+$ width is \cite{Osland:2013sla}
\begin{equation}
\Gamma(S^+\to S f \bar f^\prime)
=\frac{G_\text{F}^2}{30\pi^3}(M_{S^\pm}-M_S)^5,
\end{equation}
so the $S^+$ will be long-lived, and travel a macroscopic distance.

\subsection{The Inert Doublet Model, IDM}
\label{Sect:IDM}

The Inert Doublet Model can be defined in terms of the potential
\begin{align} \label{Eq:V_IDM}
V_\text{IDM}  
&= -\frac12\left\{m_{11}^2\Phi_1^\dagger\Phi_1
+ m_{22}^2\Phi_2^\dagger\Phi_2 \right\}
+ \frac{\lambda_1}{2}(\Phi_1^\dagger\Phi_1)^2
+ \frac{\lambda_2}{2}(\Phi_2^\dagger\Phi_2)^2 \nonumber \\
&+ \lambda_3(\Phi_1^\dagger\Phi_1)(\Phi_2^\dagger\Phi_2)
+ \lambda_4(\Phi_1^\dagger\Phi_2)(\Phi_2^\dagger\Phi_1)
+ \frac12\left[\lambda_5(\Phi_1^\dagger\Phi_2)^2 +\text{h.c.}\right].
\end{align}
This is the same potential as in Eq.~(\ref{Eq:fullpot}), but without the $Z_2$-breaking term proportional to $m_{12}^2$, and hence with $\lambda_5$ real, see Eq.~(\ref{Eq:m12}).

The charged scalar mass coming from \Eq{Eq:V_IDM} is  given by
\begin{equation}
M_{S^{\pm}}^{2} = -\frac{m_{22}^{2}}{2} + \frac{\lambda_{3} v^{2}}{2}.
\label{mch}
\end{equation}
The parameter $\lambda_3$ also governs the coupling of the $S^+$ to the Higgs particle $h$. 
\medskip

Perturbative unitarity  constraints on other lambdas, together with precision data on the electroweak parameters $S$ and $T$, limit masses of the dark scalars to less than 600~GeV for $S$ and less than 700~GeV for $A$ and $S^+$, for $|m_{22}^2|$ below $10^4~\text{GeV}^2$. Much heavier dark particles are allowed for large, negative values of $m_{22}^2$, e.g.\ $M_A, M_S, M_{S^\pm}$ can take values up to 1~TeV for $m_{22}^2=-(1~\text{TeV})^2$. 

Measurements of $R_{\gamma \gamma}$ strongly constrain masses and couplings of dark particles, the closer to 1 is this ratio, the higher the masses of dark particles, including $S^+$,  are allowed. Enhancement of this ratio above 1 would only be possible if there were no open invisible channels ($2M_S>M_h$). For example, for $R_{\gamma \gamma}> 1.3$ the range of $M_{S^\pm}$ would have to be below $\sim135~\text{GeV}$.   

If the model is required to saturate the relic DM abundance, then $M_{S^\pm}$ has to be below approximately 300~GeV, or else above $\sim 500~\text{GeV}$. In the latter case, its mass is very close to that of the DM particle.
On the other hand, if the model is not required to saturate the DM relic abundance, only not to produce too much DM, then the charged-scalar mass is less constrained \cite{Ilnicka:2015jba}.

Analyses based on an extensive set of theoretical and experimental constraints on this model have recently been performed, both at tree level \cite{Arhrib:2013ela,Ilnicka:2015jba} and at loop level \cite{Goudelis:2013uca}. Collider as well as astroparticle data limits were included, the latter in the form of dark matter relic density as well as direct detection data. A minimal scale of 45~GeV for the dark scalar mass, and a stringent mass hierarchy $M_{S^\pm} >  M_A$ are found \cite{Ilnicka:2015jba}. Parameter points and planes for dark scalar pair production $S^+S^-$ for the current LHC run are proposed, with $S^+$  masses in the range 120--450~GeV \cite{Ilnicka:2015jba}.  It is found that the decay $S^+ \to W^+ S$ dominates, and $M_{S^\pm} - M_S > 100~\text{MeV}$. 
A heavier $S^+$ benchmark ($M_{S^\pm} > 900~\text{GeV}$) is also proposed \cite{Goudelis:2013uca}.

\subsection{The CP-violating Inert Doublet Model, IDM2}
\label{Sect:IDM2}
If the above model is extended with an extra doublet, one can allow for CP violation, like in the 2HDM \cite{Grzadkowski:2009bt,Grzadkowski:2010au}.
There are then a total of three doublets, one of which is inert in the sense that it has no vacuum expectation value, and hence no coupling to fermions. This kind of model will have two charged scalars, one ($H^+$) with fermionic couplings and phenomenology similar to that of the 2HDM (but the constraints on the parameter space will be different, due to the extra degrees of freedom), and one ($S^+$) with a phenomenology similar to that of the IDM. 

The charged scalar $S^+$ could be light, down to about 70~GeV. Its phenomenology has been addressed in \cite{Osland:2013sla}. The allowed ranges for the DM particle are similar to those found for the IDM, a low-to-medium mass region up to about 120~GeV, and a high-mass region above about 500~GeV. 

The decay modes of the additional charged scalar are the same as in the IDM, it decays either to a $W$ and a DM particle, or to a $Z$  and the neutral partner of the DM particle \cite{Osland:2013sla}. In the low-to-intermediate $S^+$ mass region, the $W$ and $Z$ could be virtual.

\subsection{One or two inert doublets within 3HDM}
\label{Sect:3HDM}
In the context of 3HDMs, models with one or two inert doublets were considered, involving   new charged Higgs and/or charged inert scalars. (The IDM2 model discussed above, is one such case, allowing for CP violation.)
The richer inert particle spectrum for the case with two inert doublets enables a variety of co-annihilation channels of the DM candidate, including those with two different pairs of
charged inert bosons ~\cite{Keus:2014jha,Keus:2015xya}. This allows to relieve  the tension in current experimental constraints from Planck, LUX and the LHC. As a consequence, new DM  mass regions open up, both at the light  ($M_S\lsim 50 $ GeV)  and heavy (360 GeV $\lsim M_S\lsim 500$ GeV)  end of
the spectrum, which are precluded to the IDM and are in turn testable at the LHC.  Concerning LHC phenomenology of visible channels, a smoking gun signature of the model with two inert doublets is a new decay channel  of the next-to-lightest inert scalar into the scalar DM candidate involving (off-shell) photon(s) plus missing energy \cite{Keus:2014isa}, which is enabled by $S^+_{i}W^-$ loops.   The hallmark signal for the model with one inert doublet  would be  significantly increased $H^+\to W^+ Z$ and $W^+\gamma$
decay rates (in which a key role is played by loops involving the $S^+$ state) with respect to the IDM \cite{Moretti:2015tva}. This new phenomenology is compliant with the most up-to-date constraints on the respective parameter spaces, both experimental and theoretical \cite{Moretti:2015cwa}.

\subsection{SO(10) and the GUT-induced scalar DM scenario}
The GUT-induced scalar DM  scenario with  minimal particle content
includes a Higgs boson in a ${\bf 10}$ and  the DM in a ${\bf 16}$ representation of SO(10).
One identifies here the  $Z_2$ symmetry as the matter parity  \cite{Kadastik:2009dj,Kadastik:2009cu},
defined as
\begin{equation}
P_M=(-1)^{3(B-L)}.
\label{PM}
\end{equation}
Since the matter parity $P_M$ is directly related to the  breaking of $B-L,$ the dark sector
actually consists of scalar partners of the SM fermions which carry the same gauge quantum numbers as the MSSM squarks and sleptons.
In this scenario the origin and stability of DM, the non-vanishing neutrino masses via the seesaw mechanism \cite{Minkowski:1977sc,Yanagida:1979as,GellMann:1980vs,Glashow:1979nm,Mohapatra:1979ia} and
the baryon asymmetry of the Universe via leptogenesis \cite{Fukugita:1986hr} all spring from
the same source -- the breaking of the SO(10) gauge symmetry.

The IDM represents just one particular corner of parameter space of the
general $P_M$-odd scalar DM  scenario in which the DM is predominantly doublet.
A theoretically better motivated particle spectrum can be obtained  by renormalization group (RG) evolution of the model parameters from the GUT scale
to $m_Z$ \cite{Kadastik:2009cu}, in a direct analogy with the way the particle spectrum is obtained in the constrained minimal supersymmetric standard model (CMSSM).\footnote{To obtain successful EWSB at low energies ($\sim m_Z$) the mass parameter $m_{11}^2$ in Eq.~(\ref{Eq:V_IDM}) can become positive either by the RG evolution \cite{Kadastik:2009cu} or via the Coleman-Weinberg-like  \cite{Coleman:1973jx} mechanism \cite{Hambye:2007vf,Kadastik:2009ca}.}
In the GUT-induced scalar DM model, the EWSB may occur due to the existence of dark scalar couplings to the Higgs boson.
Moreover such couplings can lower the stability bound and accommodate also a Higgs mass around 125 GeV \cite{Kadastik:2011aa}.

Below the  GUT scale $M_\text{G}$ and above the EWSB scale
the model is described by the scalar potential for the doublets and the singlet:
\begin{equation} \label{Eq:V_SO10}
V = V_\text{IDM}
+ \text{terms bilinear and quartic in ${\cal S}$},
\end{equation}
invariant under
\begin{equation}
(\Phi_{1},\Phi_{2},{\cal S}) \leftrightarrow (\Phi_{1},-\Phi_{2},-{\cal S}),
\end{equation}
and with $V_\text{IDM}$ defined by Eq.~(\ref{Eq:V_IDM}).
The charged scalar mass of this model is determined by $V_\text{IDM}$ and given by Eq.~(\ref{mch}).

The mass degeneracy of $S$ and the next-to-lightest neutral scalar, denoted $S_{\text{NL}}$,  is a generic property of the scenario and follows from the underlying SO(10) gauge symmetry. It implies a long lifetime for
 $S_{\text{NL}}$ which provides a clear experimental signature of a displaced vertex in the decays $S_{\text{NL}}\to S \ell^+ \ell^-$ at the LHC \cite{Kadastik:2009gx}.

\subsection{Dark charged scalar phenomenology at the LHC}
\begin{figure}[htb]
\centering
\includegraphics[scale=0.9]{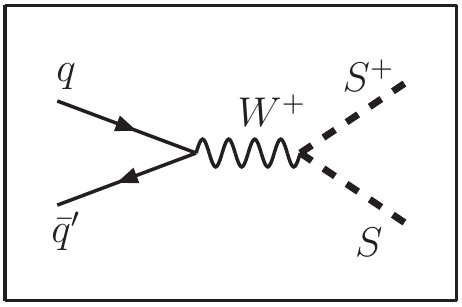}
\caption{Feynman diagram for associated dark $S^+S$ production at the LHC. \label{Fig:HSprod}}
\end{figure}

Compared to the $H^+$ of 2HDM models, dark $S^+$ production lacks some primary parton level
processes, since it has to be produced in association with a neutral dark-sector particle, as illustrated in  Fig.~\ref{Fig:HSprod}, or else pair produced via $\gamma$, $Z$ or $H_j$ in the $s$-channel (see Fig.~\ref{Fig:feyn-figures-pair}).

The early literature \cite{Barbieri:2006dq,Cao:2007rm,Dolle:2009ft,Miao:2010rg} on the IDM was mostly aimed at guiding the search for evidence on the model, via the production of the charged member, $S^+$, together with a neutral one. It focused on a DM mass of the order of 60--80~GeV, and a charged state $S^+$ with a mass of the order of 100--150~GeV. For these masses, the production cross sections are at 14~TeV of the order of $100$--$500~\text{fb}$, and the two- \cite{Dolle:2009ft} and three-lepton \cite{Miao:2010rg} channels were advocated.
For an update on the allowed parameter space and proposed benchmark points, see Ref.~\cite{Ilnicka:2015jba}.

While the $S^+S$ channel has the highest cross section, because of the larger phase space, its discovery is challenging.
The $S^+$ would decay to an (invisible) $S$, plus a virtual $W^+$, giving a two-jet or a lepton-neutrino final state. The overall signal would thus be jets or an isolated charged lepton (from the $W$) plus missing transverse energy.
If instead the heavier neutral state $A$ is produced together with the $S^+$, some of its decays (via a virtual $Z$) would lead to two-lepton and three-lepton final states \cite{Dolle:2009ft,Miao:2010rg}. Various cuts would permit the extraction of a signal against the $t\bar t$ and $WZ$ background. In the case of the very similar phenomenology of the IDM2, a study of $M_S=75~\text{GeV}$ with $100~\text{fb}^{-1}$ of data concludes that the best $S^+$ search channel is in the hadronic decay of the $W$, leading to two merging jets plus missing transverse energy \cite{Osland:2013sla},
\begin{equation}
pp \to j+\text{MET}.
\end{equation} 

Recently, also the four-lepton modes for $S^+$ masses in the range 98--160~GeV have been studied \cite{Gustafsson:2012aj} , and how to constrain the model from existing data on SUSY searches \cite{Belanger:2015kga}, considering two $S^+$ masses, 85 and 150~GeV. 
Another recent study of the $S^+S$ and $S^+A$ channels \cite{Poulose:2016lvz} concludes that the dijet channel may offer the best prospects for discovery, but that a luminosity of $500~\text{fb}^{-1}$ would be needed for an $S^+$ mass up to 150~GeV, whereas 1 and $2~\text{ab}^{-1}$ for masses of 200 and 300~GeV.

In the above SO(10) scenario the mass difference $M_{S^\pm}-M_S$ turns out to be
less than $m_W$ and at the leading order the allowed decays
are only those given by Eq.~(\ref{Eq:darkHplusdecay}) with a virtual $W$.

In some cases $S^{+}$ is so long-lived (decay length $\ell \gtrsim 1$~mm), due to an accidental  mass degeneracy
between $S^+$ and $S$ that it may decay outside the detector.
Those experimental signatures are in principle background free and allow $S^+$ to be discovered at the LHC
up to masses $M_{S^\pm}\lesssim 300$~GeV.

To study charged scalar pair production in the SO(10) scenario at the LHC, parameter points with distinctive phenomenologies are proposed \cite{Kadastik:2009dj,Kadastik:2009cu,Kadastik:2009gx,Kadastik:2009ca,Huitu:2010uc,Kadastik:2011aa}.
For some parameter points, the  $S^+$ decays inside the tracker of an LHC experiment.
The experimental signature of those points is that the charged
track of $S^+$  breaks into a  charged lepton track and missing energy. 

\section{Summary and outlook}
\label{sect:summary}
\setcounter{equation}{0}

Since the summer of 2012 we are in the final stage of confirmation of the foundation of the SM. However, so far there is no clear clue for a further direction. Various SM-like models with extra Higgs scalars exist. A charged Higgs boson ($H^+$) would be the most striking signal of a Higgs sector with more than one Higgs doublet. Such a discovery at the LHC is a distinct possibility, with or without supersymmetry. However, a charged Higgs particle might be hard to find, even if it is abundantly produced.

For masses of the charged scalar below 500~GeV, a variety of 2HDM models remain viable, with $H^+$ decaying either to heavy flavors or to $\tau^+\nu$. Some of these have Model~I-type Yukawa couplings, others arise in models that accommodate dark matter.
Above 500 GeV, also Model~II would be a possible interpretation. Here, the most ``natural'' decay modes would be to $t\bar b$ and $H_jW^+$, where $H_j$ could be any of the three neutral Higgs bosons.

If a signal were to be found, one of the first questions would be whether it is the charged Higgs of the MSSM or not.
We note that the MSSM mass spectrum is very constrained, the heavier states should be close in mass. Secondly, the Yukawa couplings would at tree level be those of
Model~II. This means that the low-mass region would be severely
constrained by $B \to X_s \gamma$, unless there is some cancellation of the $H^+$ contribution. A natural candidate would be a squark-chargino loop. But lower bounds on squark and chargino masses make this hard to arrange.

The charged Higgs boson can also be part of a higher representation. 
Additionally, in higher representations one could have doubly-charged $H^{++}$, and also more ``exotic"
decay modes.
For example, with a Higgs triplet, one could have \cite{Cheung:2002gd}
$H^+\to W^+Z$ at the tree-level. Note however, a recent ATLAS analysis \cite{Aad:2015nfa} excludes charged Higgs between 240 and 700 GeV if $H^+\to W^+Z$ is the dominant decay mode.
This could be the case for the Georgi--Machacek model \cite{Georgi:1985nv}.
This process is also possible in the 3HDM discussed in section~\ref{sect:multi-Higgs-doublet-models}, but then only generated at the one-loop level.

For the above-mentioned $H^+\to H_jW^+$ decay modes, there are two competing effects. (i) Decay to the lightest state $H_1$ (or $h$) benefits from a non-negligible phase space, but vanishes in the alignment limit, see Eq.~(\ref{Eq:CPC-gauge-couplings}) and Fig.~\ref{Fig:br-toWH1}. (ii) Decays to the heavier ones, $H_2$ and $H_3$ (or $H$ and $A$), where couplings are not suppressed, suffer from a small phase space, since various constraint ($T$, in particular) force these masses to be close to $M_{H^\pm}$.
In view of the convergence of measurements pointing to a CP-conserving Higgs sector, and alignment, the parameter space for the $H^+\to H_jW^+$ decay modes is shrinking, and at high masses the $t\bar b$ mode may be the most promising one. However, the QCD background is very challenging, so improved analysis techniques could turn out to be very beneficial.

\vspace{2cm}
\noindent
\paragraph{Acknowledgements:} 
We would like to thank the organizers of numerous workshops on
``Prospects for Charged Higgs Discovery at Colliders'', Uppsala, Sweden, in particular Prof.\ Tord Ekel\"of, for encouraging and stimulating this study.

It is a pleasure to thank M.~Misiak and T.~Peiffer for discussions on the low-energy constraints, and permission to reproduce some of their figures.
It is also a pleasure to thank the ATLAS and CMS collaborations for kind permission to reproduce some of their figures.
Furthermore, it is a pleasure to thank R. Enberg and M. Hashemi for discussions.

The  work  of  AGA, AA, RE, KH, SK,  SM, RS, EY and KY is
funded through the grant H2020-MSCA-RISE-2014 No. 645722  (NonMinimalHiggs).   
MK and IG are supported in part by the National Science  Centre, Poland,  the HARMONIA project under contract UMO-2015/18/M/ST2/00518.
SM  is  supported  in  part
through  the  NExT  Institute  and  STFC  Consolidated
Grant ST/J000396/1. 
PO is supported in part by the Research Council of Norway.
The work of PS was supported by the Australian Research Council through the ARC Center of Excellence in Particle Physics (CoEPP) at the Terascale (grant no. CE110001004).
KY's work is supported by a JSPS
Postdoctoral Fellowships for Research Abroad. 
EY is supported by the Ministry of National Education of Turkey.
This work was supported by the Estonian Research Council grants
PUT799, PUT1026, IUT23-4, IUT23-6 and through the ERDF CoE program.

\appendix
\section*{Appendix A: Field decompositions}
\label{app:decomposition}
\renewcommand{\thesection}{A}
\setcounter{subsection}{0}
\setcounter{equation}{0}

Breaking the electroweak symmetry spontaneously, we assume that the electrically neutral components of the Higgs doublets have non-zero expectation values, cf.~Eq.~\eqref{Obasis}. By assuming that they are real and positive, we define a basis in which 
\begin{equation}
\Ave{\Phi_1}=\begin{pmatrix}
0 \\
\frac{v_1}{\sqrt{2}}
\end{pmatrix}, \qquad
\Ave{\Phi_2}=\begin{pmatrix}
0 \\
\frac{v_2}{\sqrt{2}}
\end{pmatrix}
\end{equation}
with
\begin{equation}
\tan\beta=\frac{v_2}{v_1}.
\end{equation}
Note that the introduced parameter $\tan\beta$ has no \emph{a priori} connection to the Yukawa interaction.

The decompositions for $\Phi_1$ and $\tilde\Phi_1=-\mathrm{i}\left[\Phi_1^\dagger\sigma_2\right]^\text{T}
=\mathrm{i}\sigma_2\Phi_1^*$ are given by
\begin{equation}
\Phi_1=\begin{pmatrix}
\varphi_1^+ \\
\frac{1}{\sqrt{2}}(v_1+\eta_1+i\chi_1)
\end{pmatrix}, \quad
\tilde\Phi_1=\begin{pmatrix}
\frac{1}{\sqrt{2}}(v_1+\eta_1-i\chi_1) \\
-\varphi_1^-
\end{pmatrix}
\end{equation}
and similarly for $\Phi_2$ and $\tilde\Phi_2$.

The massless charged Goldstone boson, $G^+$, and the charged Higgs boson, $H^+$, are given as
\begin{align}
G^+&=\cos\beta\,\varphi_1^++\sin\beta\,\varphi_2^+, \nonumber \\
H^+&=-\sin\beta\,\varphi_1^++\cos\beta\,\varphi_2^+.
\end{align}
Inverting these relations, we find
\begin{align}
\varphi_1^+&=\cos\beta\, G^+-\sin\beta\, H^+, \nonumber \\
\varphi_2^+&=\sin\beta\, G^++\cos\beta\, H^+.
\end{align}

\section*{Appendix B: Yukawa couplings for the 2HDM}
\renewcommand{\thesection}{B}
\label{app:Yukawa}
\setcounter{subsection}{0}
\setcounter{equation}{0}

For completeness, we summarize in this appendix the definition of Yukawa couplings in the general 2HDM employed for the analysis in this paper. Below, we also give a comparison with other notations.

\subsection{Our notation}

Assuming the SM fermion content (without right-handed neutrinos), couplings of the fermions to two scalar doublets ($a=1,2$) may be written in a completely general setting as
\begin{equation}
-\mathcal{L}_{\mathrm{Yukawa}}
=\overline{Q}_{L}\Phi_aF_a^DD_R+\overline{Q}_{L}\widetilde{\Phi}_aF_a^UU_R+\overline{L}_{L}\Phi_aF_a^LL_R+\mathrm{h.c.},
\label{eq:yukL}
\end{equation}
where $\tilde \Phi_a$ are defined above as charge conjugate doublets with hypercharge opposite to $\Phi_a$. 

The Lagrangian is written in the basis of weak eigenstates, i.e.~$Q_L$ and $L_L$ are $\mathrm{SU}(2)$ doublets, while $U_R$, $D_R$, and $L_R$ are singlets. The fermions are 3-component vectors in flavor space. Consequently, the Yukawa couplings $F_a^F$ are $3\times 3$ complex matrices. 

There are various ways fermions can couple to the Higgs doublets, leading to different Yukawa couplings.
Since an extended Higgs sector naturally leads to FCNC, these would have to be suppressed. This is
normally achieved by imposing discrete symmetries in modeling the
Yukawa interactions, as for example 
$Z_2$ symmetry under the transformation $\Phi_1 \to \Phi_1$, $\Phi_2
\to -\Phi_2$.
There are four such possible models with Natural Flavor Conservation (NFC) : all fermions couple only to {\it one} doublet (conventionally taken to be $\Phi_2$), or one fermion ($U$, $D$, $L$) couples to one doublet, the other two to the other doublet\footnote{Avoiding FCNC at tree level may not be sufficient, however. One should also investigate stability of these conditions under radiative corrections \cite{Buras:2010mh}.}.
Still other Yukawa models are being considered, where  all fermions couple to both doublets (Model III), leading to tree level FCNC processes. This issue is discussed in Appendix B.3 below.    

In the three-generation case  with discrete symmetry imposed on  the Yukawa Lagrangian, such that each right-handed fermionic state interacts with only one scalar doublet,   we have for fermion mass eigenstates
\begin{equation}
{\cal L}_\text{ch} =
\frac{g}{\sqrt2m_W}\left\{\left[V_\text{CKM}\overline{U}
\left(M^\text{diag}_D P_R{\cal F}^D+M^\text{diag}_U P_L{\cal F}^U\right)D
+ \overline{N}M^\text{diag}_L P_R{\cal F}^L L \right]H^+
+\text{h.c.}\right\},
\label{Eq:lagrangian}
\end{equation}
\noindent
where we used a notation like in Eq.~(\ref{eq:yukL}), with  $N$
referring to the neutrinos. Here, $P_L$ and $P_R$ are chirality projection operators.
The couplings ${\cal F^D,\,\, F^U}$ defining
models of Yukawa interactions are given in Table~\ref{tab:couplings}
for the notation  that is used in this paper.  Note the appearance of  the
$V_{CKM}$ matrix.

\begin{table}[htb]
\begin{center}
\begin{tabular}{|c | c c | c  c | c  c |} \hline \hline
Fermion   &\multicolumn{2}{c|}{$D$}& \multicolumn{2}{c|}{$U$} 
& \multicolumn{2}{c|}{$L$}   \\ \hline 
Model  & vev & ${\cal F}^D$ & vev & ${\cal F}^U$ & vev & ${\cal F}^L$   \\ \hline \hline
{\bf I} &  2 & $-\cot\beta$ & 2 & $+\cot\beta$  & 2 & $-\cot\beta$   \\
{\bf II} &  1 & $+\tan\beta$ & 2 & $+\cot\beta$  & 1 & $+\tan\beta$   \\
{\bf X}  &  2 & $-\cot\beta$ & 2 & $+\cot\beta$ & 1 & $+\tan\beta$  \\ 
{\bf Y} &  1 & $+\tan\beta$ & 2 & $+\cot\beta$  & 2 & $-\cot\beta$  \\ 
\hline
\end{tabular}
\caption{Relevant vacuum expectation values, for $\Phi_1$ or $\Phi_2$, denoted 1 and 2, and reduced Yukawa couplings ${\cal F}$, as defined by Eq.~(\ref{Eq:lagrangian}) for models without tree-level FCNC.}
\label{tab:couplings}
\end{center}
\vskip -0.2cm
\end{table}

We can write the charged-Higgs Lagrangian for one generation in the simplified form (neglecting elements of the CKM matrix):
\begin{equation} \label{Eq:cal-F}
{\cal L}_\text{ch}
=\frac{g}{\sqrt{2}\,m_W}
\bigl\{\bigl[\overline u(
m_dP_R{\cal F}^D
+m_uP_L{\cal F}^U)
d+ \overline{\nu}m_\ell P_R{\cal F}^L\ell\bigr]H^+ +\text{h.c.}
\bigr\}.
\end{equation}

For Model II we have
\begin{equation}
{\cal L}^{II}_\text{ch}
=\frac{g}{\sqrt{2}\,m_W}
\bar u\bigl[
m_dP_R\tan\beta
+m_uP_L\cot\beta
\bigr]dH^+ +\text{h.c.}
\end{equation}
(see Eq.~(\ref{Eq:Yukawa-charged-II})).

For Model I we have
\begin{equation}
{\cal L}^{I}_\text{ch}
=\frac{g\cot\beta}{\sqrt{2}\,m_W}
\bar u\bigl[
-m_dP_R
+m_uP_L
\bigr]dH^+ +\text{h.c.}
\end{equation}
In the limit that in the above equation the second term dominates (for example, for the third generation, with $m_t\gg m_b$) these couplings are the same as for Model~II, for moderate values of $\tan\beta$.

\begin{table}[htb]
\begin{center}
\begin{tabular}{| c | c | c | c | c | c | c | c | c |} \hline \hline
$\Phi_1$ & $\Phi_2$ & {\bf This work}   & HHG &BHP& G, AS  & ARS & AKTY & BFLRSS \\ \hline 
 & $u,d,\ell$ &
 {\bf I} & I & I  & I (*)  & --- & I & I   \\
 $d,\ell$ & $u$ &
{\bf II} & II & II & II  & --- & II & II\\
$u,d,\ell$ & $u,d,\ell$ &
{\bf III} & --- & --- & --- & III & --- & III \\
$\ell$ & $u,d$ &
{\bf X} & --- & IV  & I' (*) & --- & X & lepton specific \\
$d$ & $u,\ell$ &
{\bf Y} & --- & III & II'  & --- & Y & flipped\\
\hline
\end{tabular}
\caption{Dictionary of notations. 
``HHG'': Higgs Hunter's Guide \cite{Gunion:1989we}.
``BHP'': Barger, Hewett, Phillips \cite{Barger:1989fj}.
``G'': Grossman \cite{Grossman:1994jb}, ``AS'': Akeroyd, Stirling \cite{Akeroyd:1994ga}.
The (*) denotes interchange $\Phi_1\leftrightarrow\Phi_2$.
``ARS'': Atwood, Reina, Soni \cite{Atwood:1996vj}.
``AKTY'': Aoki, Kanemura, Tsumura, Yagyu \cite{Aoki:2009ha}.
``BFLRSS'': Branco, Ferreira, Lavoura, Rebelo, Sher, Silva \cite{Branco:2011iw}.}
\label{tab:dictionary}
\end{center}
\vskip -0.2cm
\end{table}

\subsection{Various notations}
The 1981 paper by Hall and Wise \cite{Hall:1981bc} may have been the
first to introduce ``Model I'' and ``Model II''. They were introduced
in analogy with the later convention of ``The Higgs Hunter's Guide''
(see below), but with the role of $\Phi_1$ and $\Phi_2$ interchanged.
An early paper distinguishing quarks and leptons in this respect, was that of Barnett, Senjanovic, Wolfenstein, and Wyler \cite{Barnett:1983mm}. They define models IA, IB, IIA, IIB. 

The definitions of ``Model I'' and ``Model II'' presented above coincide with those of the ``Higgs Hunter's Guide'' \cite{Gunion:1989we}.
Barger, Hewett and Phillips \cite{Barger:1989fj} defined additional models, where quarks and leptons couple differently.
Also Grossman \cite{Grossman:1994jb}, Akeroyd and Stirling \cite{Akeroyd:1994ga} discussed such models, under different names.
Aoki, Kanemura, Tsumura, and Yagyu \cite{Aoki:2009ha} introduced ``Model X'' and ``Model Y'' to avoid the ambiguity previously associated with ``Model III''. We have adopted the latter notation in this paper.

In table~\ref{tab:dictionary} we present a ``dictionary'' of notations for the five models.

\subsection{Minimal flavor violation}
In the most general version of the 2HDM, the fermionic couplings of the neutral scalars are non-diagonal in flavor, leading to FCNC at the tree level. 

In Refs.~\cite{Pich:2009sp,Tuzon:2010vt}, the authors propose the so-called {\it aligned} 2HDM
by fixing  the  matrices $F_a^F$ in Eq.~(\ref{eq:yukL}), for $a=1$ and $a=2$, to be pairwise proportional,
\begin{equation}
F^D_{1} \sim F^D_{2} \sim  {{ Y^D}},\quad
F^U_{1} \sim  F^U_{2} \sim  {{ Y^U}}.
\label{MFVfirstorder}
\end{equation}
Thus, there is no FCNC at the tree level. 

The aligned 2HDM is just  the most general minimally flavor-violating (MFV) \mbox{renormalizable} 
2HDM,  with the  lowest order in the  couplings ${ Y^F}$.

Following Ref.~\cite{D'Ambrosio:2002ex}, the most general MFV ansatz is  
given by the expansion
\begin{eqnarray}
F^D_{1} &=& Y^D, \nonumber\\
F^D_{2} &=& \epsilon_{0} Y^D + 
\epsilon_{1} Y^D  (Y^D)^\dagger Y^D                    
+  \epsilon_{2}  Y^U (Y^U)^\dagger Y^D + \ldots, \nonumber\\
F^U_{1} &=& \epsilon{'}_{0} Y^U + 
\epsilon{'}_{1}  Y^U (Y^U)^\dagger Y^U  +  \epsilon{'}_{2}  Y^D (Y^D)^\dagger Y^U + \ldots~, \nonumber\\
F^U_{2} &=& Y^U.  \label{MFVall} 
\end{eqnarray}
This simple form of $F^D_{1}$ and $F^U_{2}$ can be assumed
without loss of generality.
But even if the higher-order terms in $F^D_{2}$  and $F^U_{1}$ are not included at the tree-level,
they are  generated by radiative corrections. This is assured by the RG invariance of the MFV 
hypothesis which is implemented by the {\it flavor}  $SU(3)^3$ symmetry. 
Thus, the functional form of Eq.~(\ref{MFVall}) is preserved, only the coefficients 
$\epsilon_i$ and $\epsilon{'}_i$ change and become related via the RG equations.  
In view of this, it is also clear that setting all $\epsilon$ coefficients to zero leads to heavy fine-tuning. 
Thus, in general there is no Yukawa alignment  within the MFV framework. 

In Ref.~\cite{Buras:2010mh}, the stability of the various tree-level implementations is discussed. 
In the MFV case, the FCNC induced by higher-order terms 
are under control, since even when the coefficients in Eq.~(\ref{MFVall}) are of ${\cal O}(1)$ the 
expansion is rapidly convergent due to small CKM matrix elements and 
small quark masses \cite{D'Ambrosio:2002ex}.   

The higher-dimensional operators which are $Z_2$ invariant  may still induce new FCNC and further flavor protection is needed\cite{Buras:2010mh}, e.g. via the MFV hypothesis.
This problem already occurs  
in the case of  one Higgs doublet~\cite{Giudice:2008uua,Agashe:2009di,Azatov:2009na}.


\bibliographystyle{h-physrev5}
\bibliography{biblio}

\begin{thebibliography}{100}

\bibitem{Aad:2012tfa}
ATLAS, G.~Aad {\em et~al.},
\newblock Phys.Lett. {\bf B716}, 1 (2012), arXiv:1207.7214.

\bibitem{Chatrchyan:2012ufa}
CMS, S.~Chatrchyan {\em et~al.},
\newblock Phys.Lett. {\bf B716}, 30 (2012), arXiv:1207.7235.

\bibitem{Khachatryan:2014jba}
CMS, V.~Khachatryan {\em et~al.},
\newblock Eur. Phys. J. {\bf C75}, 212 (2015), arXiv:1412.8662.

\bibitem{Aad:2015ona}
ATLAS, G.~Aad {\em et~al.},
\newblock JHEP {\bf 08}, 137 (2015), arXiv:1506.06641.

\bibitem{Aad:2015zhl}
ATLAS, CMS, G.~Aad {\em et~al.},
\newblock Phys.Rev.Lett. {\bf 114}, 191803 (2015), arXiv:1503.07589.

\bibitem{Gunion:1989we}
J.~F. Gunion, H.~E. Haber, G.~L. Kane, and S.~Dawson,
\newblock {\em The Higgs Hunter's Guide} (Addison-Wesley Publishing Company,
  1990).

\bibitem{Branco:2011iw}
G.~Branco {\em et~al.},
\newblock Phys.Rept. {\bf 516}, 1 (2012), arXiv:1106.0034.

\bibitem{Ross:1975fq}
D.~A. Ross and M.~J.~G. Veltman,
\newblock Nucl. Phys. {\bf B95}, 135 (1975).

\bibitem{Veltman:1976rt}
M.~J.~G. Veltman,
\newblock Acta Phys. Polon. {\bf B8}, 475 (1977).

\bibitem{Veltman:1977kh}
M.~J.~G. Veltman,
\newblock Nucl. Phys. {\bf B123}, 89 (1977).

\bibitem{Glashow:1976nt}
S.~L. Glashow and S.~Weinberg,
\newblock Phys.Rev. {\bf D15}, 1958 (1977).

\bibitem{Paige:1977nz}
F.~E. Paige, E.~A. Paschos, and T.~Trueman,
\newblock Phys.Rev. {\bf D15}, 3416 (1977).

\bibitem{Misiak:2015xwa}
M.~Misiak {\em et~al.},
\newblock Phys.Rev.Lett. {\bf 114}, 221801 (2015), arXiv:1503.01789.

\bibitem{Misiak:2017bgg}
M.~Misiak and M.~Steinhauser,
\newblock Eur. Phys. J. {\bf C77}, 201 (2017), arXiv:1702.04571.

\bibitem{Akeroyd:1994ga}
A.~Akeroyd and W.~Stirling,
\newblock Nucl.Phys. {\bf B447}, 3 (1995).

\bibitem{Logan:2009uf}
H.~E. Logan and D.~MacLennan,
\newblock Phys.Rev. {\bf D79}, 115022 (2009), arXiv:0903.2246.

\bibitem{Abbiendi:2013hk}
ALEPH, DELPHI, L3, OPAL, LEP, G.~Abbiendi {\em et~al.},
\newblock Eur.Phys.J. {\bf C73}, 2463 (2013), arXiv:1301.6065.

\bibitem{Lee:1973iz}
T.~D. Lee,
\newblock Phys. Rev. {\bf D8}, 1226 (1973).

\bibitem{Riotto:1999yt}
A.~Riotto and M.~Trodden,
\newblock Ann. Rev. Nucl. Part. Sci. {\bf 49}, 35 (1999), arXiv:hep-ph/9901362.

\bibitem{Accomando:2006ga}
E.~Accomando {\em et~al.},
\newblock (2006), arXiv:hep-ph/0608079,
\newblock Report of the Workshop on CP Studies and Non-standard Higgs Physics,
  CERN, Geneva, Switzerland, May 2004 - Dec 2005.

\bibitem{ElKaffas:2007rq}
A.~W. El~Kaffas, P.~Osland, and O.~M. Ogreid,
\newblock Nonlin. Phenom. Complex Syst. {\bf 10}, 347 (2007),
  arXiv:hep-ph/0702097.

\bibitem{ElKaffas:2006nt}
A.~W. El~Kaffas, W.~Khater, O.~M. Ogreid, and P.~Osland,
\newblock Nucl.Phys. {\bf B775}, 45 (2007), arXiv:hep-ph/0605142.

\bibitem{Ginzburg:2014pra}
I.~F. Ginzburg,
\newblock JETP Lett. {\bf 99}, 742 (2014), arXiv:1410.0873.

\bibitem{Nie:1998yn}
S.~Nie and M.~Sher,
\newblock Phys.Lett. {\bf B449}, 89 (1999), arXiv:hep-ph/9811234.

\bibitem{Ferreira:2004yd}
P.~Ferreira, R.~Santos, and A.~Barroso,
\newblock Phys.Lett. {\bf B603}, 219 (2004), arXiv:hep-ph/0406231.

\bibitem{Goudelis:2013uca}
A.~Goudelis, B.~Herrmann, and O.~St{\aa}l,
\newblock JHEP {\bf 1309}, 106 (2013), arXiv:1303.3010.

\bibitem{Swiezewska:2015paa}
B.~Swiezewska,
\newblock JHEP {\bf 07}, 118 (2015), arXiv:1503.07078.

\bibitem{Khan:2015ipa}
N.~Khan and S.~Rakshit,
\newblock Phys. Rev. {\bf D92}, 055006 (2015), arXiv:1503.03085.

\bibitem{Deshpande:1977rw}
N.~G. Deshpande and E.~Ma,
\newblock Phys.Rev. {\bf D18}, 2574 (1978).

\bibitem{Kanemura:1999xf}
S.~Kanemura, T.~Kasai, and Y.~Okada,
\newblock Phys.Lett. {\bf B471}, 182 (1999), arXiv:hep-ph/9903289.

\bibitem{Barroso:2013awa}
A.~Barroso, P.~Ferreira, I.~Ivanov, and R.~Santos,
\newblock JHEP {\bf 1306}, 045 (2013), arXiv:1303.5098.

\bibitem{Ginzburg:2010wa}
I.~Ginzburg, K.~Kanishev, M.~Krawczyk, and D.~Sokolowska,
\newblock Phys.Rev. {\bf D82}, 123533 (2010), arXiv:1009.4593.

\bibitem{Swiezewska:2012ej}
B.~Swiezewska,
\newblock Phys.Rev. {\bf D88}, 055027 (2013), arXiv:1209.5725.

\bibitem{Kanemura:1993hm}
S.~Kanemura, T.~Kubota, and E.~Takasugi,
\newblock Phys. Lett. {\bf B313}, 155 (1993), arXiv:hep-ph/9303263.

\bibitem{Akeroyd:2000wc}
A.~G. Akeroyd, A.~Arhrib, and E.-M. Naimi,
\newblock Phys.Lett. {\bf B490}, 119 (2000), arXiv:hep-ph/0006035.

\bibitem{Arhrib:2000is}
A.~Arhrib,
\newblock (2000), arXiv:hep-ph/0012353,
\newblock based on hep-ph/0006035.

\bibitem{Ginzburg:2003fe}
I.~Ginzburg and I.~Ivanov,
\newblock (2003), arXiv:hep-ph/0312374.

\bibitem{Ginzburg:2005dt}
I.~Ginzburg and I.~Ivanov,
\newblock Phys.Rev. {\bf D72}, 115010 (2005), arXiv:hep-ph/0508020.

\bibitem{WahabElKaffas:2007xd}
A.~Wahab El~Kaffas, P.~Osland, and O.~M. Ogreid,
\newblock Phys.Rev. {\bf D76}, 095001 (2007), arXiv:0706.2997.

\bibitem{Gorczyca:2011he}
B.~Gorczyca and M.~Krawczyk,
\newblock (2011), arXiv:1112.5086.

\bibitem{Grinstein:2015rtl}
B.~Grinstein, C.~W. Murphy, and P.~Uttayarat,
\newblock JHEP {\bf 06}, 070 (2016), arXiv:1512.04567.

\bibitem{Kennedy:1988sn}
D.~C. Kennedy and B.~W. Lynn,
\newblock Nucl. Phys. {\bf B322}, 1 (1989).

\bibitem{Peskin:1990zt}
M.~E. Peskin and T.~Takeuchi,
\newblock Phys. Rev. Lett. {\bf 65}, 964 (1990).

\bibitem{Altarelli:1990zd}
G.~Altarelli and R.~Barbieri,
\newblock Phys. Lett. {\bf B253}, 161 (1991).

\bibitem{Peskin:1991sw}
M.~E. Peskin and T.~Takeuchi,
\newblock Phys. Rev. {\bf D46}, 381 (1992).

\bibitem{Altarelli:1991fk}
G.~Altarelli, R.~Barbieri, and S.~Jadach,
\newblock Nucl. Phys. {\bf B369}, 3 (1992),
\newblock [Erratum: Nucl. Phys.B376,444(1992)].

\bibitem{Grimus:2007if}
W.~Grimus, L.~Lavoura, O.~Ogreid, and P.~Osland,
\newblock J.Phys.G {\bf G35}, 075001 (2008), arXiv:0711.4022.

\bibitem{Grimus:2008nb}
W.~Grimus, L.~Lavoura, O.~Ogreid, and P.~Osland,
\newblock Nucl.Phys. {\bf B801}, 81 (2008), arXiv:0802.4353.

\bibitem{Jung:2010ik}
M.~Jung, A.~Pich, and P.~Tuzon,
\newblock JHEP {\bf 1011}, 003 (2010), arXiv:1006.0470.

\bibitem{Gunion:1986nh}
J.~F. Gunion and H.~E. Haber,
\newblock Nucl. Phys. {\bf B278}, 449 (1986),
\newblock [Erratum: Nucl. Phys.B402,569(1993)].

\bibitem{Moretti:1994ds}
S.~Moretti and W.~Stirling,
\newblock Phys.Lett. {\bf B347}, 291 (1995), arXiv:hep-ph/9412209.

\bibitem{Djouadi:1995gv}
A.~Djouadi, J.~Kalinowski, and P.~Zerwas,
\newblock Z.Phys. {\bf C70}, 435 (1996), arXiv:hep-ph/9511342.

\bibitem{Djouadi:1997yw}
A.~Djouadi, J.~Kalinowski, and M.~Spira,
\newblock Comput.Phys.Commun. {\bf 108}, 56 (1998), arXiv:hep-ph/9704448.

\bibitem{Kanemura:2009mk}
S.~Kanemura, S.~Moretti, Y.~Mukai, R.~Santos, and K.~Yagyu,
\newblock Phys.Rev. {\bf D79}, 055017 (2009), arXiv:0901.0204.

\bibitem{Eriksson:2009ws}
D.~Eriksson, J.~Rathsman, and O.~St{\aa}l,
\newblock Comput.Phys.Commun. {\bf 181}, 189 (2010), arXiv:0902.0851.

\bibitem{Harlander:2013qxa}
R.~Harlander, M.~M{\"u}hlleitner, J.~Rathsman, M.~Spira, and O.~St{\aa}l,
\newblock (2013), arXiv:1312.5571.

\bibitem{Mendez:1990jr}
A.~Mendez and A.~Pomarol,
\newblock Phys. Lett. {\bf B252}, 461 (1990).

\bibitem{Li:1990ag}
C.-S. Li and R.~J. Oakes,
\newblock Phys. Rev. {\bf D43}, 855 (1991).

\bibitem{Djouadi:1994gf}
A.~Djouadi and P.~Gambino,
\newblock Phys. Rev. {\bf D51}, 218 (1995), arXiv:hep-ph/9406431,
\newblock [Erratum: Phys. Rev.D53,4111(1996)].

\bibitem{Fontes:2015xva}
D.~Fontes, J.~C. Romão, R.~Santos, and J.~P. Silva,
\newblock Phys. Rev. {\bf D92}, 055014 (2015), arXiv:1506.06755.

\bibitem{Keus:2015hva}
V.~Keus, S.~F. King, S.~Moretti, and K.~Yagyu,
\newblock JHEP {\bf 04}, 048 (2016), arXiv:1510.04028.

\bibitem{Basso:2015dka}
L.~Basso, P.~Osland, and G.~M. Pruna,
\newblock JHEP {\bf 06}, 083 (2015), arXiv:1504.07552.

\bibitem{Dicus:1989vf}
D.~Dicus, J.~Hewett, C.~Kao, and T.~Rizzo,
\newblock Phys.Rev. {\bf D40}, 787 (1989).

\bibitem{BarrientosBendezu:1998gd}
A.~Barrientos~Bendezu and B.~A. Kniehl,
\newblock Phys.Rev. {\bf D59}, 015009 (1999), arXiv:hep-ph/9807480.

\bibitem{Moretti:1998xq}
S.~Moretti and K.~Odagiri,
\newblock Phys.Rev. {\bf D59}, 055008 (1999), arXiv:hep-ph/9809244.

\bibitem{BarrientosBendezu:1999vd}
A.~Barrientos~Bendezu and B.~A. Kniehl,
\newblock Phys.Rev. {\bf D61}, 097701 (2000), arXiv:hep-ph/9909502.

\bibitem{Brein:2000cv}
O.~Brein, W.~Hollik, and S.~Kanemura,
\newblock Phys.Rev. {\bf D63}, 095001 (2001), arXiv:hep-ph/0008308.

\bibitem{Hollik:2001hy}
W.~Hollik and S.-h. Zhu,
\newblock Phys.Rev. {\bf D65}, 075015 (2002), arXiv:hep-ph/0109103.

\bibitem{Asakawa:2005nx}
E.~Asakawa, O.~Brein, and S.~Kanemura,
\newblock Phys.Rev. {\bf D72}, 055017 (2005), arXiv:hep-ph/0506249.

\bibitem{Eriksson:2006yt}
D.~Eriksson, S.~Hesselbach, and J.~Rathsman,
\newblock Eur.Phys.J. {\bf C53}, 267 (2008), arXiv:hep-ph/0612198.

\bibitem{Hashemi:2010ce}
M.~Hashemi,
\newblock Phys.Rev. {\bf D83}, 055004 (2011), arXiv:1008.3785.

\bibitem{Gunion:1986pe}
J.~Gunion, H.~Haber, F.~Paige, W.-K. Tung, and S.~Willenbrock,
\newblock Nucl.Phys. {\bf B294}, 621 (1987).

\bibitem{DiazCruz:1992gg}
J.~Diaz-Cruz and O.~Sampayo,
\newblock Phys.Rev. {\bf D50}, 6820 (1994).

\bibitem{Moretti:1996ra}
S.~Moretti and K.~Odagiri,
\newblock Phys.Rev. {\bf D55}, 5627 (1997), arXiv:hep-ph/9611374.

\bibitem{Miller:1999bm}
D.~Miller, S.~Moretti, D.~Roy, and W.~Stirling,
\newblock Phys.Rev. {\bf D61}, 055011 (2000), arXiv:hep-ph/9906230.

\bibitem{Moretti:1999bw}
S.~Moretti and D.~Roy,
\newblock Phys.Lett. {\bf B470}, 209 (1999), arXiv:hep-ph/9909435.

\bibitem{Zhu:2001nt}
S.-h. Zhu,
\newblock Phys.Rev. {\bf D67}, 075006 (2003), arXiv:hep-ph/0112109.

\bibitem{Plehn:2002vy}
T.~Plehn,
\newblock Phys.Rev. {\bf D67}, 014018 (2003), arXiv:hep-ph/0206121.

\bibitem{Berger:2003sm}
E.~L. Berger, T.~Han, J.~Jiang, and T.~Plehn,
\newblock Phys. Rev. {\bf D71}, 115012 (2005), arXiv:hep-ph/0312286.

\bibitem{Kidonakis:2004ib}
N.~Kidonakis,
\newblock JHEP {\bf 05}, 011 (2005), arXiv:hep-ph/0412422.

\bibitem{Weydert:2009vr}
C.~Weydert {\em et~al.},
\newblock Eur. Phys. J. {\bf C67}, 617 (2010), arXiv:0912.3430.

\bibitem{Kidonakis:2010ux}
N.~Kidonakis,
\newblock Phys. Rev. {\bf D82}, 054018 (2010), arXiv:1005.4451.

\bibitem{Flechl:2014wfa}
M.~Flechl, R.~Klees, M.~Kramer, M.~Spira, and M.~Ubiali,
\newblock Phys. Rev. {\bf D91}, 075015 (2015), arXiv:1409.5615.

\bibitem{Degrande:2015vpa}
C.~Degrande, M.~Ubiali, M.~Wiesemann, and M.~Zaro,
\newblock JHEP {\bf 10}, 145 (2015), arXiv:1507.02549.

\bibitem{Kidonakis:2016eeu}
N.~Kidonakis,
\newblock Phys. Rev. {\bf D94}, 014010 (2016), arXiv:1605.00622.

\bibitem{Degrande:2016hyf}
C.~Degrande {\em et~al.},
\newblock (2016), arXiv:1607.05291.

\bibitem{Borzumati:1999th}
F.~Borzumati, J.-L. Kneur, and N.~Polonsky,
\newblock Phys. Rev. {\bf D60}, 115011 (1999), arXiv:hep-ph/9905443.

\bibitem{Alwall:2004xw}
J.~Alwall and J.~Rathsman,
\newblock JHEP {\bf 0412}, 050 (2004), arXiv:hep-ph/0409094.

\bibitem{He:1998ie}
H.-J. He and C.~Yuan,
\newblock Phys.Rev.Lett. {\bf 83}, 28 (1999), arXiv:hep-ph/9810367.

\bibitem{DiazCruz:2001gf}
J.~Diaz-Cruz, H.-J. He, and C.~Yuan,
\newblock Phys.Lett. {\bf B530}, 179 (2002), arXiv:hep-ph/0103178.

\bibitem{Slabospitsky:2002gw}
S.~Slabospitsky,
\newblock (2002), arXiv:hep-ph/0203094.

\bibitem{Dittmaier:2007uw}
S.~Dittmaier, G.~Hiller, T.~Plehn, and M.~Spannowsky,
\newblock Phys.Rev. {\bf D77}, 115001 (2008), arXiv:0708.0940.

\bibitem{Harlander:2011aa}
R.~Harlander, M.~Kramer, and M.~Schumacher,
\newblock (2011), arXiv:1112.3478.

\bibitem{Barnett:1987jw}
R.~M. Barnett, H.~E. Haber, and D.~E. Soper,
\newblock Nucl. Phys. {\bf B306}, 697 (1988).

\bibitem{Olness:1987ep}
F.~I. Olness and W.-K. Tung,
\newblock Nucl. Phys. {\bf B308}, 813 (1988).

\bibitem{deFlorian:2016spz}
LHC Higgs Cross Section Working Group, D.~de~Florian {\em et~al.},
\newblock (2016), arXiv:1610.07922.

\bibitem{Belyaev:2001qm}
A.~Belyaev, D.~Garcia, J.~Guasch, and J.~Sola,
\newblock Phys. Rev. {\bf D65}, 031701 (2002), arXiv:hep-ph/0105053.

\bibitem{Belyaev:2002eq}
A.~Belyaev, D.~Garcia, J.~Guasch, and J.~Sola,
\newblock JHEP {\bf 06}, 059 (2002), arXiv:hep-ph/0203031.

\bibitem{Guchait:2001pi}
M.~Guchait and S.~Moretti,
\newblock JHEP {\bf 0201}, 001 (2002), arXiv:hep-ph/0110020.

\bibitem{Alwall:2003tc}
J.~Alwall, C.~Biscarat, S.~Moretti, J.~Rathsman, and A.~Sopczak,
\newblock Eur.Phys.J. {\bf C39S1}, 37 (2005), arXiv:hep-ph/0312301.

\bibitem{Assamagan:2004gv}
K.~Assamagan, M.~Guchait, and S.~Moretti,
\newblock (2004), arXiv:hep-ph/0402057.

\bibitem{Kanemura:2001hz}
S.~Kanemura and C.~Yuan,
\newblock Phys.Lett. {\bf B530}, 188 (2002), arXiv:hep-ph/0112165.

\bibitem{Akeroyd:2003bt}
A.~G. Akeroyd and M.~A. Diaz,
\newblock Phys.Rev. {\bf D67}, 095007 (2003), arXiv:hep-ph/0301203.

\bibitem{Akeroyd:2003jp}
A.~Akeroyd,
\newblock Phys.Rev. {\bf D68}, 077701 (2003), arXiv:hep-ph/0306045.

\bibitem{Cao:2003tr}
Q.-H. Cao, S.~Kanemura, and C.~Yuan,
\newblock Phys.Rev. {\bf D69}, 075008 (2004), arXiv:hep-ph/0311083.

\bibitem{Belyaev:2006rf}
A.~Belyaev, Q.-H. Cao, D.~Nomura, K.~Tobe, and C.-P. Yuan,
\newblock Phys.Rev.Lett. {\bf 100}, 061801 (2008), arXiv:hep-ph/0609079.

\bibitem{Miao:2010rg}
X.~Miao, S.~Su, and B.~Thomas,
\newblock Phys.Rev. {\bf D82}, 035009 (2010), arXiv:1005.0090.

\bibitem{Eichten:1984eu}
E.~Eichten, I.~Hinchliffe, K.~D. Lane, and C.~Quigg,
\newblock Rev.Mod.Phys. {\bf 56}, 579 (1984).

\bibitem{Willenbrock:1986ry}
S.~S. Willenbrock,
\newblock Phys.Rev. {\bf D35}, 173 (1987).

\bibitem{Glover:1987nx}
E.~Glover and J.~van~der Bij,
\newblock Nucl.Phys. {\bf B309}, 282 (1988).

\bibitem{Dicus:1987ic}
D.~A. Dicus, C.~Kao, and S.~S. Willenbrock,
\newblock Phys.Lett. {\bf B203}, 457 (1988).

\bibitem{Jiang:1997cg}
Y.~Jiang, L.~Han, W.-G. Ma, Z.-H. Yu, and M.~Han,
\newblock J.Phys.G {\bf G23}, 385 (1997), arXiv:hep-ph/9703275.

\bibitem{Krause:1997rc}
A.~Krause, T.~Plehn, M.~Spira, and P.~Zerwas,
\newblock Nucl.Phys. {\bf B519}, 85 (1998), arXiv:hep-ph/9707430.

\bibitem{BarrientosBendezu:1999gp}
A.~Barrientos~Bendezu and B.~A. Kniehl,
\newblock Nucl.Phys. {\bf B568}, 305 (2000), arXiv:hep-ph/9908385.

\bibitem{Brein:1999sy}
O.~Brein and W.~Hollik,
\newblock Eur.Phys.J. {\bf C13}, 175 (2000), arXiv:hep-ph/9908529.

\bibitem{Moretti:2001pp}
S.~Moretti,
\newblock J.Phys.G {\bf G28}, 2567 (2002), arXiv:hep-ph/0102116.

\bibitem{Moretti:2003px}
S.~Moretti and J.~Rathsman,
\newblock Eur.Phys.J. {\bf C33}, 41 (2004), arXiv:hep-ph/0308215.

\bibitem{Alves:2005kr}
A.~Alves and T.~Plehn,
\newblock Phys.Rev. {\bf D71}, 115014 (2005), arXiv:hep-ph/0503135.

\bibitem{Pumplin:2002vw}
J.~Pumplin {\em et~al.},
\newblock JHEP {\bf 0207}, 012 (2002), arXiv:hep-ph/0201195.

\bibitem{Spira:1995rr}
M.~Spira, A.~Djouadi, D.~Graudenz, and P.~M. Zerwas,
\newblock Nucl. Phys. {\bf B453}, 17 (1995), arXiv:hep-ph/9504378.

\bibitem{Basso:2012st}
L.~Basso {\em et~al.},
\newblock JHEP {\bf 1211}, 011 (2012), arXiv:1205.6569.

\bibitem{Basso:2013wna}
L.~Basso {\em et~al.},
\newblock PoS {\bf Corfu2012}, 029 (2013), arXiv:1305.3219.

\bibitem{Beneke:2011mq}
M.~Beneke, P.~Falgari, S.~Klein, and C.~Schwinn,
\newblock Nucl. Phys. {\bf B855}, 695 (2012), arXiv:1109.1536.

\bibitem{Cacciari:2011hy}
M.~Cacciari, M.~Czakon, M.~Mangano, A.~Mitov, and P.~Nason,
\newblock Phys. Lett. {\bf B710}, 612 (2012), arXiv:1111.5869.

\bibitem{Czakon:2011xx}
M.~Czakon and A.~Mitov,
\newblock Comput. Phys. Commun. {\bf 185}, 2930 (2014), arXiv:1112.5675.

\bibitem{Baernreuther:2012ws}
P.~Bärnreuther, M.~Czakon, and A.~Mitov,
\newblock Phys. Rev. Lett. {\bf 109}, 132001 (2012), arXiv:1204.5201.

\bibitem{Czakon:2012zr}
M.~Czakon and A.~Mitov,
\newblock JHEP {\bf 12}, 054 (2012), arXiv:1207.0236.

\bibitem{Czakon:2012pz}
M.~Czakon and A.~Mitov,
\newblock JHEP {\bf 01}, 080 (2013), arXiv:1210.6832.

\bibitem{Czakon:2013goa}
M.~Czakon, P.~Fiedler, and A.~Mitov,
\newblock Phys. Rev. Lett. {\bf 110}, 252004 (2013), arXiv:1303.6254.

\bibitem{Czarnecki:1998qc}
A.~Czarnecki and K.~Melnikov,
\newblock Nucl. Phys. {\bf B544}, 520 (1999), arXiv:hep-ph/9806244.

\bibitem{Chetyrkin:1999ju}
K.~G. Chetyrkin, R.~Harlander, T.~Seidensticker, and M.~Steinhauser,
\newblock Phys. Rev. {\bf D60}, 114015 (1999), arXiv:hep-ph/9906273.

\bibitem{Blokland:2004ye}
I.~R. Blokland, A.~Czarnecki, M.~Slusarczyk, and F.~Tkachov,
\newblock Phys. Rev. Lett. {\bf 93}, 062001 (2004), arXiv:hep-ph/0403221.

\bibitem{Blokland:2005vq}
I.~R. Blokland, A.~Czarnecki, M.~Slusarczyk, and F.~Tkachov,
\newblock Phys. Rev. {\bf D71}, 054004 (2005), arXiv:hep-ph/0503039,
\newblock [Erratum: Phys. Rev.D79,019901(2009)].

\bibitem{Czarnecki:2010gb}
A.~Czarnecki, J.~G. Korner, and J.~H. Piclum,
\newblock Phys. Rev. {\bf D81}, 111503 (2010), arXiv:1005.2625.

\bibitem{Gao:2012ja}
J.~Gao, C.~S. Li, and H.~X. Zhu,
\newblock Phys. Rev. Lett. {\bf 110}, 042001 (2013), arXiv:1210.2808.

\bibitem{Brucherseifer:2013iv}
M.~Brucherseifer, F.~Caola, and K.~Melnikov,
\newblock JHEP {\bf 04}, 059 (2013), arXiv:1301.7133.

\bibitem{Czarnecki:1992zm}
A.~Czarnecki and S.~Davidson,
\newblock Phys. Rev. {\bf D48}, 4183 (1993), arXiv:hep-ph/9301237.

\bibitem{Mahmoudi:2009zx}
F.~Mahmoudi and O.~St{\aa}l,
\newblock Phys.Rev. {\bf D81}, 035016 (2010), arXiv:0907.1791.

\bibitem{Crivellin:2013wna}
A.~Crivellin, A.~Kokulu, and C.~Greub,
\newblock Phys. Rev. {\bf D87}, 094031 (2013), arXiv:1303.5877.

\bibitem{Krawczyk:1987zj}
P.~Krawczyk and S.~Pokorski,
\newblock Phys.Rev.Lett. {\bf 60}, 182 (1988).

\bibitem{Cline:2015lqp}
J.~M. Cline,
\newblock Phys. Rev. {\bf D93}, 075017 (2016), arXiv:1512.02210.

\bibitem{Crivellin:2015hha}
A.~Crivellin, J.~Heeck, and P.~Stoffer,
\newblock Phys. Rev. Lett. {\bf 116}, 081801 (2016), arXiv:1507.07567.

\bibitem{Abbiendi:2001fi}
OPAL Collaboration, G.~Abbiendi {\em et~al.},
\newblock Phys.Lett. {\bf B520}, 1 (2001), arXiv:hep-ex/0108031.

\bibitem{Krawczyk:2004na}
M.~Krawczyk and D.~Temes,
\newblock Eur.Phys.J. {\bf C44}, 435 (2005), arXiv:hep-ph/0410248.

\bibitem{Agashe:2014kda}
Particle Data Group, K.~A. Olive {\em et~al.},
\newblock Chin. Phys. {\bf C38}, 090001 (2014).

\bibitem{Amhis:2014hma}
Heavy Flavor Averaging Group (HFAG), Y.~Amhis {\em et~al.},
\newblock (2014), arXiv:1412.7515.

\bibitem{Charles:2004jd}
CKMfitter Group, J.~Charles {\em et~al.},
\newblock Eur. Phys. J. {\bf C41}, 1 (2005), arXiv:hep-ph/0406184.

\bibitem{Hou:1992sy}
W.-S. Hou,
\newblock Phys.Rev. {\bf D48}, 2342 (1993).

\bibitem{Grossman:1994ax}
Y.~Grossman and Z.~Ligeti,
\newblock Phys.Lett. {\bf B332}, 373 (1994), arXiv:hep-ph/9403376.

\bibitem{Grossman:1995yp}
Y.~Grossman, H.~E. Haber, and Y.~Nir,
\newblock Phys.Lett. {\bf B357}, 630 (1995), arXiv:hep-ph/9507213.

\bibitem{Aubert:2007dsa}
BABAR Collaboration, B.~Aubert {\em et~al.},
\newblock Phys.Rev.Lett. {\bf 100}, 021801 (2008), arXiv:0709.1698.

\bibitem{Nierste:2008qe}
U.~Nierste, S.~Trine, and S.~Westhoff,
\newblock Phys.Rev. {\bf D78}, 015006 (2008), arXiv:0801.4938.

\bibitem{Lees:2012xj}
BaBar, J.~P. Lees {\em et~al.},
\newblock Phys. Rev. Lett. {\bf 109}, 101802 (2012), arXiv:1205.5442.

\bibitem{Lees:2013uzd}
BaBar, J.~Lees {\em et~al.},
\newblock Phys.Rev. {\bf D88}, 072012 (2013), arXiv:1303.0571.

\bibitem{Huschle:2015rga}
Belle, M.~Huschle {\em et~al.},
\newblock Phys. Rev. {\bf D92}, 072014 (2015), arXiv:1507.03233.

\bibitem{Abdesselam:2016cgx}
Belle, A.~Abdesselam {\em et~al.},
\newblock (2016), arXiv:1603.06711.

\bibitem{Aaij:2015yra}
LHCb, R.~Aaij {\em et~al.},
\newblock Phys. Rev. Lett. {\bf 115}, 111803 (2015), arXiv:1506.08614,
\newblock [Addendum: Phys. Rev. Lett.115,no.15,159901(2015)].

\bibitem{Freytsis:2015qca}
M.~Freytsis, Z.~Ligeti, and J.~T. Ruderman,
\newblock Phys. Rev. {\bf D92}, 054018 (2015), arXiv:1506.08896.

\bibitem{Fajfer:2012vx}
S.~Fajfer, J.~F. Kamenik, and I.~Nisandzic,
\newblock Phys. Rev. {\bf D85}, 094025 (2012), arXiv:1203.2654.

\bibitem{Lattice:2015rga}
MILC s, J.~A. Bailey {\em et~al.},
\newblock Phys. Rev. {\bf D92}, 034506 (2015), arXiv:1503.07237.

\bibitem{Na:2015kha}
HPQCD, H.~Na, C.~M. Bouchard, G.~P. Lepage, C.~Monahan, and J.~Shigemitsu,
\newblock Phys. Rev. {\bf D92}, 054510 (2015), arXiv:1505.03925.

\bibitem{Akeroyd:2009tn}
A.~Akeroyd and F.~Mahmoudi,
\newblock JHEP {\bf 0904}, 121 (2009), arXiv:0902.2393.

\bibitem{Chetyrkin:1996vx}
K.~G. Chetyrkin, M.~Misiak, and M.~Munz,
\newblock Phys.Lett. {\bf B400}, 206 (1997), arXiv:hep-ph/9612313.

\bibitem{Buras:1997bk}
A.~J. Buras, A.~Kwiatkowski, and N.~Pott,
\newblock Phys.Lett. {\bf B414}, 157 (1997), arXiv:hep-ph/9707482.

\bibitem{Bauer:1997fe}
C.~W. Bauer,
\newblock Phys.Rev. {\bf D57}, 5611 (1998), arXiv:hep-ph/9710513.

\bibitem{Bobeth:1999mk}
C.~Bobeth, M.~Misiak, and J.~Urban,
\newblock Nucl.Phys. {\bf B574}, 291 (2000), arXiv:hep-ph/9910220.

\bibitem{Buras:2002tp}
A.~J. Buras, A.~Czarnecki, M.~Misiak, and J.~Urban,
\newblock Nucl.Phys. {\bf B631}, 219 (2002), arXiv:hep-ph/0203135.

\bibitem{Misiak:2004ew}
M.~Misiak and M.~Steinhauser,
\newblock Nucl.Phys. {\bf B683}, 277 (2004), arXiv:hep-ph/0401041.

\bibitem{Neubert:2004dd}
M.~Neubert,
\newblock Eur.Phys.J. {\bf C40}, 165 (2005), arXiv:hep-ph/0408179.

\bibitem{Melnikov:2005bx}
K.~Melnikov and A.~Mitov,
\newblock Phys.Lett. {\bf B620}, 69 (2005), arXiv:hep-ph/0505097.

\bibitem{Misiak:2006zs}
M.~Misiak {\em et~al.},
\newblock Phys.Rev.Lett. {\bf 98}, 022002 (2007), arXiv:hep-ph/0609232.

\bibitem{Misiak:2006ab}
M.~Misiak and M.~Steinhauser,
\newblock Nucl.Phys. {\bf B764}, 62 (2007), arXiv:hep-ph/0609241.

\bibitem{Asatrian:2006rq}
H.~M. Asatrian, T.~Ewerth, H.~Gabrielyan, and C.~Greub,
\newblock Phys. Lett. {\bf B647}, 173 (2007), arXiv:hep-ph/0611123.

\bibitem{Czakon:2006ss}
M.~Czakon, U.~Haisch, and M.~Misiak,
\newblock JHEP {\bf 03}, 008 (2007), arXiv:hep-ph/0612329.

\bibitem{Boughezal:2007ny}
R.~Boughezal, M.~Czakon, and T.~Schutzmeier,
\newblock JHEP {\bf 09}, 072 (2007), arXiv:0707.3090.

\bibitem{Ewerth:2008nv}
T.~Ewerth,
\newblock Phys. Lett. {\bf B669}, 167 (2008), arXiv:0805.3911.

\bibitem{Misiak:2010sk}
M.~Misiak and M.~Steinhauser,
\newblock Nucl. Phys. {\bf B840}, 271 (2010), arXiv:1005.1173.

\bibitem{Asatrian:2010rq}
H.~M. Asatrian, T.~Ewerth, A.~Ferroglia, C.~Greub, and G.~Ossola,
\newblock Phys. Rev. {\bf D82}, 074006 (2010), arXiv:1005.5587.

\bibitem{Ferroglia:2010xe}
A.~Ferroglia and U.~Haisch,
\newblock Phys. Rev. {\bf D82}, 094012 (2010), arXiv:1009.2144.

\bibitem{Misiak:2010tk}
M.~Misiak and M.~Poradzinski,
\newblock Phys. Rev. {\bf D83}, 014024 (2011), arXiv:1009.5685.

\bibitem{Kaminski:2012eb}
M.~Kaminski, M.~Misiak, and M.~Poradzinski,
\newblock Phys. Rev. {\bf D86}, 094004 (2012), arXiv:1209.0965.

\bibitem{Czakon:2015exa}
M.~Czakon {\em et~al.},
\newblock JHEP {\bf 04}, 168 (2015), arXiv:1503.01791.

\bibitem{Ciafaloni:1997un}
P.~Ciafaloni, A.~Romanino, and A.~Strumia,
\newblock Nucl. Phys. {\bf B524}, 361 (1998), arXiv:hep-ph/9710312.

\bibitem{Ciuchini:1997xe}
M.~Ciuchini, G.~Degrassi, P.~Gambino, and G.~F. Giudice,
\newblock Nucl. Phys. {\bf B527}, 21 (1998), arXiv:hep-ph/9710335.

\bibitem{Borzumati:1998tg}
F.~Borzumati and C.~Greub,
\newblock Phys.Rev. {\bf D58}, 074004 (1998), arXiv:hep-ph/9802391.

\bibitem{Bobeth:1999ww}
C.~Bobeth, M.~Misiak, and J.~Urban,
\newblock Nucl. Phys. {\bf B567}, 153 (2000), arXiv:hep-ph/9904413.

\bibitem{Gambino:2001ew}
P.~Gambino and M.~Misiak,
\newblock Nucl.Phys. {\bf B611}, 338 (2001), arXiv:hep-ph/0104034.

\bibitem{Trabelsi:2015}
K.~Trabelsi,
\newblock (2015), Talk given at EPS conference.

\bibitem{Saito:2014das}
Belle, T.~Saito {\em et~al.},
\newblock Phys. Rev. {\bf D91}, 052004 (2015), arXiv:1411.7198.

\bibitem{Grossman:1994jb}
Y.~Grossman,
\newblock Nucl.Phys. {\bf B426}, 355 (1994), arXiv:hep-ph/9401311.

\bibitem{Abbott:1979dt}
L.~Abbott, P.~Sikivie, and M.~B. Wise,
\newblock Phys.Rev. {\bf D21}, 1393 (1980).

\bibitem{Inami:1980fz}
T.~Inami and C.~Lim,
\newblock Prog.Theor.Phys. {\bf 65}, 297 (1981).

\bibitem{Athanasiu:1985ie}
G.~G. Athanasiu, P.~J. Franzini, and F.~J. Gilman,
\newblock Phys.Rev. {\bf D32}, 3010 (1985).

\bibitem{Glashow:1987qe}
S.~L. Glashow and E.~E. Jenkins,
\newblock Phys.Lett. {\bf B196}, 233 (1987).

\bibitem{Geng:1988bq}
C.~Geng and J.~N. Ng,
\newblock Phys.Rev. {\bf D38}, 2857 (1988).

\bibitem{Urban:1997gw}
J.~Urban, F.~Krauss, U.~Jentschura, and G.~Soff,
\newblock Nucl.Phys. {\bf B523}, 40 (1998), arXiv:hep-ph/9710245.

\bibitem{Deschamps:2009rh}
O.~Deschamps {\em et~al.},
\newblock Phys. Rev. {\bf D82}, 073012 (2010), arXiv:0907.5135.

\bibitem{Gorbahn:2015gxa}
M.~Gorbahn, J.~M. No, and V.~Sanz,
\newblock JHEP {\bf 10}, 036 (2015), arXiv:1502.07352.

\bibitem{Cheung:2003pw}
K.~Cheung and O.~C. Kong,
\newblock Phys.Rev. {\bf D68}, 053003 (2003), arXiv:hep-ph/0302111.

\bibitem{Chang:2000ii}
D.~Chang, W.-F. Chang, C.-H. Chou, and W.-Y. Keung,
\newblock Phys.Rev. {\bf D63}, 091301 (2001), arXiv:hep-ph/0009292.

\bibitem{Krawczyk:2002df}
M.~Krawczyk,
\newblock Acta Phys.Polon. {\bf B33}, 2621 (2002), arXiv:hep-ph/0208076,
\newblock Dedicated to Stefan Pokorski on the occasion of his 60th birthday.

\bibitem{Regan:2002ta}
B.~Regan, E.~Commins, C.~Schmidt, and D.~DeMille,
\newblock Phys.Rev.Lett. {\bf 88}, 071805 (2002).

\bibitem{Pilaftsis:2002fe}
A.~Pilaftsis,
\newblock Nucl.Phys. {\bf B644}, 263 (2002), arXiv:hep-ph/0207277.

\bibitem{Baron:2013eja}
ACME, J.~Baron {\em et~al.},
\newblock Science {\bf 343}, 269 (2014), arXiv:1310.7534.

\bibitem{Barr:1990vd}
S.~M. Barr and A.~Zee,
\newblock Phys.Rev.Lett. {\bf 65}, 21 (1990).

\bibitem{Flacher:2008zq}
H.~Flacher {\em et~al.},
\newblock Eur. Phys. J. {\bf C60}, 543 (2009), arXiv:0811.0009,
\newblock [Erratum: Eur. Phys. J.C71,1718(2011)].

\bibitem{Bona:2009cj}
UTfit, M.~Bona {\em et~al.},
\newblock Phys. Lett. {\bf B687}, 61 (2010), arXiv:0908.3470.

\bibitem{Enomoto:2015wbn}
T.~Enomoto and R.~Watanabe,
\newblock JHEP {\bf 05}, 002 (2016), arXiv:1511.05066.

\bibitem{Denner:1991ie}
A.~Denner, R.~Guth, W.~Hollik, and J.~H. Kuhn,
\newblock Z.Phys. {\bf C51}, 695 (1991).

\bibitem{:2009zh}
D0 Collaboration, V.~Abazov {\em et~al.},
\newblock Phys.Lett. {\bf B682}, 278 (2009), arXiv:0908.1811.

\bibitem{Logan:2010ag}
H.~E. Logan and D.~MacLennan,
\newblock Phys.Rev. {\bf D81}, 075016 (2010), arXiv:1002.4916.

\bibitem{Aaltonen:2009ke}
CDF Collaboration, T.~Aaltonen {\em et~al.},
\newblock Phys.Rev.Lett. {\bf 103}, 101803 (2009), arXiv:0907.1269.

\bibitem{Abazov:2008rn}
D0 Collaboration, V.~Abazov {\em et~al.},
\newblock Phys.Rev.Lett. {\bf 102}, 191802 (2009), arXiv:0807.0859.

\bibitem{Khachatryan:2015uua}
CMS, V.~Khachatryan {\em et~al.},
\newblock JHEP {\bf 12}, 178 (2015), arXiv:1510.04252.

\bibitem{Khachatryan:2015qxa}
CMS, V.~Khachatryan {\em et~al.},
\newblock JHEP {\bf 11}, 018 (2015), arXiv:1508.07774.

\bibitem{Aad:2013hla}
ATLAS, G.~Aad {\em et~al.},
\newblock Eur. Phys. J. {\bf C73}, 2465 (2013), arXiv:1302.3694.

\bibitem{Aad:2014kga}
ATLAS, G.~Aad {\em et~al.},
\newblock JHEP {\bf 03}, 088 (2015), arXiv:1412.6663.

\bibitem{Aad:2015typ}
ATLAS, G.~Aad {\em et~al.},
\newblock JHEP {\bf 03}, 127 (2016), arXiv:1512.03704.

\bibitem{Chatrchyan:2012vca}
CMS, S.~Chatrchyan {\em et~al.},
\newblock JHEP {\bf 07}, 143 (2012), arXiv:1205.5736.

\bibitem{Dorsch:2016tab}
G.~C. Dorsch, S.~J. Huber, K.~Mimasu, and J.~M. No,
\newblock (2016), arXiv:1601.04545.

\bibitem{Aad:2015pla}
ATLAS, G.~Aad {\em et~al.},
\newblock JHEP {\bf 11}, 206 (2015), arXiv:1509.00672.

\bibitem{ATLAS:2012ae}
ATLAS Collaboration, G.~Aad {\em et~al.},
\newblock Phys.Lett. {\bf B710}, 49 (2012), arXiv:1202.1408.

\bibitem{Chatrchyan:2012tx}
CMS, S.~Chatrchyan {\em et~al.},
\newblock Phys. Lett. {\bf B710}, 26 (2012), arXiv:1202.1488.

\bibitem{Aad:2015gba}
ATLAS, G.~Aad {\em et~al.},
\newblock Eur. Phys. J. {\bf C76}, 6 (2016), arXiv:1507.04548.

\bibitem{Aoki:2011wd}
M.~Aoki {\em et~al.},
\newblock Phys.Rev. {\bf D84}, 055028 (2011), arXiv:1104.3178.

\bibitem{Arhrib:2009hc}
A.~Arhrib, R.~Benbrik, C.-H. Chen, R.~Guedes, and R.~Santos,
\newblock JHEP {\bf 0908}, 035 (2009), arXiv:0906.0387.

\bibitem{Moretti:2000yg}
S.~Moretti,
\newblock Phys.Lett. {\bf B481}, 49 (2000), arXiv:hep-ph/0003178.

\bibitem{Ghosh:2004wr}
D.~K. Ghosh and S.~Moretti,
\newblock Eur.Phys.J. {\bf C42}, 341 (2005), arXiv:hep-ph/0412365.

\bibitem{Enberg:2014pua}
R.~Enberg, W.~Klemm, S.~Moretti, S.~Munir, and G.~Wouda,
\newblock Nucl. Phys. {\bf B893}, 420 (2015), arXiv:1412.5814.

\bibitem{Moretti:2016jkp}
S.~Moretti, R.~Santos, and P.~Sharma,
\newblock Phys. Lett. {\bf B760}, 697 (2016), arXiv:1604.04965.

\bibitem{Dev:2014yca}
P.~S. Bhupal~Dev and A.~Pilaftsis,
\newblock JHEP {\bf 12}, 024 (2014), arXiv:1408.3405,
\newblock [Erratum: JHEP11,147(2015)].

\bibitem{Lester:1999tx}
C.~G. Lester and D.~J. Summers,
\newblock Phys. Lett. {\bf B463}, 99 (1999), arXiv:hep-ph/9906349.

\bibitem{Allanach:2006fy}
B.~C. Allanach {\em et~al.},
\newblock {Les Houches physics at TeV colliders 2005 beyond the standard model
  working group: Summary report},
\newblock in {\em {Physics at TeV colliders. Proceedings, Workshop, Les
  Houches, France, May 2-20, 2005}}, 2006, arXiv:hep-ph/0602198.

\bibitem{Godbole:2006tq}
R.~M. Godbole, S.~D. Rindani, and R.~K. Singh,
\newblock JHEP {\bf 12}, 021 (2006), arXiv:hep-ph/0605100.

\bibitem{Huitu:2010ad}
K.~Huitu, S.~Kumar~Rai, K.~Rao, S.~D. Rindani, and P.~Sharma,
\newblock JHEP {\bf 04}, 026 (2011), arXiv:1012.0527.

\bibitem{Godbole:2011vw}
R.~M. Godbole, L.~Hartgring, I.~Niessen, and C.~D. White,
\newblock JHEP {\bf 01}, 011 (2012), arXiv:1111.0759.

\bibitem{Rindani:2013mqa}
S.~D. Rindani, R.~Santos, and P.~Sharma,
\newblock JHEP {\bf 11}, 188 (2013), arXiv:1307.1158.

\bibitem{Haber:2015pua}
H.~E. Haber and O.~Stål,
\newblock Eur. Phys. J. {\bf C75}, 491 (2015), arXiv:1507.04281.

\bibitem{Kling:2016opi}
F.~Kling, J.~M. No, and S.~Su,
\newblock JHEP {\bf 09}, 093 (2016), arXiv:1604.01406.

\bibitem{Coleppa:2014cca}
B.~Coleppa, F.~Kling, and S.~Su,
\newblock JHEP {\bf 12}, 148 (2014), arXiv:1408.4119.

\bibitem{Kling:2015uba}
F.~Kling, A.~Pyarelal, and S.~Su,
\newblock JHEP {\bf 11}, 051 (2015), arXiv:1504.06624.

\bibitem{Arhrib:2015gra}
A.~Arhrib, K.~Cheung, J.~S. Lee, and C.-T. Lu,
\newblock JHEP {\bf 05}, 093 (2016), arXiv:1509.00978.

\bibitem{Cree:2011uy}
G.~Cree and H.~E. Logan,
\newblock Phys.Rev. {\bf D84}, 055021 (2011), arXiv:1106.4039.

\bibitem{Trott:2010iz}
M.~Trott and M.~B. Wise,
\newblock JHEP {\bf 1011}, 157 (2010), arXiv:1009.2813.

\bibitem{Akeroyd:2016ssd}
A.~G. Akeroyd, S.~Moretti, K.~Yagyu, and E.~Yildirim,
\newblock (2016), arXiv:1605.05881.

\bibitem{Akeroyd:1998dt}
A.~Akeroyd,
\newblock Nucl.Phys. {\bf B544}, 557 (1999), arXiv:hep-ph/9806337.

\bibitem{Akeroyd:2012yg}
A.~Akeroyd, S.~Moretti, and J.~Hernandez-Sanchez,
\newblock Phys.Rev. {\bf D85}, 115002 (2012), arXiv:1203.5769.

\bibitem{Patt:2006fw}
B.~Patt and F.~Wilczek,
\newblock (2006), arXiv:hep-ph/0605188.

\bibitem{Pierce:2007ut}
A.~Pierce and J.~Thaler,
\newblock JHEP {\bf 08}, 026 (2007), arXiv:hep-ph/0703056.

\bibitem{Ma:2006km}
E.~Ma,
\newblock Phys.Rev. {\bf D73}, 077301 (2006), arXiv:hep-ph/0601225.

\bibitem{Barbieri:2006dq}
R.~Barbieri, L.~J. Hall, and V.~S. Rychkov,
\newblock Phys.Rev. {\bf D74}, 015007 (2006), arXiv:hep-ph/0603188.

\bibitem{LopezHonorez:2006gr}
L.~Lopez~Honorez, E.~Nezri, J.~F. Oliver, and M.~H. Tytgat,
\newblock JCAP {\bf 0702}, 028 (2007), arXiv:hep-ph/0612275.

\bibitem{Grzadkowski:2009bt}
B.~Grzadkowski, O.~Ogreid, and P.~Osland,
\newblock Phys.Rev. {\bf D80}, 055013 (2009), arXiv:0904.2173.

\bibitem{Grzadkowski:2010au}
B.~Grzadkowski, O.~Ogreid, P.~Osland, A.~Pukhov, and M.~Purmohammadi,
\newblock JHEP {\bf 1106}, 003 (2011), arXiv:1012.4680.

\bibitem{Keus:2014jha}
V.~Keus, S.~F. King, S.~Moretti, and D.~Sokolowska,
\newblock JHEP {\bf 11}, 016 (2014), arXiv:1407.7859.

\bibitem{Keus:2015xya}
V.~Keus, S.~F. King, S.~Moretti, and D.~Sokolowska,
\newblock JHEP {\bf 11}, 003 (2015), arXiv:1507.08433.

\bibitem{Bonilla:2014xba}
C.~Bonilla, D.~Sokolowska, N.~Darvishi, J.~L. Diaz-Cruz, and M.~Krawczyk,
\newblock J. Phys. {\bf G43}, 065001 (2016), arXiv:1412.8730.

\bibitem{McDonald:1993ex}
J.~McDonald,
\newblock Phys.Rev. {\bf D50}, 3637 (1994), arXiv:hep-ph/0702143.

\bibitem{Burgess:2000yq}
C.~Burgess, M.~Pospelov, and T.~ter Veldhuis,
\newblock Nucl.Phys. {\bf B619}, 709 (2001), arXiv:hep-ph/0011335.

\bibitem{Barger:2007im}
V.~Barger, P.~Langacker, M.~McCaskey, M.~J. Ramsey-Musolf, and G.~Shaughnessy,
\newblock Phys.Rev. {\bf D77}, 035005 (2008), arXiv:0706.4311.

\bibitem{Barger:2008jx}
V.~Barger, P.~Langacker, M.~McCaskey, M.~Ramsey-Musolf, and G.~Shaughnessy,
\newblock Phys.Rev. {\bf D79}, 015018 (2009), arXiv:0811.0393.

\bibitem{Kadastik:2009dj}
M.~Kadastik, K.~Kannike, and M.~Raidal,
\newblock Phys.Rev. {\bf D81}, 015002 (2010), arXiv:0903.2475.

\bibitem{Kadastik:2009cu}
M.~Kadastik, K.~Kannike, and M.~Raidal,
\newblock Phys.Rev. {\bf D80}, 085020 (2009), arXiv:0907.1894.

\bibitem{Fritzsch:1974nn}
H.~Fritzsch and P.~Minkowski,
\newblock Annals Phys. {\bf 93}, 193 (1975).

\bibitem{Krauss:1988zc}
L.~M. Krauss and F.~Wilczek,
\newblock Phys.Rev.Lett. {\bf 62}, 1221 (1989).

\bibitem{Martin:1992mq}
S.~P. Martin,
\newblock Phys.Rev. {\bf D46}, 2769 (1992), arXiv:hep-ph/9207218.

\bibitem{DeMontigny:1993gy}
M.~De~Montigny and M.~Masip,
\newblock Phys.Rev. {\bf D49}, 3734 (1994), arXiv:hep-ph/9309312.

\bibitem{Cao:2007rm}
Q.-H. Cao, E.~Ma, and G.~Rajasekaran,
\newblock Phys.Rev. {\bf D76}, 095011 (2007), arXiv:0708.2939.

\bibitem{Dolle:2009ft}
E.~Dolle, X.~Miao, S.~Su, and B.~Thomas,
\newblock Phys.Rev. {\bf D81}, 035003 (2010), arXiv:0909.3094.

\bibitem{Huitu:2010uc}
K.~Huitu, K.~Kannike, A.~Racioppi, and M.~Raidal,
\newblock JHEP {\bf 1101}, 010 (2011), arXiv:1005.4409.

\bibitem{Osland:2013sla}
P.~Osland, A.~Pukhov, G.~Pruna, and M.~Purmohammadi,
\newblock JHEP {\bf 1304}, 040 (2013), arXiv:1302.3713.

\bibitem{Ilnicka:2015jba}
A.~Ilnicka, M.~Krawczyk, and T.~Robens,
\newblock Phys. Rev. {\bf D93}, 055026 (2016), arXiv:1508.01671.

\bibitem{Arhrib:2013ela}
A.~Arhrib, Y.-L.~S. Tsai, Q.~Yuan, and T.-C. Yuan,
\newblock JCAP {\bf 1406}, 030 (2014), arXiv:1310.0358.

\bibitem{Keus:2014isa}
V.~Keus, S.~F. King, and S.~Moretti,
\newblock Phys. Rev. {\bf D90}, 075015 (2014), arXiv:1408.0796.

\bibitem{Moretti:2015tva}
S.~Moretti, D.~Rojas, and K.~Yagyu,
\newblock JHEP {\bf 08}, 116 (2015), arXiv:1504.06432.

\bibitem{Moretti:2015cwa}
S.~Moretti and K.~Yagyu,
\newblock Phys. Rev. {\bf D91}, 055022 (2015), arXiv:1501.06544.

\bibitem{Minkowski:1977sc}
P.~Minkowski,
\newblock Phys.Lett. {\bf B67}, 421 (1977).

\bibitem{Yanagida:1979as}
T.~Yanagida,
\newblock Conf.Proc. {\bf C7902131}, 95 (1979).

\bibitem{GellMann:1980vs}
M.~Gell-Mann, P.~Ramond, and R.~Slansky,
\newblock Conf.Proc. {\bf C790927}, 315 (1979),
\newblock To be published in Supergravity, P. van Nieuwenhuizen \& D.Z.
  Freedman (eds.), North Holland Publ. Co., 1979.

\bibitem{Glashow:1979nm}
S.~Glashow,
\newblock NATO Adv.Study Inst.Ser.B Phys. {\bf 59}, 687 (1980),
\newblock Preliminary version given at Colloquium in Honor of A. Visconti,
  Marseille-Luminy Univ., Jul 1979.

\bibitem{Mohapatra:1979ia}
R.~N. Mohapatra and G.~Senjanovic,
\newblock Phys.Rev.Lett. {\bf 44}, 912 (1980).

\bibitem{Fukugita:1986hr}
M.~Fukugita and T.~Yanagida,
\newblock Phys.Lett. {\bf B174}, 45 (1986).

\bibitem{Coleman:1973jx}
S.~R. Coleman and E.~J. Weinberg,
\newblock Phys.Rev. {\bf D7}, 1888 (1973).

\bibitem{Hambye:2007vf}
T.~Hambye and M.~H. Tytgat,
\newblock Phys.Lett. {\bf B659}, 651 (2008), arXiv:0707.0633.

\bibitem{Kadastik:2009ca}
M.~Kadastik, K.~Kannike, A.~Racioppi, and M.~Raidal,
\newblock Phys.Rev.Lett. {\bf 104}, 201301 (2010), arXiv:0912.2729.

\bibitem{Kadastik:2011aa}
M.~Kadastik, K.~Kannike, A.~Racioppi, and M.~Raidal,
\newblock JHEP {\bf 1205}, 061 (2012), arXiv:1112.3647.

\bibitem{Kadastik:2009gx}
M.~Kadastik, K.~Kannike, A.~Racioppi, and M.~Raidal,
\newblock Phys.Lett. {\bf B694}, 242 (2010), arXiv:0912.3797.

\bibitem{Gustafsson:2012aj}
M.~Gustafsson, S.~Rydbeck, L.~Lopez-Honorez, and E.~Lundstrom,
\newblock Phys. Rev. {\bf D86}, 075019 (2012), arXiv:1206.6316.

\bibitem{Belanger:2015kga}
G.~Belanger {\em et~al.},
\newblock Phys. Rev. {\bf D91}, 115011 (2015), arXiv:1503.07367.

\bibitem{Poulose:2016lvz}
P.~Poulose, S.~Sahoo, and K.~Sridhar,
\newblock Phys. Lett. {\bf B765}, 300 (2017), arXiv:1604.03045.

\bibitem{Cheung:2002gd}
K.~Cheung and D.~K. Ghosh,
\newblock JHEP {\bf 11}, 048 (2002), arXiv:hep-ph/0208254.

\bibitem{Aad:2015nfa}
ATLAS, G.~Aad {\em et~al.},
\newblock Phys. Rev. Lett. {\bf 114}, 231801 (2015), arXiv:1503.04233.

\bibitem{Georgi:1985nv}
H.~Georgi and M.~Machacek,
\newblock Nucl. Phys. {\bf B262}, 463 (1985).

\bibitem{Buras:2010mh}
A.~J. Buras, M.~V. Carlucci, S.~Gori, and G.~Isidori,
\newblock JHEP {\bf 1010}, 009 (2010), arXiv:1005.5310.

\bibitem{Barger:1989fj}
V.~D. Barger, J.~Hewett, and R.~Phillips,
\newblock Phys.Rev. {\bf D41}, 3421 (1990).

\bibitem{Atwood:1996vj}
D.~Atwood, L.~Reina, and A.~Soni,
\newblock Phys.Rev. {\bf D55}, 3156 (1997), arXiv:hep-ph/9609279.

\bibitem{Aoki:2009ha}
M.~Aoki, S.~Kanemura, K.~Tsumura, and K.~Yagyu,
\newblock Phys.Rev. {\bf D80}, 015017 (2009), arXiv:0902.4665.

\bibitem{Hall:1981bc}
L.~J. Hall and M.~B. Wise,
\newblock Nucl.Phys. {\bf B187}, 397 (1981).

\bibitem{Barnett:1983mm}
R.~Barnett, G.~Senjanovic, L.~Wolfenstein, and D.~Wyler,
\newblock Phys.Lett. {\bf B136}, 191 (1984).

\bibitem{Pich:2009sp}
A.~Pich and P.~Tuzon,
\newblock Phys.Rev. {\bf D80}, 091702 (2009), arXiv:0908.1554.

\bibitem{Tuzon:2010vt}
P.~Tuzon and A.~Pich,
\newblock Acta Phys.Polon.Supp. {\bf 3}, 215 (2010), arXiv:1001.0293.

\bibitem{D'Ambrosio:2002ex}
G.~D'Ambrosio, G.~Giudice, G.~Isidori, and A.~Strumia,
\newblock Nucl.Phys. {\bf B645}, 155 (2002), arXiv:hep-ph/0207036.

\bibitem{Giudice:2008uua}
G.~F. Giudice and O.~Lebedev,
\newblock Phys.Lett. {\bf B665}, 79 (2008), arXiv:0804.1753.

\bibitem{Agashe:2009di}
K.~Agashe and R.~Contino,
\newblock Phys.Rev. {\bf D80}, 075016 (2009), arXiv:0906.1542.

\bibitem{Azatov:2009na}
A.~Azatov, M.~Toharia, and L.~Zhu,
\newblock Phys.Rev. {\bf D80}, 035016 (2009), arXiv:0906.1990.

\end{thebibliography}
\end{document}